\documentclass[fleqn,usenatbib]{mnras}

\usepackage{newtxtext,newtxmath}

\usepackage[T1]{fontenc}
\usepackage{ae,aecompl}

\usepackage{graphicx}	

\newcommand{\qEul}{{\it qEul}}
\newcommand{\qEulDiez}{{\it \textcolor{qEul10}{qEul10}}}
\newcommand{\qEulVeinte}{{\it \textcolor{qEul20}{qEul20}}}

\newcommand{\qLag}{{\it qLag}}
\newcommand{\qLweak}{{\it qLweak}}
\newcommand{\qLagDiez}{{\it \textcolor{qLag10}{qLag10}}}
\newcommand{\qLagVeinte}{{\it \textcolor{qLag20}{qLag20}}}

\newcommand{\qLweakDiez}{{\it \textcolor{qLweak10}{qLweak10}}}
\newcommand{\qLweakVeinte}{{\it \textcolor{qLweak20}{qLweak20}}}

\newcommand{\K}{\mathrm{K}}
\newcommand{\erg}{\mathrm{erg}}
\newcommand{\s}{\mathrm{s}}
\newcommand{\g}{\mathrm{g}}
\newcommand{\cm}{\mathrm{cm}}
\newcommand{\pc}{\mathrm{pc}}
\newcommand{\kpc}{\mathrm{kpc}}
\newcommand{\muG}{\mu\mathrm{G}}
\newcommand{\Msun}{\mathrm{M}_\odot}
\newcommand{\galregion}{$r_\text{gal} < 0.2\,r_\text{vir}$}

\newcommand{\Dres}{{\Delta x}_\text{max}}
\newcommand{\Cmag}{\mathcal{C}_\text{mag}}
\newcommand{\emag}{\varepsilon_\text{mag}}

\title[Turbulent dynamo amplification in galaxies]{Towards convergence of turbulent dynamo amplification in cosmological simulations of galaxies}

\author[Martin-Alvarez et al.]{
Sergio Martin-Alvarez,$^{1}$\thanks{E-mail: \href{mailto:smartin@ast.cam.ac.uk}{smartin@ast.cam.ac.uk} (SMA)}
Julien Devriendt,$^{2,3}$
Adrianne Slyz,$^{2}$ \newauthor
Debora Sijacki,$^{1}$
Mark L.A. Richardson,$^{4,5,2}$ and
Harley Katz$^{2}$\\
$^{1}$Institute of Astronomy and Kavli Institute for Cosmology, University of Cambridge, Madingley Road, Cambridge CB3 0HA, UK\\
$^{2}$Subdepartment of Astrophysics, University of Oxford, Keble Road, Oxford, OX1 3RH, UK\\
$^{3}$Observatoire de Lyon, UMR 5574, 9 avenue Charles Andr\'e, F-69561 Saint Genis Laval, France\\
$^{4}$Arthur B. McDonald Canadian Astroparticle Physics Research Institute, 64 Bader Lane, Kingston, ON, Canada,K7L 3N6\\
$^{5}$Department of Physics, Engineering Physics, and Astronomy, Queen's University, 64 Bader Lane, Kingston, ON, Canada, K7L 3N6\\
}

\date{Accepted XXX. Received YYY; in original form ZZZ}

\pubyear{2021}

\begin{document}
\label{firstpage}
\pagerange{\pageref{firstpage}--\pageref{lastpage}}
\maketitle

\definecolor{qEul10}{rgb}{0.704333, 0.0814833, 0.0079118}
\definecolor{qEul20}{rgb}{0.871407, 0.206844, 0.0134}
\definecolor{qEul40}{rgb}{0.969963, 0.376081, 0.0322881}
\definecolor{qEul80}{rgb}{0.989988, 0.556646, 0.06421}
\definecolor{qEul160}{rgb}{1., 0.704662, 0.0957656}
\definecolor{qEul320}{rgb}{1., 0.820127, 0.126955}

\definecolor{qLweak10}{rgb}{0.259846, 0.0405081, 0.50186}
\definecolor{qLweak20}{rgb}{0.282353, 0.149725, 0.679062}
\definecolor{qLweak40}{rgb}{0.235431, 0.32765, 0.833291}
\definecolor{qLweak80}{rgb}{0.266694, 0.550462, 0.926485}
\definecolor{qLweak160}{rgb}{0.44557, 0.749452, 0.981622}
\definecolor{qLweak320}{rgb}{0.772061, 0.92462, 0.998703}

\definecolor{qLag10}{rgb}{0.0, 0.295239, 0.0499504}
\definecolor{qLag20}{rgb}{0.0, 0.354287, 0.0599405}
\definecolor{qLag40}{rgb}{0.0, 0.442859, 0.0749256}
\definecolor{qLag80}{rgb}{0.096442, 0.523608, 0.0862034}
\definecolor{qLag160}{rgb}{0.289326, 0.685107, 0.108759}
\definecolor{qLag320}{rgb}{1.0, 0.984375, 0.230411}

\definecolor{reviewcol}{rgb}{0.6,0.1,0.1}

\begin{abstract}
Our understanding of the process through which magnetic fields reached their observed strengths in present-day galaxies remains incomplete. One of the advocated solutions is a turbulent dynamo mechanism that rapidly amplifies weak magnetic field seeds to the order of ${\sim}\mu$G. However, simulating the turbulent dynamo is a very challenging computational task due to the demanding span of spatial scales and the complexity of the required numerical methods. In particular, turbulent velocity and magnetic fields are extremely sensitive to the spatial discretisation of simulated domains. To explore how refinement schemes affect galactic turbulence and amplification of magnetic fields in cosmological simulations, we compare two refinement strategies. A traditional quasi-Lagrangian adaptive mesh refinement approach focusing spatial resolution on dense regions, and a new refinement method that resolves the entire galaxy with a high resolution quasi-uniform grid. Our new refinement strategy yields much faster magnetic energy amplification than the quasi-Lagrangian method, which is also significantly greater than the adiabatic compressional estimate indicating that the extra amplification is produced through stretching of magnetic field lines. Furthermore, with our new refinement the magnetic energy growth factor scales with resolution following $\propto \Dres^{-1/2}$, in much better agreement with small-scale turbulent box simulations. Finally, we find evidence suggesting most magnetic amplification in our simulated galaxies occurs in the warm phase of their interstellar medium, which has a better developed turbulent field with our new refinement strategy.
\end{abstract}

\section{Introduction}
The interstellar medium (ISM) of every galaxy observed in the local Universe is permeated by magnetic fields, with their magnetic energy typically measured to be in approximate equipartition with the thermal and turbulent components \citep{Beck2007,Beck2015}. Due to their considerable contribution to the energy budget, magnetic fields play a fundamental role in regulating the structure and dynamics of the ISM \citep{Iffrig2017,Ji2018} as well as the distribution of gas across its different phases \citep{Kortgen2019}. Furthermore, magnetic fields intervene in other galaxy formation processes such as gas fragmentation \citep{Inoue2019}, star formation \citep[e.g.][]{Padoan2011,Zamora-Aviles2018}, galactic winds \citep{Bendre2015,Gronnow2018,Steinwandel2019}, and can even affect their global size and kinematic properties \citep{Martin-Alvarez2020}.

However, the origin of galactic magnetic fields remains unknown, with two main alternatives prevailing: either an astrophysical or a primordial origin. The first possibility assumes the formation of galaxies with extremely weak magnetic fields. These are then amplified to the $\muG$ strengths measured by observations through a combination of ISM-scale processes such as dynamo amplification \citep{Dubois2010,Gressel2013,Pakmor2014,Martin-Alvarez2018} and/or highly magnetised ejecta from stars and/or AGN \citep{Butsky2017,Vazza2017,KMA2019}. The second alternative is that magnetic fields are of primordial nature. In this scenario strong primordial magnetic fields (PMFs) are produced during the early stages of our Universe \citep[a classic review on the subject is given by][]{Widrow2002}, remaining strong until the post-recombination era. These magnetic fields finally reach the order of ${\sim}\muG$ during the collapse of matter perturbations at the onset of galaxy formation \citep[$z \gtrsim 10$, ][]{Lesch1995,Kandus2011}. Current observational constraints cannot rule out strong PMFs, but provide a conservative comoving magnetic field strength upper limit of $B_0 < 10^{-9}$~G, obtained by \citet[][]{PlanckCollaboration2015}\footnote{Note how alternative observational probes such as ultra-high energy cosmic rays \citep{Bray2018,Alves-Batista2021} or ultra-faint dwarf galaxies \citep{Safarzadeh2019} may provide more constraining limits, although these still remain on the same approximate order of magnitude.}. Numerical studies offer a promising avenue to further understand the origin and evolution of galactic magnetic fields, for example, by studying the evolution of magnetic fields with different origins \citep{Martin-Alvarez2021,Garaldi2021} and predicting observables upon which strong PMFs may have an impact on, such as galactic properties \citep{Martin-Alvarez2020}, supermassive black hole masses \citep{Pillepich2018a}, the population of galaxies \citep{Marinacci2016} or cosmic reionization \citep{Sanati2020,KMA2021}.

It is worth noting that lower limits for PMFs sufficiently high to accommodate the magnetic fields observed in galaxies would eliminate the need for astrophysical amplification. While some lower limits have been proposed, e.g. based on the secondary GeV emission from blazars ($B_0 \gtrsim 10^{-16}$~G; \citealt{Neronov2010}) various raised caveats to their validity still require further investigation \citep[e.g.][]{Broderick2012,Broderick2018}. Additional skepticism regarding the existence of strong PMFs emerges from theoretical models generally favouring the generation of weaker PMFs $B_0 \lesssim 10^{-20}$~G \citep[see a review by][]{Subramanian2016}. Nonetheless, most of the proposed lower limits still provide seed magnetic fields too weak to directly attain the $\muG$ values expected for galaxies. In conjunction, all of this hints toward an astrophysical origin of magnetic fields in galaxies. One of the most popular mechanisms to amplify such seeds to the observed strengths is the small-scale turbulent dynamo \citep{Beresnyak2019}. This dynamo has two particularly attractive features. Firstly, the ubiquity of turbulence across galaxies, specially at high redshift \citep{ForsterSchreiber2009}, facilitating amplification immediately after galaxy formation \citep{Martin-Alvarez2018}. Secondly, its rapid amplification of the magnetic field, on timescales much shorter than the lifetime of galaxies \citep{Bhat2013}. Turbulent dynamo fast amplification has been demonstrated numerically in turbulent box simulations of the ISM \citep[e.g.][]{Schekochihin2002,Federrath2016}. Additional simulations of a magnetised ISM have also shown that dynamo activity will affect and, in turn, be affected by the properties of the ISM multi-phase medium \citep[e.g.][]{Evirgen2017,Evirgen2019}. However, galaxy formation simulations trying to resolve this dynamo face the virtually impossible task of capturing a representative range of the turbulent and magnetic dynamical spatial scales of interest. Accurately simulating galaxy formation demands accounting for cosmological environmental effects and accretion (i.e. $\sim$Mpc scales) while resolving the galactic ISM below the viscous scale (i.e. sub-pc scales) and ideally down to the resistive scale. However, even convergence of turbulence in the ISM cold phase may require spatial resolutions below 0.1 pc \citep{Kortgen2017}, far beyond the current capabilities of galaxy formation simulations. Thus, present-day galaxy formation simulations cannot yet capture the physical viscosity and diffusivity of the real ISM.

To better understand this mechanism of galactic magnetisation, various studies have reviewed the growth of magnetic fields in isolated \citep{Pakmor2013,Rieder2016,Steinwandel2019} and cosmological zoom-in galaxy simulations \citep{Pakmor2017,Rieder2017b,Martin-Alvarez2018}. In addition to the well-known dependence on resolution, galaxy formation simulations find the growth of the magnetic energy to depend on the stellar feedback employed, with stronger prescriptions providing faster amplification rates, particularly at late times of galaxy evolution \citep[e.g.][]{Rieder2016,Su2017,Martin-Alvarez2018}. Furthermore, a disparity of amplification timescales are found when using different numerical methods. This is somewhat expected, as different methods of solving the magnetic component do not necessarily converge to the same result \citep{Toth2000,Balsara2004}. This disagreement is apparent both for simple and complex problems \citep[see a comparison of different methods by][applied to multiple astrophysical problems]{Hopkins2016}. On one side, Powell divergence cleaning \citep{Powell1999} methods appear to yield fast amplification in galaxy formation simulations \citep{Pakmor2014}, whereas Dedner \citep{Dedner2002} and constrained transport \citep[CT; ][]{Teyssier2006} find lower growth rates  \citep{Wang2009,Rieder2017b,Martin-Alvarez2018}. In its implementation of CT, \citet{Mocz2016} explores how Powell and CT compare for an isolated galaxy. Divergence cleaning (DC) methods have the advantage of not requiring the use of a magnetic vector potential \citep{Mocz2016} or the memory load associated with storing the magnetic field across each cell interface \citep{Teyssier2006}. Instead of strictly fulfilling the divergence constraint $\vec{\nabla} \cdot \vec{B} = 0$, DC relies on maintaining the magnetic divergence under control, with the reported per-cell average ratios of magnetic divergence to magnetic field strength being on the order of $\sim 10^{-1}-10^{-2}$. However, these divergence ratios become locally higher in shocks and turbulent flows, with divergence across resolution element sizes frequently higher than the local magnetic field. Alternatively, CT methods have the advantage of fulfilling $\vec{\nabla} \cdot \vec{B} = 0$ to numerical precision, which avoids spurious modifications of thermodynamical quantities \citep{Toth2000}.
However, CT methods have a significantly higher computational cost and produce considerably lower amplification rates. This is the result of the high numerical diffusivity introduced by the numerical solver used by these methods when discretising the spatial domain. Multi-level refinement strategies in particular introduce artificial numerical diffusivity and viscosity at various scales, each associated to a level of refinement. Consequently, it is crucial to explore magnetic fields in galaxies both through the less resistive DC methods as well as the divergence-less CT methods, and to investigate whether both may ultimately provide similar answers regarding the evolution of magnetic fields in galaxies.

To better understand the shortcomings of amplification in CT galaxy formation simulations we need to analyse not solely the effects of spatial resolution, but also the employed refinement strategies. Adaptive mesh refinement (AMR) has been a revolutionary development for astrophysical simulations in the last decades, allowing these to simultaneously resolve galaxies and their environment. AMR is commonly configured to naturally focus refinement on the densest structures, thus operating in a {\it quasi-Lagrangian} fashion. This approach under-resolves the dynamics of the diffuse gas in order to provide a lower computational cost. As an example, quasi-Lagrangian refinement neglects a significant fraction of turbulence and substructure that naturally occurs in the circumgalactic medium (CGM) when employing a Mach number threshold refinement strategy \citep{Bennett2020}. A natural consequence of such adaptive refinement schemes is that, as resolution is increased to resolve denser structures, multiple dissipation scales are introduced in both the magnetic and kinetic energy cascades. These dampen turbulent and magnetic vector fields at several scales, which further complicates analysis in resolution terms. Similarly, as refined gas traverses refinement boundaries towards lower resolutions, information regarding its small-scale structure is lost and cannot be recovered. Contrarily to this approach, numerical simulations aiming to study small-scale magnetohydrodynamical (MHD) turbulence often employ uniform discretisation of their spatial domain and thus circumvent the aforementioned disadvantages \citep{Ji2018,Schober2018,Evirgen2019}.

In this work we study the kinematic turbulence and turbulent magnetic field amplification in high-resolution magnetohydrodynamical (MHD) cosmological zoom-in simulations of a Milky Way-like galaxy. We contrast the quasi-Lagrangian AMR method with a new refinement strategy which resolves the majority of the galactic volume using a fixed spatial resolution, a.k.a {\it quasi-uniform} or {\it quasi-Eulerian} refinement. This reproduces an almost Eulerian picture in which the region of interest is resolved with a (virtually) uniform grid irrespectively of the Lagrangian behaviour of the ideal MHD gas flow. We describe our new refinement strategy and the simulation setup in Section~\ref{s:Methods}. Our main results are explored in Section~\ref{s:Results}. Finally, we conclude this manuscript in Section~\ref{s:Conclusions} with a summary of our work.

\section{Numerical methods}
\label{s:Methods}

\begin{figure*}
    \centering
    \includegraphics[width=2.1\columnwidth]{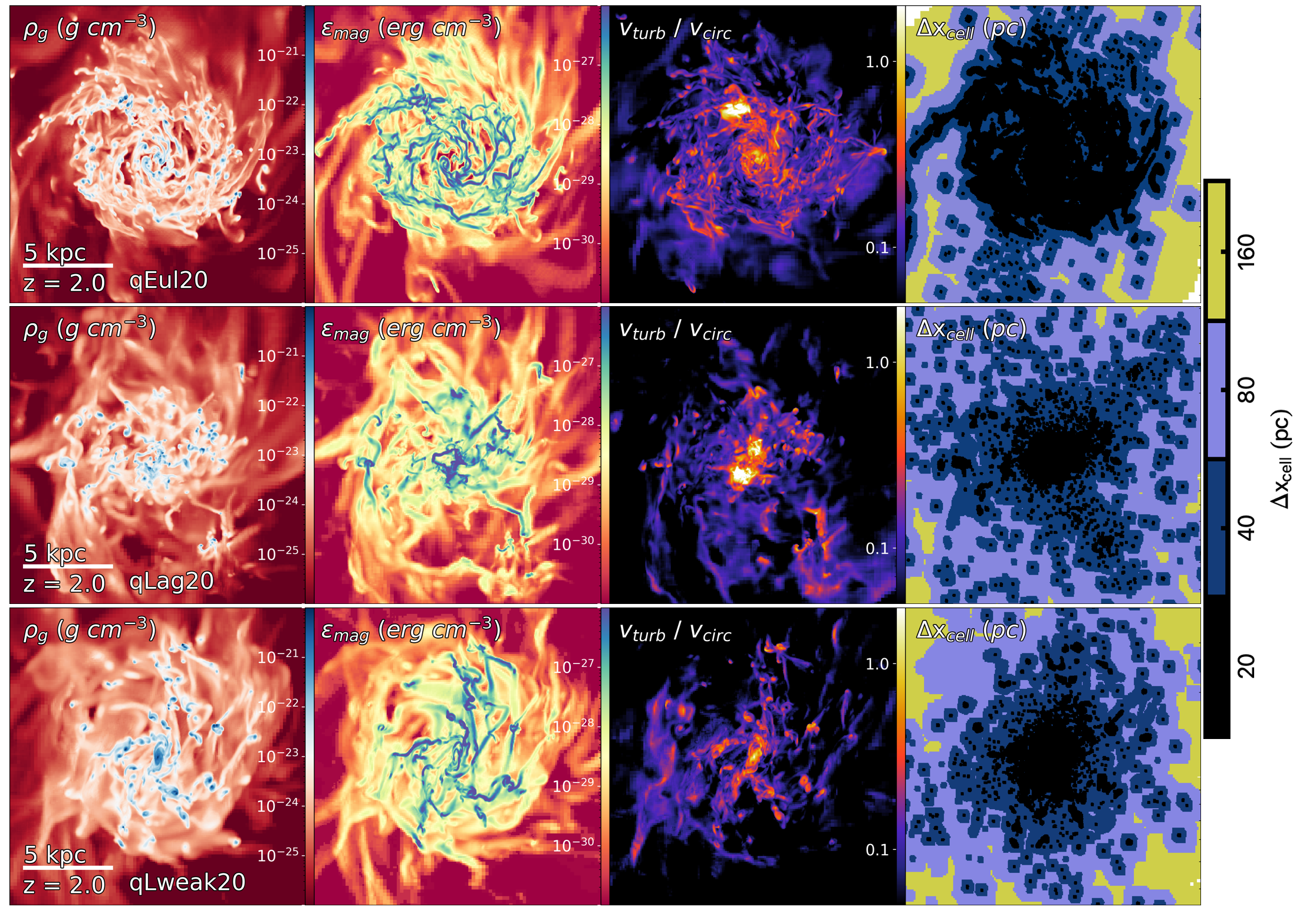}\\
    \caption{Face-on views centred on the galaxy at $z \sim 2$. Panels are gas-density weighted projections of a cube with $16.5$~kpc physical side. The rows compare the quasi-Eulerian refinement run (\qEulVeinte, top) with the two quasi-Lagrangian refinement runs (\qLagVeinte, centre; \qLweakVeinte, bottom), all performed with maximum physical resolution $\Dres = 20\,\pc$. From left to right, these panels show gas density $\rho_\text{gas}$, magnetic energy density $\epsilon_\text{mag}$, ratio of small-scale turbulent velocity $v_\text{turb}$ over $v_\text{circ}$ (see equation~\ref{eq:vCirc}), and maximum cell resolution $\Delta x_\text{cell}$. The computation of small-scale turbulent velocity is described in Section~\ref{ss:SSturbulence}. The \qEulVeinte~simulation appears more diffuse and turbulent in its gas density and turbulent velocity ratio views than the corresponding quasi-Lagrangian strategy simulations. Furthermore, more substructures can be found in its diffuse-warm phase. By $z \sim 2$, the ISM in the \qEulVeinte~also displays a higher magnetic energy density pervading a larger fraction of the galaxy.}
    \label{fig:GalaxyViews}
\end{figure*}

\begin{figure*}
    \centering
    \includegraphics[width=0.667\columnwidth]{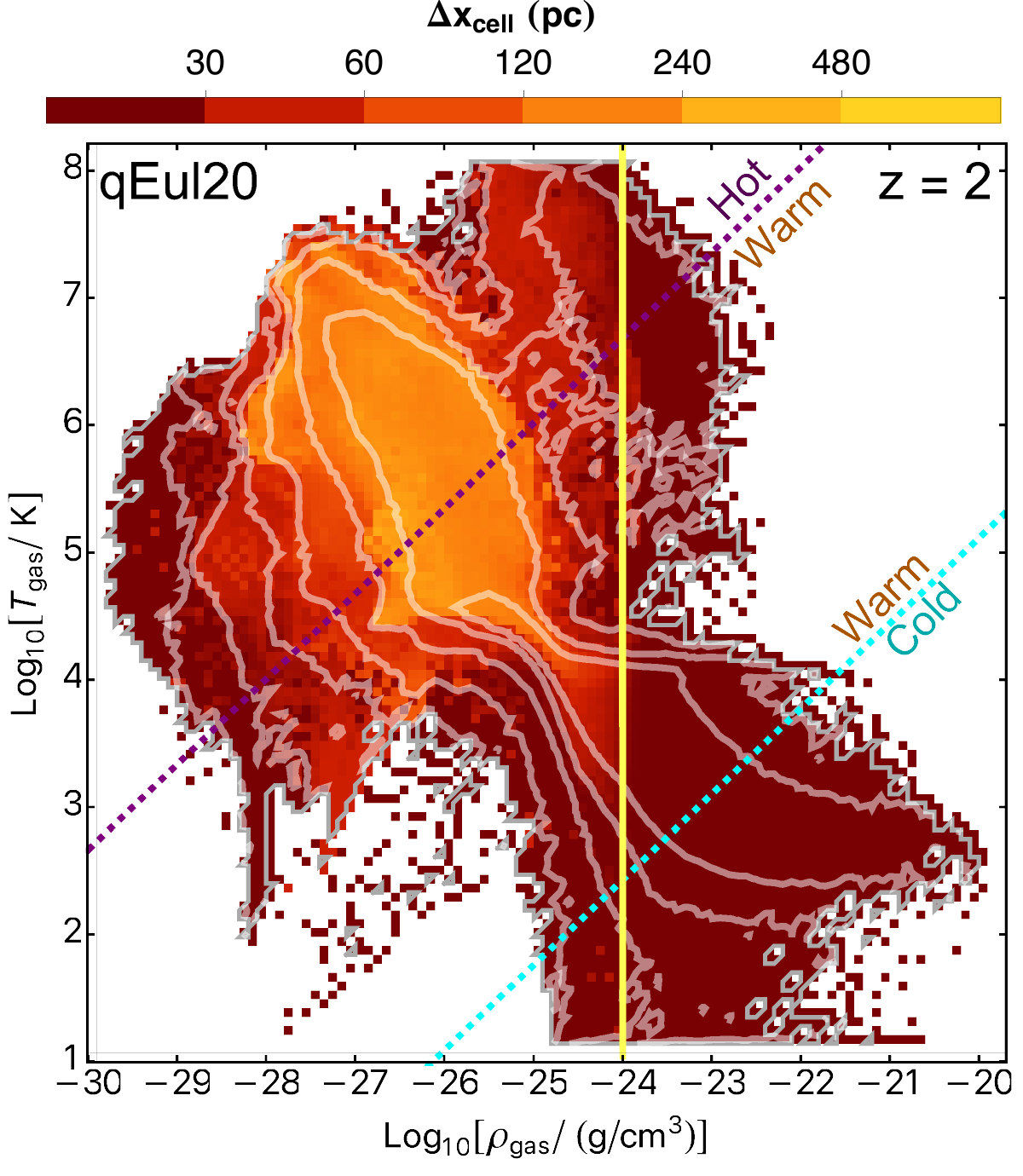}%
    \includegraphics[width=0.667\columnwidth]{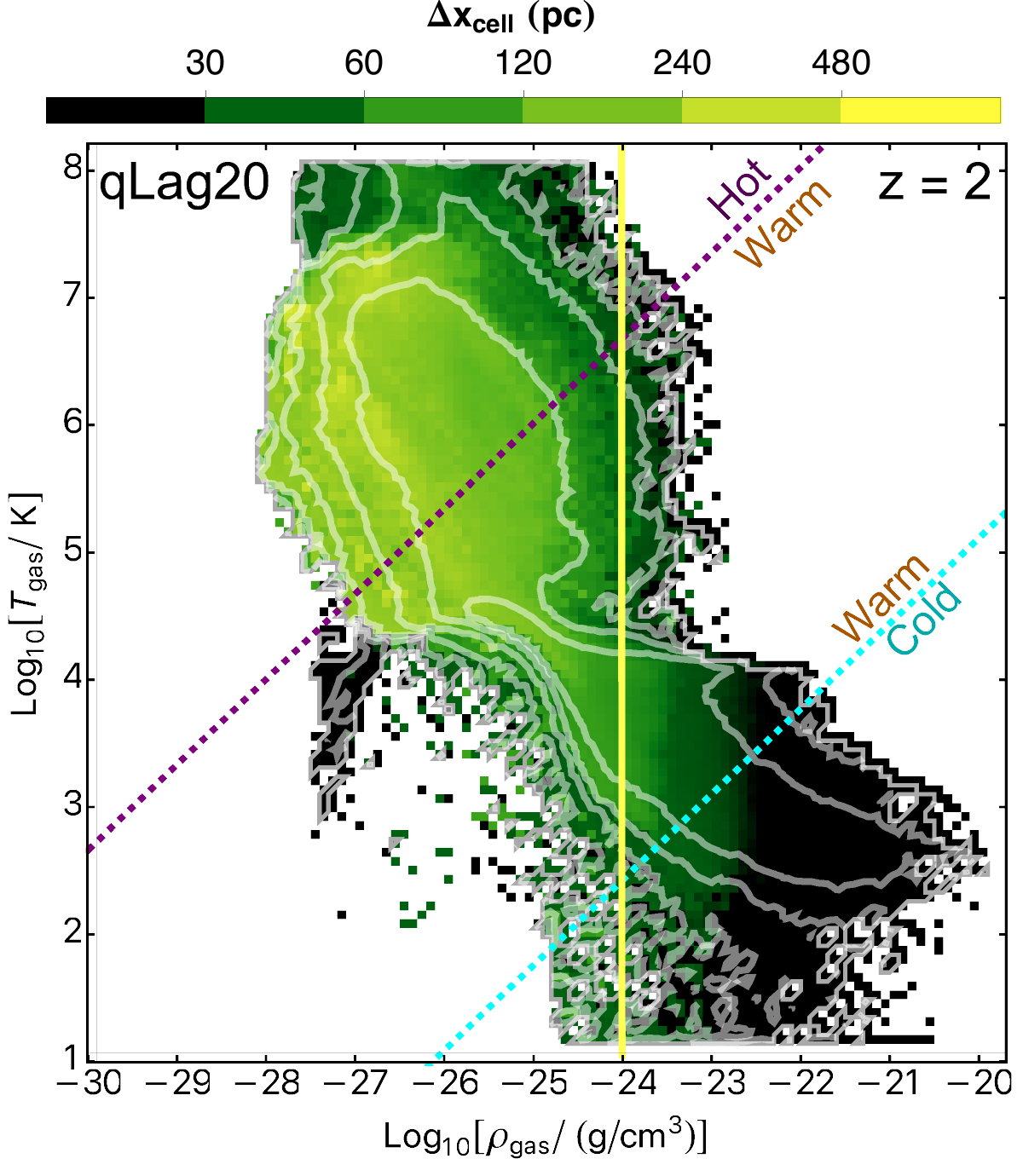}%
    \includegraphics[width=0.667\columnwidth]{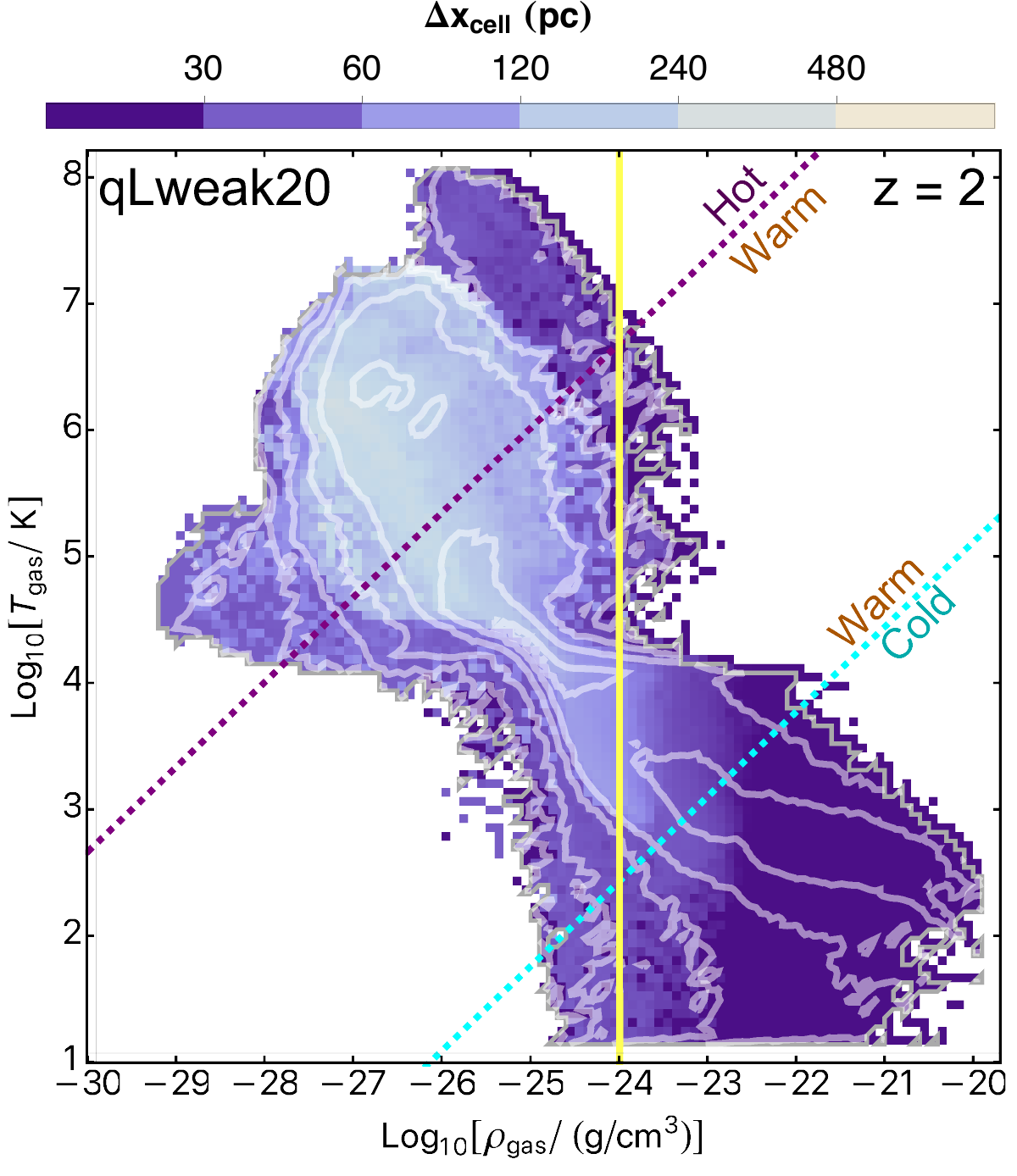}\\
    \caption{Volume-weighted average cell size $\Delta x_\text{cell}$ across the gas temperature ($T_\text{gas}$) - density ($\rho_\text{gas}$) phase space for \qEulVeinte~(left), \qLagVeinte~(centre)  and \qLweakVeinte~(right). Contours correspond to gas mass percentiles. Diagonal lines separate different ISM phases (following the definition described in Section~\ref{s:Results}). The vertical yellow line marks $\rho_\text{gas} = \rho_\text{th}$. The major difference between the two refinement strategies is a larger portion of the warm phase in the \qEul~runs being refined to the highest resolution $\Dres$.}
    \label{fig:PhaseRefine}
\end{figure*}

The MHD simulations studied in this manuscript are generated using our own modified version of the {\sc ramses} code\footnote{\href{https://bitbucket.org/rteyssie/ramses/}{https://bitbucket.org/rteyssie/ramses/}} \citep{Teyssier2002}. {\sc ramses} discretises the simulation domain into an octree AMR grid. Simulations are evolved in time employing an Eulerian solver for the baryonic gas coupled with the N-body components for the dark matter and stellar components through gravity. Magnetic fields are implemented in {\sc ramses} using a CT method that models them as cell face-centred quantities \citep{Teyssier2006, Fromang2006}. As the typical magnetic diffusivity of the interstellar and intergalactic mediums are negligible compared to their numerical counterparts, we set the magnetic diffusivity in the induction equation
\begin{equation}
\frac{\partial \vec{B}}{\partial t} = \vec{\nabla} \times \left(\vec{v} \times \vec{B}\right) + \eta \vec{\nabla}^2 \vec{B}
\label{eq:Induction}
\end{equation}
to $\eta = 0$. As a result of domain discretisation, CT schemes naturally introduce some degree of numerical resistivity in their solution of the induction equation. Consequently, all diffusive effects in our simulations are of numerical nature.

Our initial conditions \citep[{\sc nut},][]{Powell2011} feature a cosmological cubic box with 12.5 comoving Mpc per side with a spherical zoom region of length 4.5 comoving Mpc across. Inside the zoom region, we allow our refinement strategies (presented in Section~\ref{ss:EurRefine}) to resolve the grid down to a maximum physical resolution $\Dres$, which we vary for different runs (our suite of simulations is summarised in Table~\ref{table:TAsetups}). We study a Milky Way-like galaxy forming approximately at the centre of the zoom, with a dark matter halo of virial mass $M_\text{vir} (z = 0) \simeq 5 \cdot 10^{11}\,\Msun$. This galaxy is the same system studied by \citet{Martin-Alvarez2018}. Stellar and dark matter particles have mass resolutions of $M_\text{DM} \simeq 5 \cdot 10^4\,\Msun$ and $M_{*} \simeq 5 \cdot 10^3\,\Msun$, respectively. Finally, our cosmological parameters are selected following the WMAP5 cosmology \citep{Dunkley2009}.

Along the lines of most galaxy formation simulations, we include various sub-grid prescriptions that capture the most important physical processes for galaxy formation. We model reionization as a UV background initiated at $z = 10$ \citep{Haardt1996} as well as metal cooling both above (interpolating {\sc cloudy} tables; \citealt{Ferland1998}) and below \citep{Rosen1995} a temperature of $10^4$~K. During the simulation and subsequent analysis, all gas is assumed ideal and mono-atomic (i.e. with thermodynamical specific heat ratio 5/3).

We model star formation using our magneto-thermo-turbulent star formation prescription \citep{Kimm2017,Trebitsch2017,Martin-Alvarez2020}. This prescription transforms gas at the highest level of refinement \citep{Rasera2006} into stellar particles, only in regions where the gravitational pull exceeds the combined turbulent, magnetic and thermal pressure. Where this requirement is fulfilled, gas is converted into stars following a Schmidt law \citep{Schmidt1959}. We employ a local efficiency computed according to magneto-thermo-turbulent properties of the gas neighbouring the forming stellar particle. The functional form of this efficiency is extracted and calibrated to simulations which model star formation at significantly smaller scales \citep{Padoan2011,Federrath2012}. Stellar particles produce mechanical supernova (SN) stellar feedback \citep[presented in][]{Kimm2014, Kimm2015} assuming a Kroupa initial mass function \citep{Kroupa2001}. Each SN injects back to its host cell a corresponding gas and metal mass fraction of $\eta_\text{SN} = 0.213$ and $\eta_\text{metals} = 0.075$, respectively.

We aim to investigate the dynamo amplification of the magnetic field $\vec{B} (t)$ occurring in the simulated galaxy due to the induction term of the induction equation (equation~\ref{eq:Induction}). Due to the absence of source terms, this equation requires introducing alternative sources or an initially weak seed magnetic field in order to trigger any subsequent dynamical amplification. For the sake of simplicity, and following common practice \citep[some of our previous work or by e.g.][]{Stasyszyn2013, Vazza2014, Pakmor2017}, we initialise the magnetic field as a uniform and homogeneous field. We choose a comoving strength of $B_0 = 3 \cdot 10^{-20}$~G, which is in agreement with a magnetic field typically produced by a Biermann battery \citep{Pudritz1989,Attia2021}. We stress that this magnetic field will not have enough time in our simulations to be amplified to a kinematically important magnetic field. As a result, we will review here the case of highest amplification possible (i.e. no magnetic backreaction) for our numerical configuration. As a word or caution, we do not dismiss the possibility that alternative initial magnetic field configurations (e.g. one following a spectrum of perturbations) may influence magnetic amplification, especially during the early stages of the formation of our galaxy.

\subsection{Quasi-Eulerian refinement} 
\label{ss:EurRefine}

Domain discretisation into finite resolution elements leads the magnetohydrodynamical solver to introduce numerical diffusivity and viscosity in the solution of the magnetised fluid evolution \citep{Teyssier2006}. Multi-level AMR refinement strategies in particular artificially introduce numerical diffusivity and viscosity at various scales, each associated with a new respective level of refinement. A target mass-per-cell {\it quasi-Lagrangian} refinement strategy implies that in the grid constructed by the AMR refinement across the simulated galaxy, different gas phases and portions of the galaxy are discretised using different spatial resolutions depending on the mass distribution of gas, dark matter, and stars. The various resolution levels in our galaxies are shown at $z \sim 2$ in the rightmost panels of Fig.~\ref{fig:GalaxyViews}. As a result of including multiple refinement levels, and thus not employing a spatially uniform resolution, these simulations combine different numerical viscous and resistive scales. These complicate an analysis of the magnetic and turbulent properties of the system, and imply that turbulent and magnetic energy is dissipated significantly more than expected for the $\Dres$ employed. In our quasi-Lagrangian strategy runs (second and third row in Fig.~\ref{fig:GalaxyViews}), we apply the common criterion of flagging a cell for refinement whenever its total mass is higher than that corresponding to 8 dark matter particles.

Due to our interest in turbulence and magnetic field amplification, we implement an alternative refinement strategy that aims to refine the entire galaxy to the target spatial resolution $\Dres$. As a result, the studied galaxy is resolved with an almost fully Eulerian resolution grid, reducing the number of relevant viscous and diffusion scales. We identify regions of interest through a gas density threshold $\rho_\text{th} = 10^{-24}\, \g\, \cm^{-3}$, and enforce refinement to the maximum level (i.e. cell side $\Delta x_\text{cell}$ is $\Delta x_\text{cell} = \Dres$) whenever the gas density of a given cell fulfills $\rho_\text{gas} > \rho_\text{th}$. Note that {\sc ramses} expands the refinement by $n_\text{expand}$ times (where we conserve the default $n_\text{expand}=1$) around cells marked for refinement and requires no refinement discontinuities of more than one grid level per boundary \citep{Teyssier2002}. The selected $\rho_\text{th}$ provides a reasonable compromise between fully resolving the galaxy at $\Delta x_\text{cell}$ and having a pervasive refinement of our computational domain. The galaxy resulting from this new refinement strategy with $\Dres = 20\,\pc$ (\qEulVeinte) is shown in the first row of Fig.~\ref{fig:GalaxyViews} for $z \sim 2$. The central and bottom rows display our standard quasi-Lagrangian AMR refinement runs \qLagVeinte~and \qLweakVeinte, respectively. The rightmost panel of this figure shows how the gas density refinement strategy resolves the majority of the galaxy with the target resolution $\Dres$. Using the fiducial quasi-Lagrangian AMR strategy provides a clumpier cell distribution, with various boundaries between the different levels of refinement, which are well-correlated with the mass distribution of the galaxy. Consequently, the AMR focuses refinement on the cold and dense phase while under-refining the warm and hot phases of the gas. Our new scheme resolves the warm and cold phases within the galaxy with approximately the same resolution. Outside the galaxy, mass clumps are refined using the same quasi-Lagrangian strategy as the simulations shown in the other panels.  

\subsection{Simulation suite} 
\label{ss:Simulations}

\begin{table}
    {\centering
    \caption{Compilation of all simulations studied in this manuscript. From left to right, columns indicate the simulation label, cell size at the maximum level of refinement $\Dres$, refinement strategy, SN feedback employed, $\text{N}^\text{o}$ of active cells ($N_\text{c}$) at $z = 9$ and $z = 2$ with $\Delta x_\text{cell} < 320\,\pc$.}
    \label{table:TAsetups}
    \begin{tabular}{l | l l l l r}
    \hline
    Simulation & $\Dres$ & Refinement & SN & $\text{N}_\text{c}\,(10^6)$\\
    \hline
    {\bf \textcolor{qLweak10}{qLweak10}} &  $10\,\pc$ & \qLag & Weak & 6, 11 \\
    {\bf \textcolor{qLweak20}{qLweak20}} &  $20\,\pc$ & \qLag & Weak & 5, 10 \\
    {\bf \textcolor{qLweak40}{qLweak40}} &  $40\,\pc$ & \qLag & Weak & 4, 9 \\
    {\bf \textcolor{qLweak80}{qLweak80}} &  $80\,\pc$ & \qLag & Weak & 3, 4 \\
    {\bf \textcolor{qLweak160}{qLweak160}} &  $160\,\pc$ & \qLag & Weak & 0.6, 0.5 \\
    {\bf \textcolor{qLweak320}{qLweak320}} &  $320\,\pc$ & \qLag & Weak & 0, 0 \\
    \hline
    {\bf \textcolor{qLag10}{qLag10}} &  $10\,\pc$ & \qLag & Boost & 6, 10 \\
    {\bf \textcolor{qLag20}{qLag20}} &  $20\,\pc$ & \qLag & Boost & 5, 10 \\
    {\bf \textcolor{qLag40}{qLag40}} &  $40\,\pc$ & \qLag & Boost & 4, 6 \\
    {\bf \textcolor{qLag80}{qLag80}} &  $80\,\pc$ & \qLag & Boost & 3, 4 \\
    {\bf \textcolor{qLag160}{qLag160}} &  $160\,\pc$ & \qLag & Boost & 0.6, 0.5 \\
    {\bf \textcolor{qLag320}{qLag320}} &  $320\,\pc$ & \qLag & Boost & 0, 0 \\
    \hline
    {\bf \textcolor{qEul10}{qEul10}}$^\text{\ddag}$ &  $10\,\pc$ & \qLag + \qEul & Boost & 127, {---} \\
    {\bf \textcolor{qEul20}{qEul20}} &  $20\,\pc$ & \qLag + \qEul & Boost & 31, 21 \\
    {\bf \textcolor{qEul40}{qEul40}} &  $40\,\pc$ & \qLag + \qEul & Boost & 9, 8 \\
    {\bf \textcolor{qEul80}{qEul80}} &  $80\,\pc$ & \qLag + \qEul & Boost & 4, 4 \\
    {\bf \textcolor{qEul160}{qEul160}} &  $160\,\pc$ & \qLag + \qEul & Boost & 0.6, 0.5 \\
    {\bf \textcolor{qEul320}{qEul320}} &  $320\,\pc$ & \qLag + \qEul & Boost & 0, 0 \\
    \hline
    \end{tabular}\\}
    \ddag: Due to ifz elevated computational cost, \qEulDiez~is only evolved down to $z \sim 8.5$.\\
\end{table}

All simulations in this manuscript are summarised in Table~\ref{table:TAsetups}. For the low density gas every simulation employs the quasi-Lagrangian, standard AMR refinement strategy. The runs that exclusively employ this mass-targeted AMR refinement are labelled as either \qLag~or \qLweak~in Table~\ref{table:TAsetups} and throughout this work. These two groups of runs employ two different strengths of the stellar feedback: the \qLweak~runs have the fiducial SN feedback specific energy $\varepsilon_\text{SN} = E_\text{SNII} / M_\text{SNII} \sim 10^{50}\, \erg\,\Msun^{-1}$, while the \qLag~runs have a moderately boosted feedback $\varepsilon_\text{SN} \sim 2 \cdot 10^{50}\, \erg\,\Msun^{-1}$. The second subset of simulations includes the previously described density threshold, quasi-Eulerian refinement strategy. These are labelled \qEul~runs and employ the same feedback strength as the \qLag~runs. Depending on their maximal physical spatial resolution reached, $\Dres = X$~pc, runs are named \qLag X, \qLweak X or \qEul X. Our study spans numerical resolutions from 320~pc down to 10~pc. We note that the \qEulDiez~simulation is only evolved down to $z \sim 8.5$ due to its expensive computational cost. We find the additional computational cost of \qEul~runs to be minor for resolutions below $80~pc$. The cost of \qEulVeinte~is approximately double that of \qLagVeinte. Finally, evolving \qEulDiez~to $z \sim 9$ has an extreme cost ($\sim4$ times the cost of running the \qLagVeinte~simulation to $z = 2$). Consequently, we focus our refinement strategy comparison on the $\Dres = 20\,\pc$ runs: \qLweakVeinte~, \qLagVeinte~and \qEulVeinte. Fig.~\ref{fig:PhaseRefine} compares the volume-weighted average cell size across the gas temperature - density phase space for the two different refinement strategies. As we will show in Section~\ref{s:Results}, the most important differences for magnetic amplification emerge from \qEulVeinte~resolving a larger fraction of the warm phase with higher resolution than the AMR runs. We review the magnetic divergence ($\vec{\nabla} \cdot \vec{B} = 0$) for all our simulations in Appendix~\ref{ap:Divergence}.

\begin{figure*}
    \centering
    \includegraphics[width=2.0\columnwidth]{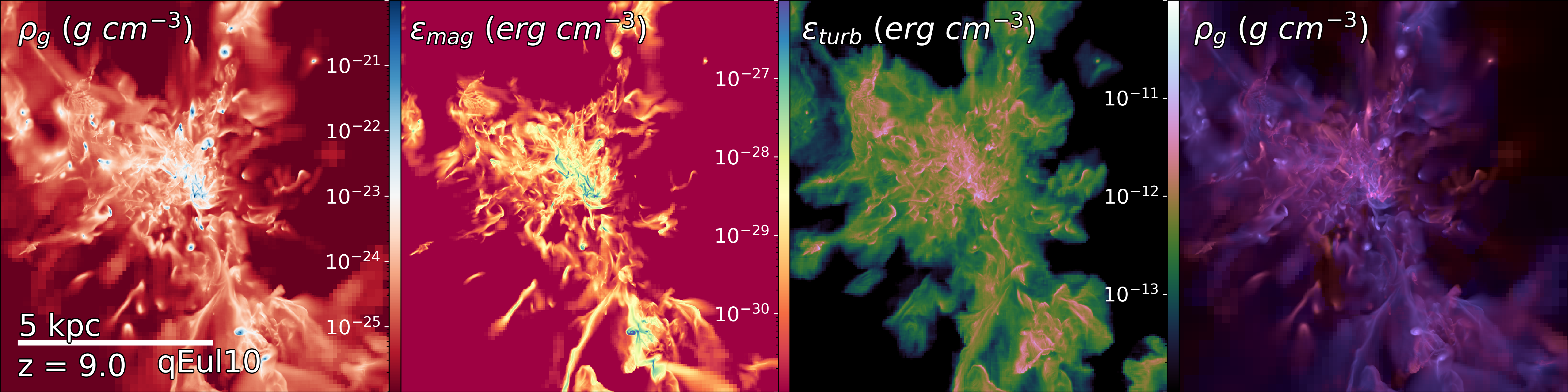}\\
    \includegraphics[width=2.0\columnwidth]{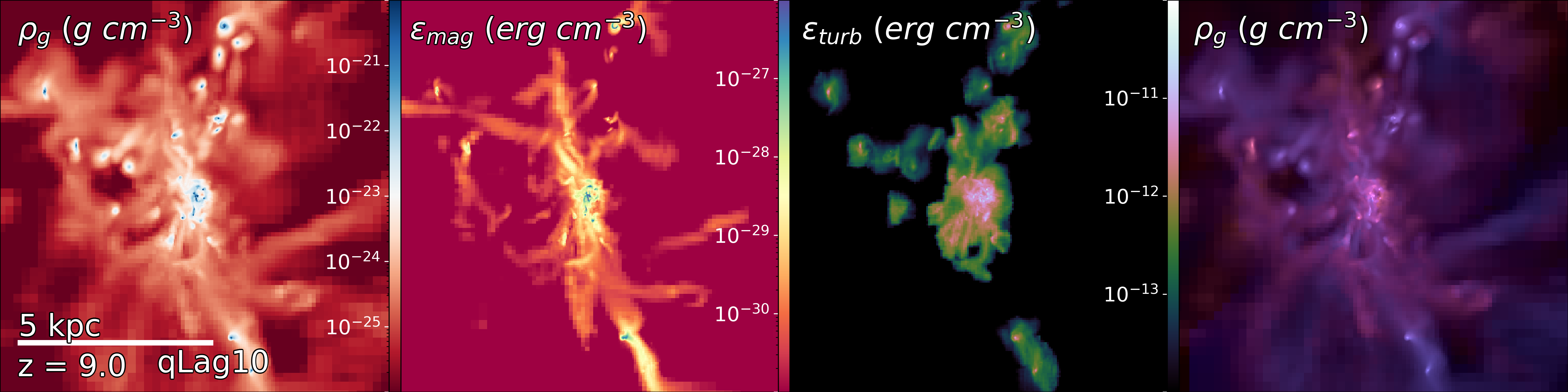}\\
    \includegraphics[width=2.0\columnwidth]{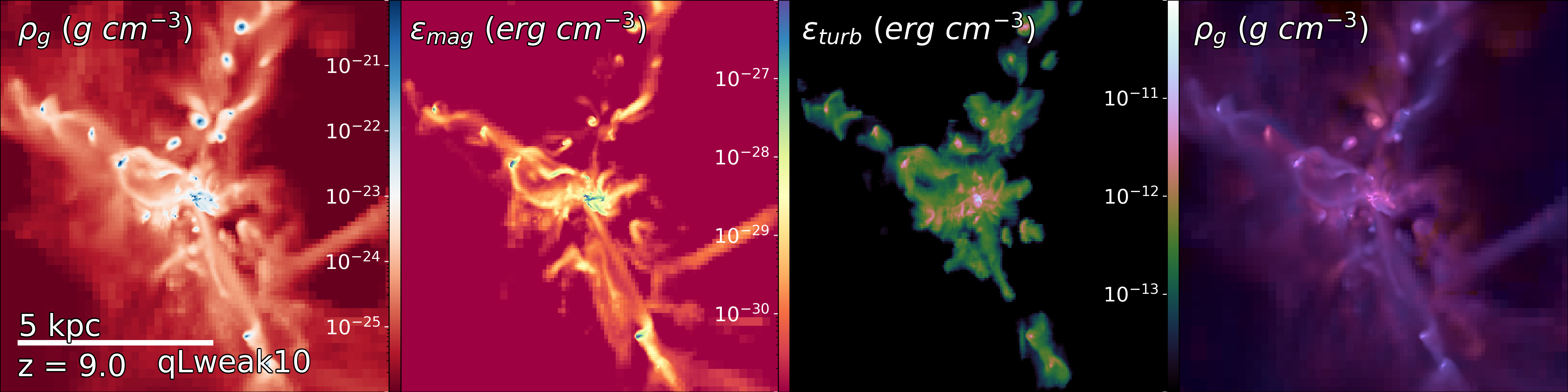}\\
    \caption{Face-on views centred at the galaxy at $z \sim 9$. All panels are gas-density weighted projections of a cube with $10$~kpc physical side. The rows compare the quasi-Eulerian refinement run (\qEulDiez, top) with the two quasi-Lagrangian runs (\qLagDiez, centre; \qLweakDiez, bottom), all with maximum physical resolution $\Dres = 10$~pc. Panels show from left to right gas density, magnetic energy density, turbulent energy density and gas density colour-coded as inflowing (blue) or outflowing (red). In the \qEulDiez~simulation, filamentary inflows are much better resolved and additional turbulent sub-structure is found in the circumgalactic region of the forming galaxy, generally identified as outflowing.}
    \label{fig:GalaxyViewsHighZ}
\end{figure*}

\subsection{Small scale turbulence computation}
\label{ss:SSturbulence}
Turbulence is a multi-scale, vectorial motion intrinsically present in galactic gas flows, but interrelated with other kinematic quantities such as bulk motions and organised rotation through the velocity field. As a result, measuring turbulence is an intricate process. A reasonable estimate of turbulence is to use the velocity dispersion. Depending on the size of the region employed to compute this velocity dispersion, the resulting estimate is associated with turbulence at an specific spatial scale.

As we are mostly interested in small-scale turbulence, we fix the spatial scale at which this measurement is done ($\mathcal{L}_\text{turb}$) to a value comparable with our resolution $\mathcal{L}_\text{turb} = 4 \Dres$. For each resolution element and for each component $i$, we obtain the local mean velocity $\bar{v_i}$ inside a sphere of radius $\mathcal{L}_\text{turb}$, and then compute its turbulent velocity field as the deviation from this mean. Therefore, we estimate the small-scale turbulence velocity in each cell as 
\begin{equation}
v_\text{turb}\,(\mathcal{L}_\text{turb}) = \sqrt{v_\text{turb,x}^2 + v_\text{turb,y}^2 + v_\text{turb,z}^2}\,,
\label{eq:SSTurb}
\end{equation}
where for each component $i$ of the fluid velocity in the cell $v_i$ 
\begin{equation}
v_\text{turb,i}\,(\mathcal{L}_\text{turb}) = \left| v_i - \bar{v_i}\, (\mathcal{L}_\text{turb}) \right|\,,
\label{eq:TurbCell}
\end{equation}
where $\bar{v_i}$ is the volume-weighted average velocity, unless explicitly indicated.

The importance of the turbulent velocity can be gauged by comparing it with the circular velocity $v_\text{circ}\,(r)$, where the circular velocity provides information about the depth of the gravitational potential. The circular velocity at a distance $r$ from the centre of the galaxy is defined as
\begin{equation}	
	v_\text{circ}\,(r) = \sqrt{\frac{G M\,(r)}{r}}\,.
	\label{eq:vCirc}
\end{equation}
In this expression, $M (r)$ is the total mass (gas, dark matter and stars) contained within the sphere of radius $r$. This approach facilitates our comparison between small-scale turbulence in two different refinement strategies. Finally, we note that all velocities employed throughout this work are measured in the frame of the galaxy.

\subsection{Halo and galaxy finder}
\label{ss:HaloMaker}
We locate the studied galaxy and its halo with the {\sc HaloMaker} software \citep{Tweed2009}. We apply this halo finder to the dark matter component to obtain the location and properties of the dark matter halo, and then find the centre and angular momentum of the galaxy by re-running the halo finder on the baryonic mass instead (i.e. gas and stars). To accurately position the galactic centre, we employ the shrinking spheres method proposed by \citet{Power2003}.  For most of our study, we will focus on the {\it galactic region} which we define as the spherical volume centred on the galaxy with a radius $r_\text{gal}$ determined by the virial size of the halo of \galregion.
\section{Results} 
\label{s:Results}

We review how a quasi-uniform refinement across a galaxy and its multiple gas phases affects its turbulent and magnetic properties compared with an exclusively quasi-Lagrangian refinement strategy. Therefore, we first review its general properties and appearance in our two refinement strategies. Fig.~\ref{fig:GalaxyViews} displays face-on views of the \qEulVeinte, \qLagVeinte~and \qLweakVeinte~galaxies at $z = 2$. A similar set of projections are shown at high redshift ($z = 9$) in Fig.~\ref{fig:GalaxyViewsHighZ} for our highest resolution runs ($\Dres = 10\,\pc$; \qEulDiez, \qLagDiez~and \qLweakDiez). A first glance shows a larger amount of small-scale structure and turbulence in the \qEul~runs compared with the quasi-Lagrangian ones. This is both within the ISM of the galaxy as well as for the inflowing gas and filaments at high redshift.  

\subsection{The ISM with quasi-Eulerian refinement}
\label{ss:QErefinementISM}
The leftmost columns of Figs.~\ref{fig:GalaxyViews} and~\ref{fig:GalaxyViewsHighZ} present gas density projections. At $z = 2$, the quasi-Eulerian refinement run has a large degree of intermediate and small scale structure that is not well-resolved by the quasi-Lagrangian simulations, especially at densities below $\rho_\text{gas} < 10^{-22} {-} 10^{-21}\,\g\,\cm^{-3}$. One of the most striking differences is the absence of a circum-nuclear gas disk in \qEulVeinte, clearly present in its \qLweakVeinte~counterpart, and to a lesser extent in \qLagVeinte. This circum-nuclear disk is found in the quasi-Lagrangian runs from the time of the formation of the extended gas disk at $z \sim 4$ onward, whereas for \qEulVeinte~it briefly forms at $z \sim 3.5$ after which it is destroyed by mergers and SN feedback. \qLweakVeinte~has a larger amount of dense clumps and a higher fraction of dense gas, as expected for a weaker SN feedback prescription. At $z = 9$, our refinement provides a dramatically different view of the galaxy and its environment. The filaments feeding the galaxy are much better resolved, displaying a well defined inflowing core. Disorganised filamentary sub-structure surrounds the forming galaxy, where visual inspection across snapshots reveals their formation takes place after SN-driven outflows. This increased amount of sub-structure and apparent turbulence at high and low redshift should have an effect on the magnetic energy budget.

\begin{figure}
    \centering
    \includegraphics[width=\columnwidth]{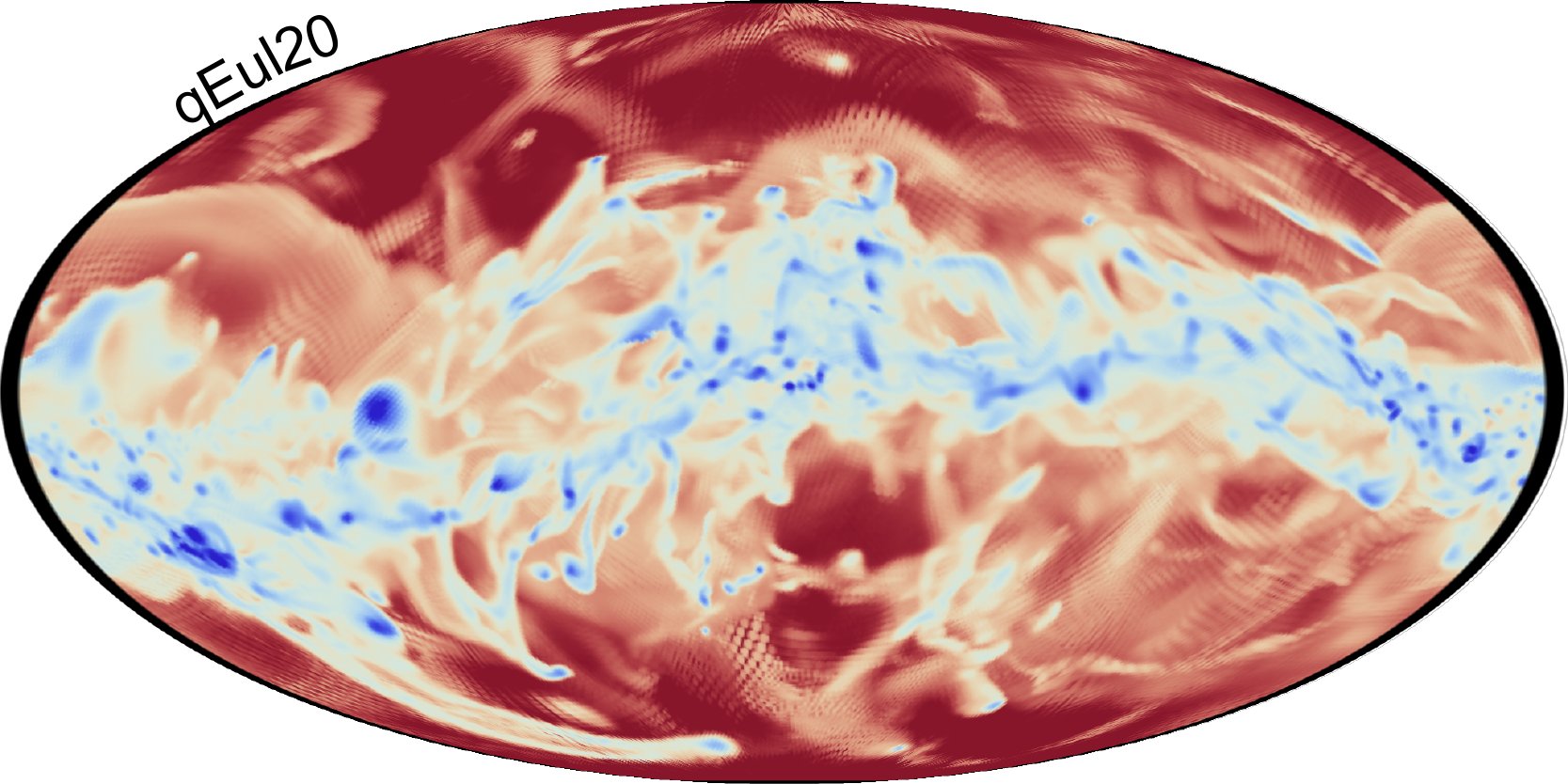}\\
    \includegraphics[width=\columnwidth]{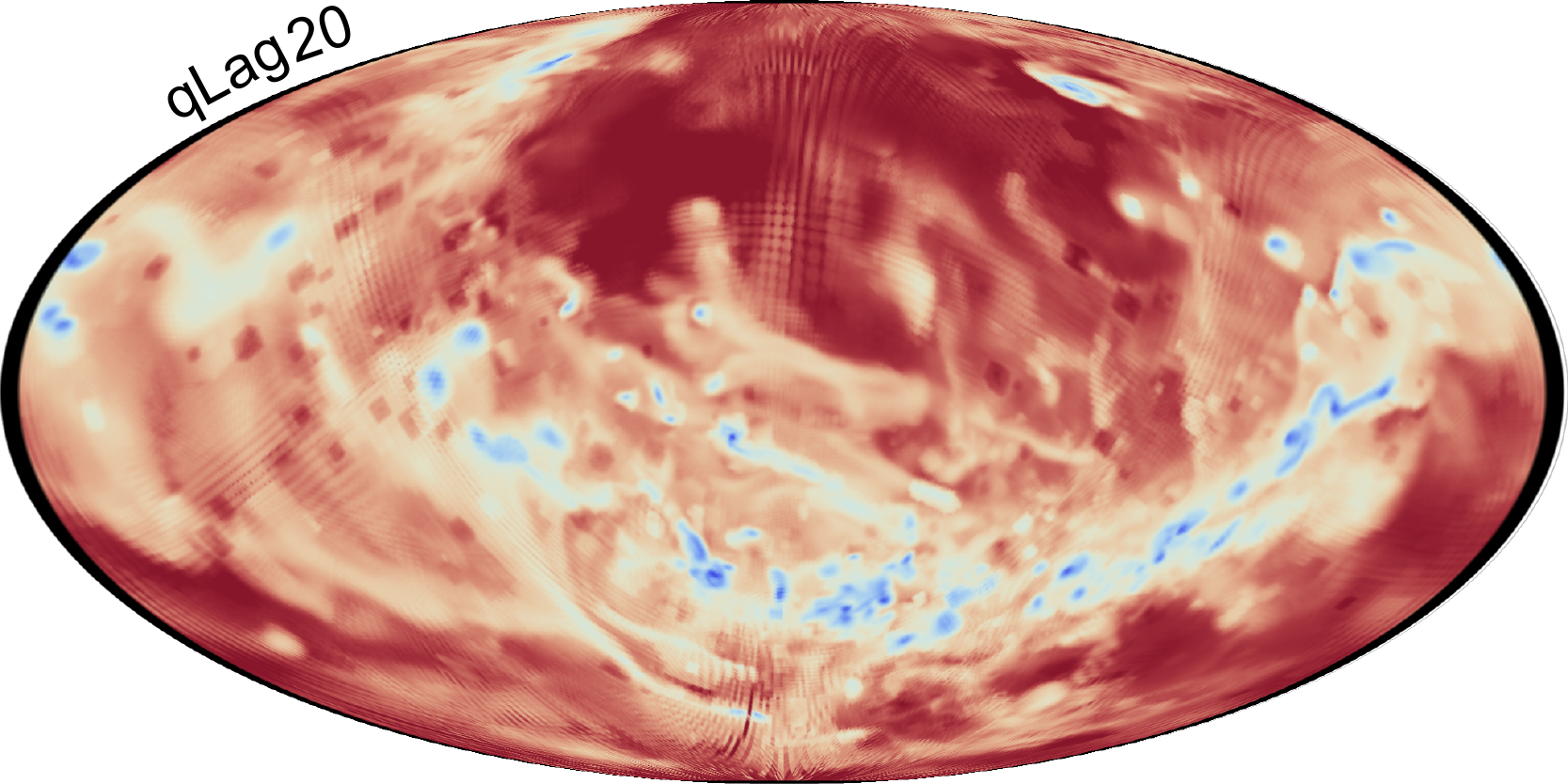}\\
    \includegraphics[width=\columnwidth]{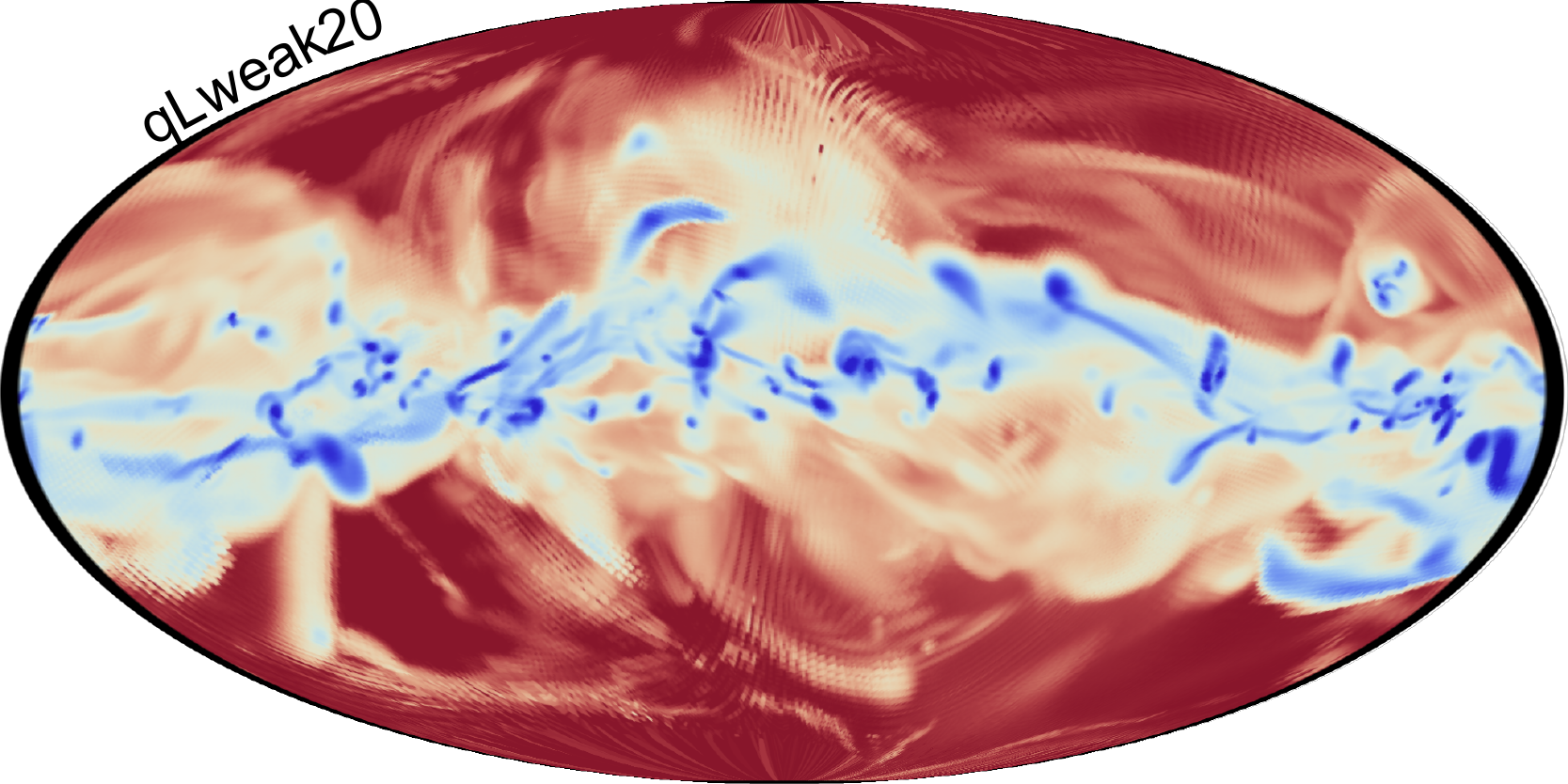}\\
    \vspace{0.1cm}
    \includegraphics[width=0.66\columnwidth]{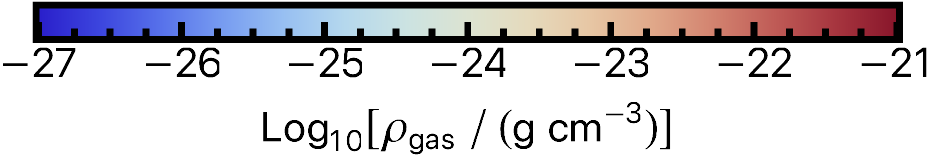}\\
    \caption{Gas density Mollweide projections as observed from the centre of the galaxy at $z = 2$ for \qEulVeinte~(top), \qLagVeinte~(middle) and \qLweakVeinte~(bottom), with the north pole aligned with the galactic angular momentum. \qEulVeinte~run displays smaller clumps connected by a more intricate filamentary network, whereas the ISM for \qLweakVeinte~is smoother. Due to its weaker SN feedback, \qLweakVeinte~is permeated with more massive clumps than \qEulVeinte~and \qLagVeinte.}
    \label{fig:Mollweide}
\end{figure}

The second column presents the magnetic energy density $\epsilon_\text{mag}$. At $z = 2$, the \qEulVeinte~run has a larger proportion of the galaxy permeated by high magnetic energies than the quasi-Lagrangian runs. Furthermore, the maps for this run also present a larger amount of high magnetic energy densities at lower gas densities. For the quasi-Lagrangian runs, the largest values of magnetic energy are concentrated in the centre of the galaxy as well as in the densest gas. At $z = 9$, we observe that the additional sub-structure in the circumgalactic medium also has a higher magnetic energy density. To do a qualitative review of the interaction between magnetic energy and turbulence, we show our estimate of the small-scale turbulent velocity $v_\text{turb}$ divided by the circular velocity in the third column of Fig.~\ref{fig:GalaxyViews}. While comparable in the inner ${\sim} 1\;\kpc$, the velocity ratio is ${\sim} 0.5$ dex higher in the outskirts of the \qEulVeinte~galaxy gas disk than in \qLagVeinte. Dense gas clumps have emphasised turbulent velocities in all three runs (particularly clear for \qLweakVeinte), \qEulVeinte~has a notably higher turbulent velocity in its diffuse ISM. Some regions in these panels present saturated turbulent velocities (e.g. north-west in \qEulVeinte~and centre of \qLagVeinte). These correspond to SN-driven outflows, correlating with temperatures $T_\text{gas} > 10^7\,\K$ (not shown). In Fig.~\ref{fig:GalaxyViewsHighZ}, we show the turbulent energy in the third column panels, as the use of the circular velocity at large distances from the galaxy becomes less appropriate. The projections clearly show how the new sub-structure observed in the density panel of \qEulDiez~has a higher turbulent energy, which cannot be captured with the quasi-Lagrangian refinement strategy. Finally, in the fourth panel of Fig.~\ref{fig:GalaxyViewsHighZ} we separate the gas into inflowing (blue coloured, radial velocity $v_r < -10\,\text{km}\,\text{s}^{-1}$) and outflowing (red coloured, $v_r > 10\,\text{km}\,\text{s}^{-1}$). This separation reveals how the inflows in \qEul~have a better resolved structure and how most of the additional structure observed in the circum-galactic medium of this run is outflowing.

\begin{figure*}
    \centering
    \includegraphics[width=\columnwidth]{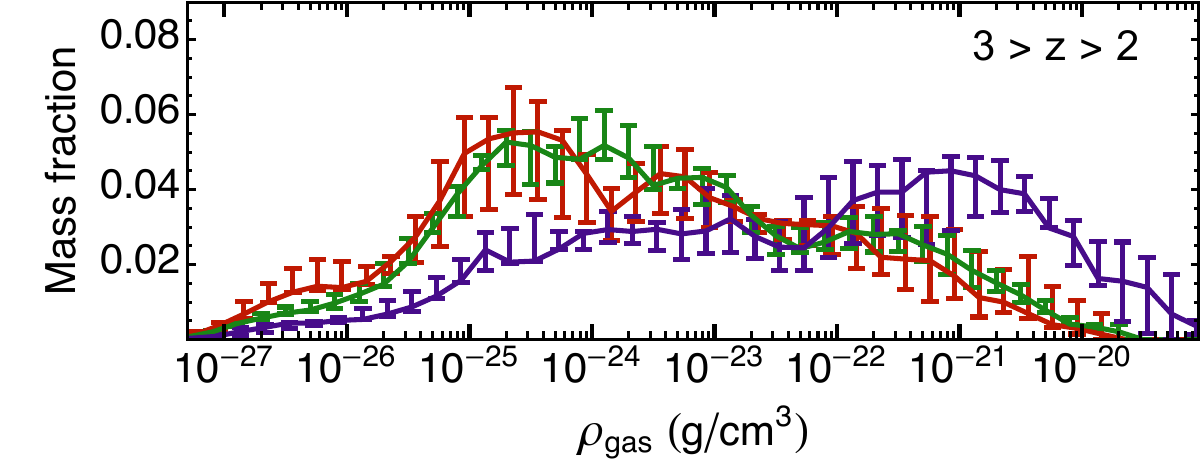}%
    \includegraphics[width=\columnwidth]{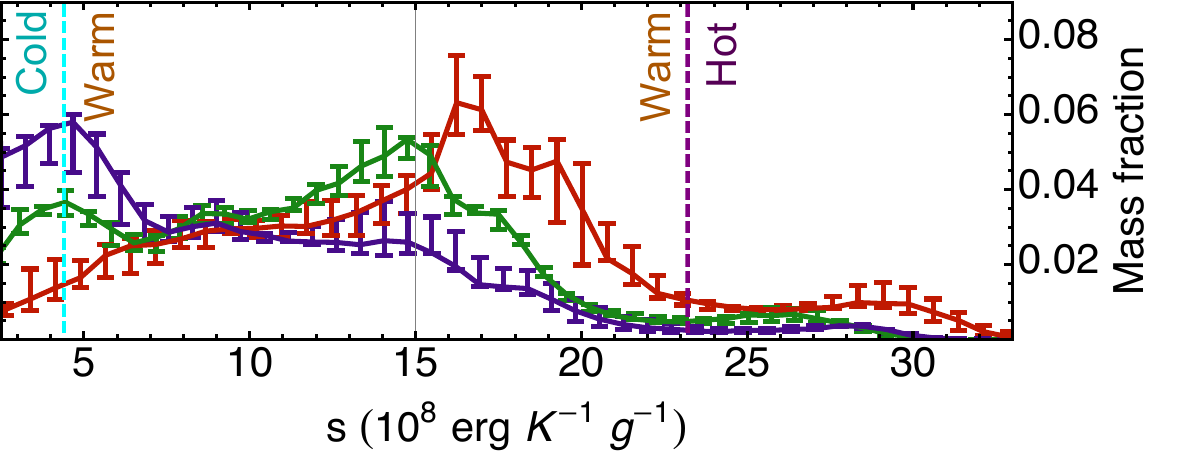}\\
    \includegraphics[width=\columnwidth]{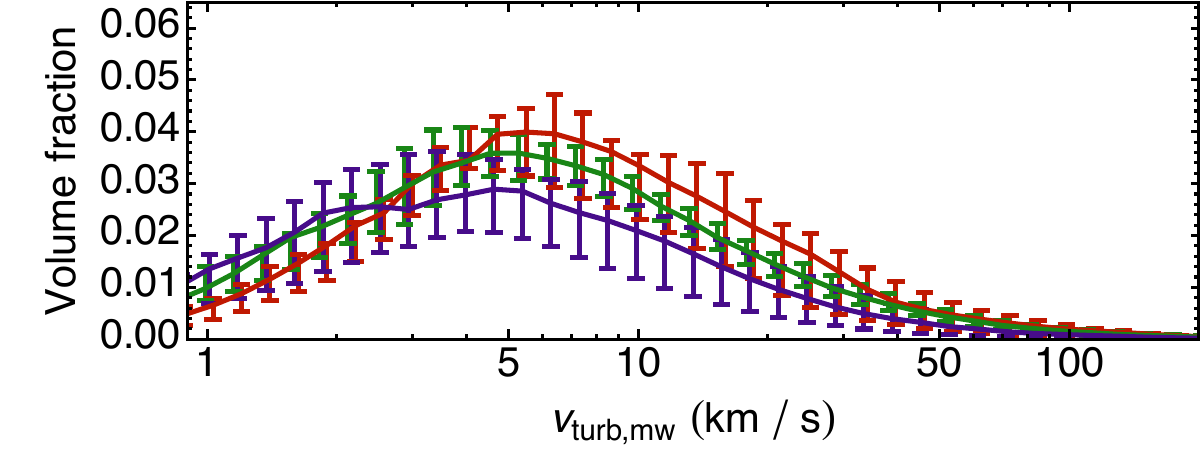}%
    \includegraphics[width=\columnwidth]{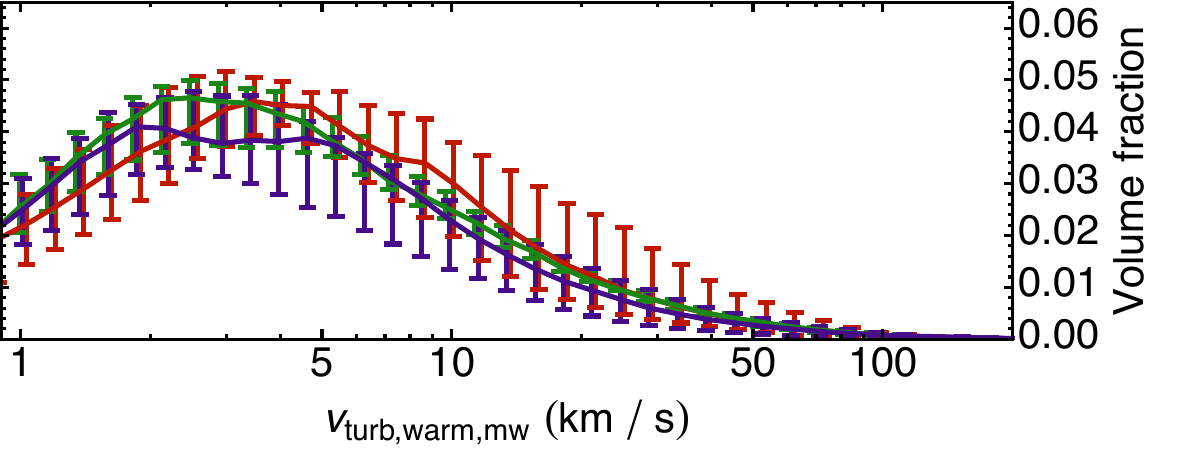}\\
    \includegraphics[width=\columnwidth]{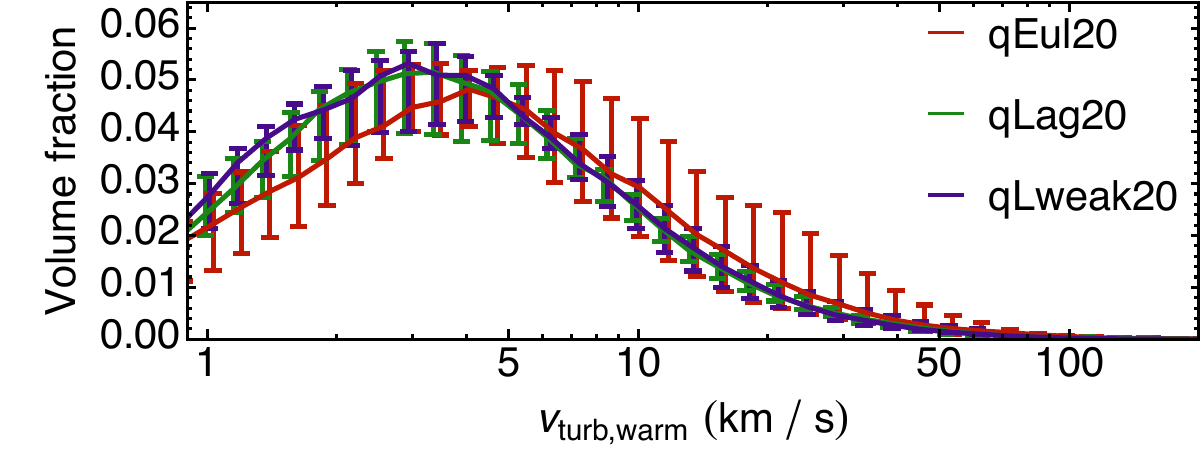}%
    \includegraphics[width=\columnwidth]{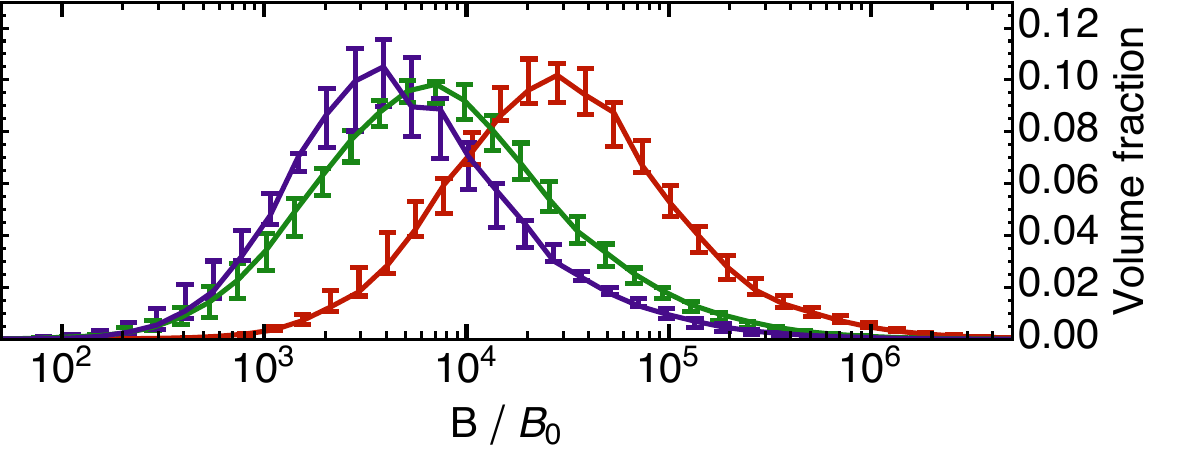}\\
    \caption{Median probability distribution function for various physical quantities in the galactic regions (\galregion) of \qEulVeinte~(red), \qLagVeinte~(green) and \qLweakVeinte~(blue) during the last 1 Gyr of the simulation ($3 \lesssim z \lesssim 2$). Panels show the gas density mass fraction $\rho_\text{gas}$ (top left), specific entropy $s$ (top right), mass-weighted small-scale turbulent velocity volume fraction (center left), warm phase mass-weighted small-scale turbulent velocity volume fraction (center right), warm phase small-scale turbulent velocity volume fraction $v_\text{turb, warm}$ (bottom left), and magnetic field volume fraction $B$ (bottom right). Error bars correspond to the first and third quartiles of the distribution over the studied time period. The specific entropy panel also shows our entropy-based division of the ISM phases as dashed vertical lines (see text for details). The \qEul~simulation shows a higher volume fraction of small-scale turbulence (particularly in its variance) and mass fraction of warm phase gas than \qLweakVeinte, and \qLagVeinte~to a lesser extent. The presence of enhanced turbulence in the warm phase is expected to produce more efficient turbulent amplification of the magnetic field. This is in agreement with the magnetic field in the bottom left panel displaying a higher strength for \qEulVeinte.}
    \label{fig:PDFs}
\end{figure*} 

The substructure in the diffuse gas of the disk at $z = 2$ is better appreciated in the Mollweide projections shown in Fig.~\ref{fig:Mollweide}. These present projections of a 18~kpc radius sphere ($r \sim 0.23 r_\text{DM}$) centred on the galaxy onto its external surface. We exclude the inner 1~kpc to remove the circumnuclear disk, particularly prominent in the projection of \qLweakVeinte. We align the north pole of the projections with the baryonic angular momentum in the sphere. Gas in the \qEulVeinte~run appears more turbulent and undergoing a higher degree of mixing, with intermediate densities displaying a more filamentary structure. In contrast, \qLweakVeinte~and to a lesser extent, \qLagVeinte, contain a higher number dense clumps, and display smoother gas in the disk.

We examine more quantitatively the effects of different refinement choices on ISM quantities in Fig.~\ref{fig:PDFs}. Panels display, from top left to bottom right, gas density ($\rho_\text{gas}$) mass fraction, specific entropy ($s$) mass fraction, mass-weighted turbulent velocity ($v_\text{turb,mw}$), mass-weighted warm phase turbulent velocity ($v_\text{turb,warm,mw}$), warm phase turbulent velocity ($v_\text{turb,warm}$) volume fraction, and magnetic field ($B$) volume fraction. For each quantity the solid line shows the median probability distribution function (PDF), with the error bars indicating the first and third quartile of data covering the redshift range $3 < z < 2$.

The gas density PDF has a similar shape in the two boosted feedback simulations. All runs show a decreasing tail towards lower densities caused by heating from SN feedback, with the corresponding cells mostly located in the inner halo region, surrounding the galaxy. The stronger feedback runs (\qEulVeinte~and \qLagVeinte) peak at $\rho_\text{gas} \sim 10^{-24} - 10^{-23}\,\g \,\cm^{-3}$. \citet{Iffrig2017} studied an isolated disk with a uniform grid of even higher resolution (${\sim} 2$~pc), finding for their run without dynamically important magnetic fields (B0 run) a lognormal distribution that peaks approximately at $\rho_\text{gas} \sim 10^{-23} \,\g\,\cm^{-3}$, comparable to \qEulVeinte. The \qLweakVeinte~run instead peaks at $\rho_\text{gas} \sim 10^{-21} \,\g\,\cm^{-3}$. This reflects the aforementioned increased presence of high density clumps in this simulation. Both runs with stronger SN feedback have a second density peak at the lower density of $\rho_\text{gas} \sim 3\cdot10^{-25}\,\g\,\cm^{-3}$, not present for \qLweakVeinte.  
We now analyze the specific entropy $s$ to explore how our prescription affects the distribution of gas across the phases of the ISM. As the warm phase of the ISM is expected to be the preferential phase of residence of the magnetic field in numerical MHD simulations of galaxies \citep{Evirgen2017}, we are interested on whether our prescription affects this phase. We divide the ISM into three phases following \citet{Gent2012} and \citet{Evirgen2017}: cold and dense ($4.4 \cdot 10^8\, \erg\, \K^{-1}\, \s^{-1} = s_\text{cold} > s$), warm ($s_\text{cold} < s < s_\text{hot} $), and hot and diffuse ($s > s_\text{hot} = 23.2 \cdot 10^8 \, \erg\, \K^{-1}\, \s^{-1}$). These divisions are shown as dashed vertical lines in the specific entropy panel of Fig.~\ref{fig:PDFs}. The mass PDF of the specific entropy illustrates some relatively small ISM changes for the warm and hot phases. The \qEulVeinte~run has a slight increase of its warm phase volume compared with \qLagVeinte, at the expense of the hot and cold phase. Both of the runs with stronger feedback show an expected drastic reduction of the cold gas phase mass with respect to \qLweakVeinte. Interestingly, the amount of gas in the cold phase is slightly decreased with our quasi-Eulerian refinement scheme. Most of the cold gas mass in \qLagVeinte~and \qEulVeinte~is transferred to the warm phase, with a minor contribution to the hot phase. 

Finally, mass-weighted turbulence and mass and volume-weighted warm phase turbulent velocity panels also reflect a shift for \qEulVeinte~towards higher turbulent velocities. While the median of turbulent velocities in the are only midly higher for the quasi-Eulerian run at all turbulent velocities $v_\text{turb} > 5\,\text{km}\,\text{s}^{-1}$, the error bars reveal how turbulence is frequently significantly higher in this run compared with the quasi-Lagrangian cases. We review turbulence in more depth in Sections~\ref{sss:TurbISM} and~\ref{sss:WarmphaseTurb}. This higher degree of turbulence in \qEulVeinte~as well as the higher mass ISM fractions for the warm phase, pose this run as more favourable for turbulent amplification. Recall, Fig.~\ref{fig:GalaxyViews} showed a higher magnetic energy density throughout the volume of the galaxy, even tough there were fewer dense gas clumps, which through gas compression amplification ($B \propto \rho_\text{gas}^{2/3}$) could lead to a larger magnetic field. 
The bottom-right panel of Fig.~\ref{fig:PDFs} confirms a more efficient amplification in \qEulVeinte. The quasi-Lagrangian \qLweakVeinte~and \qLagVeinte~both have a narrow peak at $B / B_0 \sim 5 \cdot 10^{3}$ with a tail towards higher magnetic fields. On the other hand, the distribution for \qEulVeinte~peaks approximately at $B / B_0 \sim 2 \cdot 10^{4}$, with a less pronounced tail towards higher field strength than the previous runs. The fact that \qEulVeinte~has reached a higher magnetic field by $z \sim 2$ suggests that the new refinement has a higher impact on the magnetic field amplification than the strength of stellar feedback on our particular setup.

\subsection{Magnetic energy growth}
\label{ss:EnergyGrowth}

\begin{figure}
    \centering
    \includegraphics[width=\columnwidth]{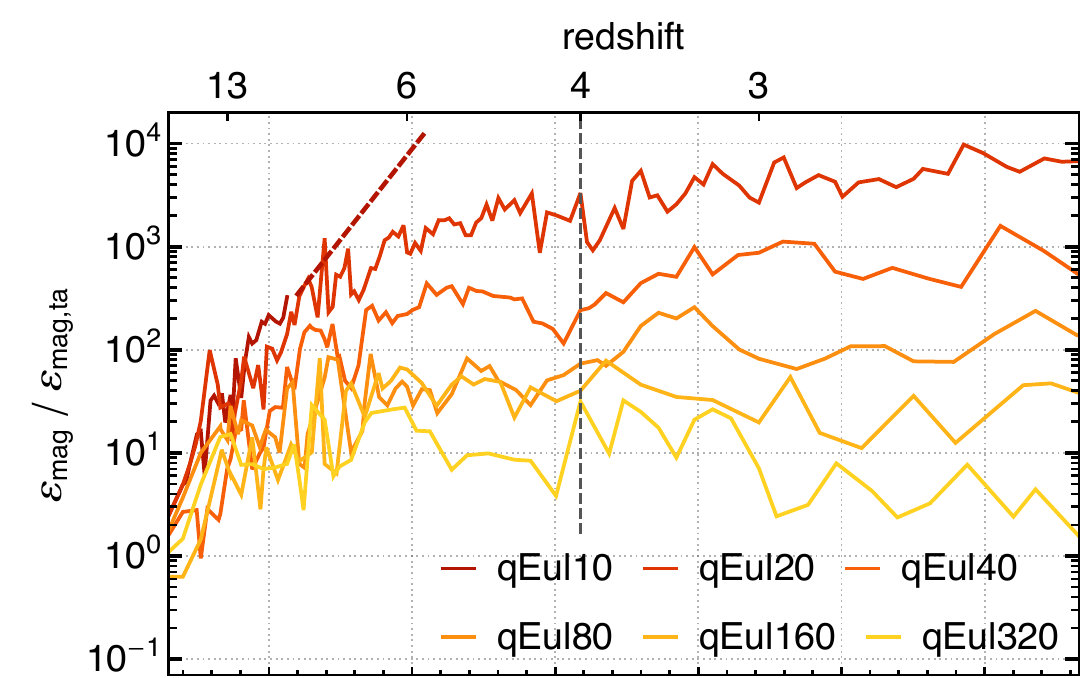}\\
    \includegraphics[width=\columnwidth]{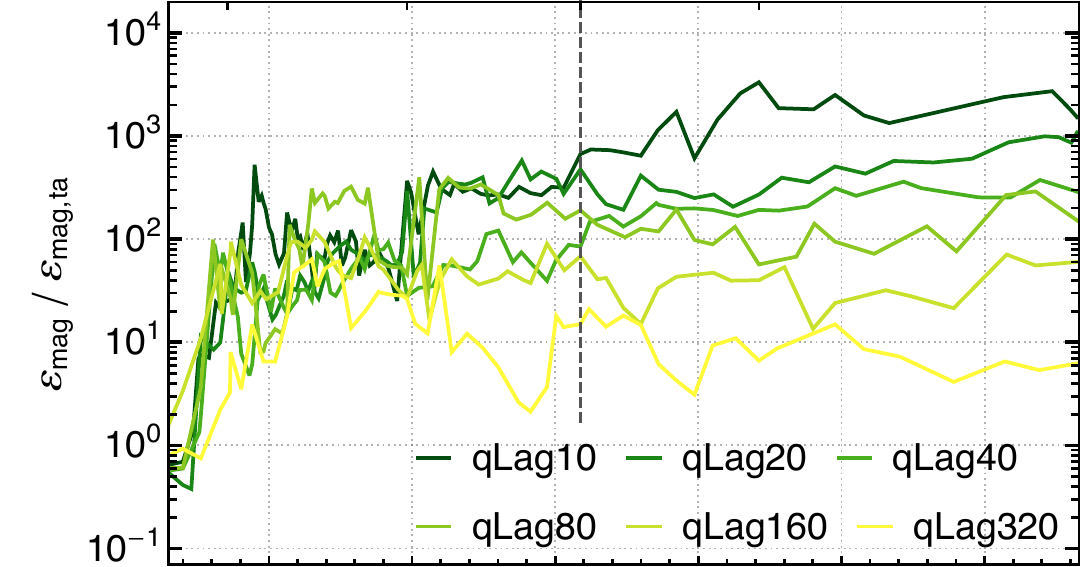}\\
    \includegraphics[width=\columnwidth]{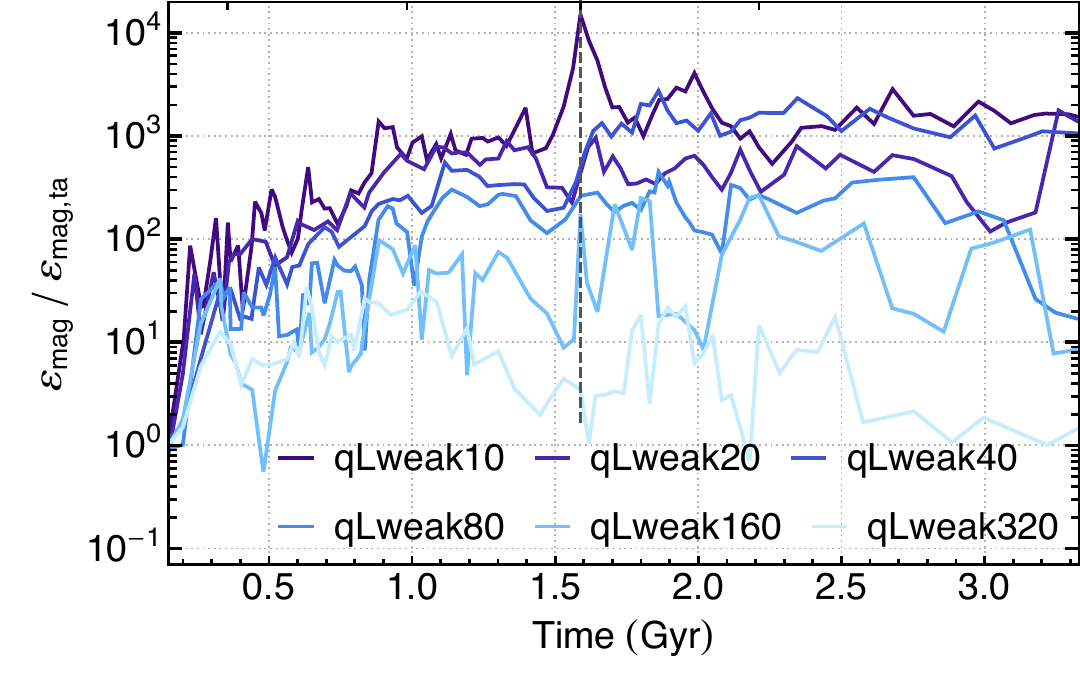}\\
    \caption{Specific magnetic energy growth in the galactic region (\galregion) of the \qEul~(top), \qLag~(middle), and \qLweak~(bottom) runs normalised to its value at the density perturbation turn-around point ($t \sim 0.15$~Gyr). Note that the middle and bottom panels also show $\Dres = 10$~pc simulations that reach $z = 2$, whereas \qEulDiez~is only evolved until $z = 8.5$. \qEul~simulations display a higher amplification of magnetic energy per unit gas mass at each given resolution, but especially once $\Dres \leq 80$~pc. Furthermore, amplification increases dramatically faster with $\Dres$ for \qEul~runs compared with the scaling observed for quasi-Lagrangian simulations. This is notable when comparing \qEulVeinte~with the higher $\Dres$ \qLagDiez~and \qLweakDiez.}
    \label{fig:sEmComparison}
\end{figure}

We now explore the magnetic field amplification in more detail by studying the time evolution of the specific magnetic energy ($\emag$) in our two refinement schemes for different maximal resolutions $\Dres$. Throughout this work, specific energies are simply $\varepsilon_\text{X} = E_\text{X} / M_\text{gas}$ with $M_\text{gas}$ the total gas mass and $E_\text{X}$ the total amount of energy X in the studied region. Employing a specific energy allows us to easily decouple the growth of an extensive quantity (such as the energy) from the natural mass growth of galaxies over time in cosmological simulations as well as the volume fraction of the galaxy within the studied region. 

Fig.~\ref{fig:sEmComparison} shows the growth of the specific magnetic energy in the galactic region. The most notable aspect is that \qEulVeinte~displays a more pronounced growth, even when compared to higher resolution runs \qLagDiez~and \qLweakDiez. During the accretion phase (approximately $0.3\, \text{Gyr} < t < 1.5\, \text{Gyr}$), there is a clear correlation between the growth of $\emag$ and higher resolutions for $\Dres$ regardless of the SN feedback strength. \qEulDiez~in particular (which was only evolved to $z \sim 8$ due to its elevated computational cost), shows a remarkably fast growth of magnetic energy. \citet{Martin-Alvarez2018} discusses the correlation between $\emag$ peaks and the main progenitor undergoing merger events. However, such peaks are less obvious in our \qEul~runs. For the \qEulVeinte~case, we only find some evidence during the accretion phase for a extraordinarily chaotic merger at $z \gtrsim 8$. Instead, the refinement strategy appears to have the largest impact on the amplification in the accretion phase. However, the two refinement approaches provide comparable energy growth during the feedback phase ($t > 1.5\, \text{Gyr}$). In this second period, the amount of growth observed seems to have a higher dependence on the strength of the SN feedback prescription. The \qEul~runs show a clear increase of $\emag$ and $\emag$ growth with smaller cell sizes for $\Dres$.

\begin{figure*}
    \centering
    \includegraphics[width=\columnwidth]{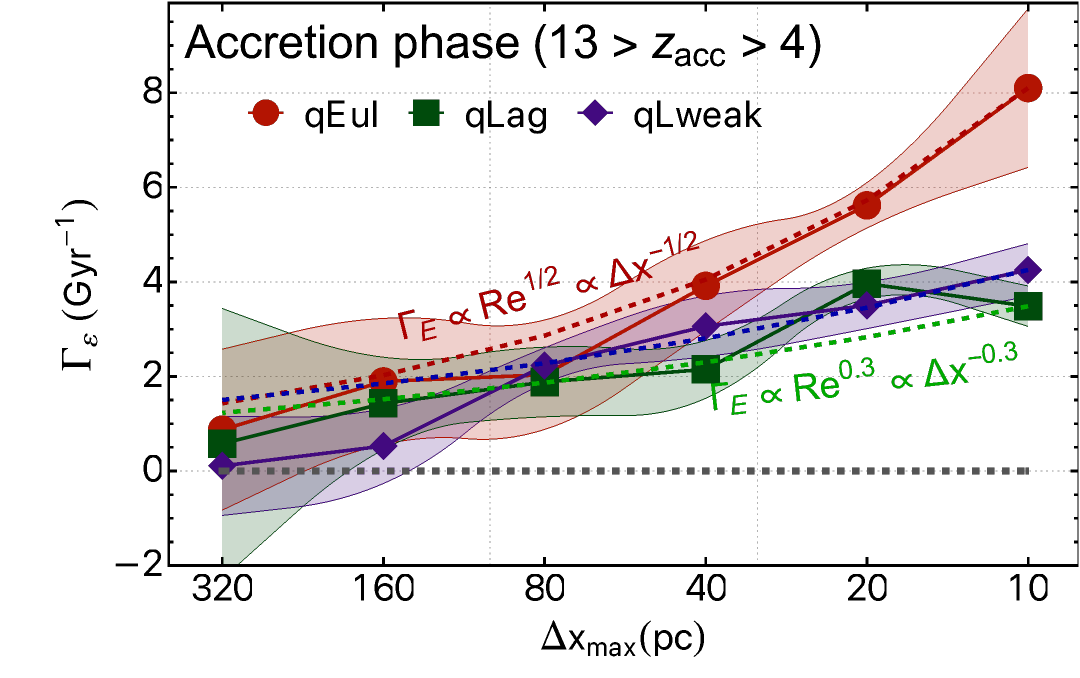}%
    \includegraphics[width=\columnwidth]{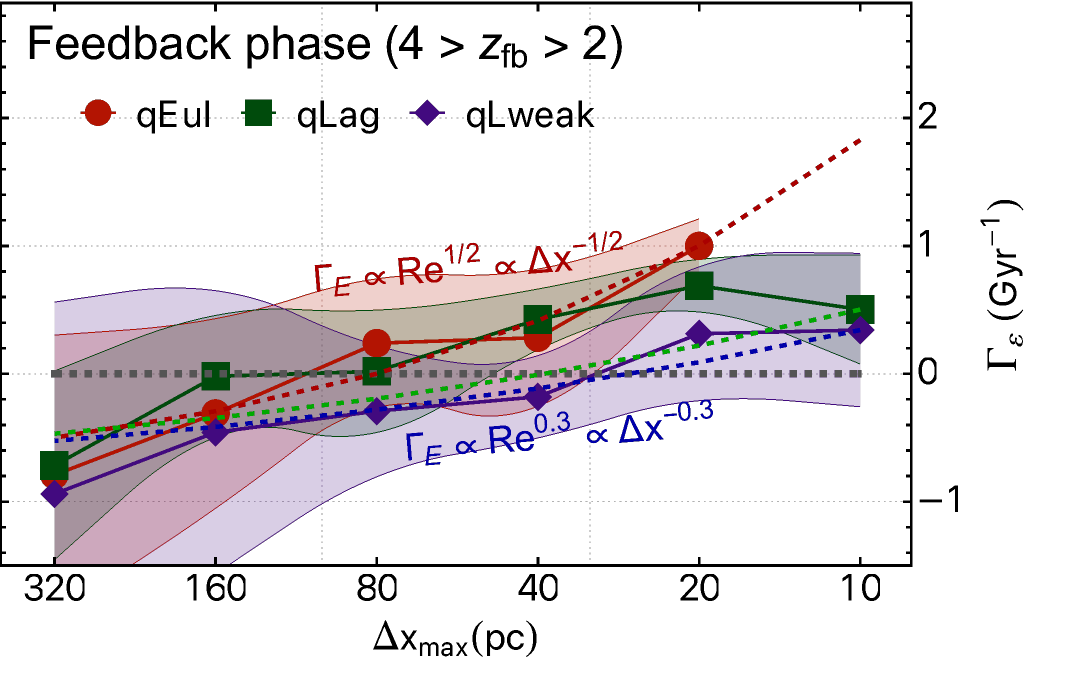}\\
    \caption{Specific magnetic energy exponential growth factors $\Gamma_\varepsilon$ versus maximum resolution $\Dres$ during the accretion (left) and feedback (right) phases for the runs shown in Fig.~\ref{fig:sEmComparison}. The period over which the growth factor is measured spans approximately the redshift range indicated in its associated panel. Points and shaded bands correspond to the fit and associated $3 \sigma$ error. \qEul~simulations (red lines) display higher growth factors at a given $\Dres$ than quasi-Lagrangian runs (green and blue lines) for the accretion phase. The effect of the refinement scheme is more important during the accretion phase, whereas it becomes comparable to the importance of boosting SN feedback during the feedback phase. Growth factors for \qEul~runs scale with increasing Reynolds number following approximately $\Gamma_\varepsilon \propto \Dres^{-1/2}$ (red dashed lines), compatible with Kolmogorov turbulence with viscosity $\nu \propto \Dres$. The quasi-Lagrangian \qLag~and \qLweak~runs are better matched by $\Gamma_\varepsilon \propto \Dres^{-0.3}$ (green and blue dashes lines, respectively), as obtained by \citet{Federrath2011}.}
    \label{fig:GrowthComparison}
\end{figure*}

To better quantify the specific magnetic energy growth of our different runs, we study the growth rates ($\Gamma_\varepsilon$) for each of our simulations. $\Gamma_\varepsilon$ is computed by fitting $\emag (t)$ to the function $\alpha \exp{\left[\Gamma_\varepsilon \left(t - t_0\right)\right]}$ over some period of time, where $t_0$ corresponds to the time a given phase commences. The normalisation parameter $\alpha$ is fixed to $\emag (t_0)$ at the start of the accretion phase to discard the amplification due to collapse, and left free during the feedback phase. We adopt $t \in [0.3, 1.4]$~Gyr and $t \in [1.7, 3.3]$~Gyr for the accretion and feedback phases, respectively. This ensures that the values of $\emag$ used are well within each phase. Furthermore, it removes the sharp spike observed in $\emag$ at $z \sim 4$ in \qLweakDiez, attributed to a merger \citep{Martin-Alvarez2018}. Our measurements of $\Gamma_\varepsilon$ (and their 3 $\sigma$ errors as shaded bands) are shown in Fig.~\ref{fig:GrowthComparison}. In the accretion phase, our \qEul~runs have higher growth rates for virtually all resolutions studied. Furthermore, \qLag~and \qLweak~have approximately equal growth rates at each resolution, as we find refinement strategy rather than SN feedback strength to be more crucial in the amplification process at this stage. Values for $\Gamma_\varepsilon$ start to separate between the different sets of runs for $\Dres \lesssim 40$~pc, suggesting that the standard quasi-Lagrangian AMR refinement provides a reasonably homogeneous coverage of the galaxy down to this resolution. This agrees with Fig.~\ref{fig:GalaxyViews}, where all three galaxies are almost entirely covered by the light blue shade corresponding to $80$~pc; and Fig.~\ref{fig:PhaseRefine}, where most gas at $T_\text{gas} \lesssim 3 \cdot 10^4$~K has $\Delta_\text{cell} \leq 80$~pc. The feedback phase measurements show negative $\Gamma_\varepsilon$ values for the lowest resolutions, indicating that the amplification is not enough to sustain the growth of magnetic fields. Therefore, magnetic energy per gas mass is now decreasing due to numerical resistivity and reconnection, as well as the accretion of pristine gas with lower magnetisation. While the difference is modest, runs with the stronger SN feedback provide higher growth rates during this so-called feedback phase. Meanwhile, the \qEul~refinement provides some additional amplification above \qLag, albeit minor and secondary when compared to the apparent effect of boosting feedback strength.   

One of the predictions for numerical simulations of turbulent magnetic amplification is a scaling of $\Gamma_\varepsilon \propto \text{Re}^{\gamma}$, where $\gamma$ depends on the properties of the velocity field. The maximum and minimum values for $\gamma$ are obtained for incompressible \citep[$\gamma=0.5$,][]{Kolmogorov1941} and compressible turbulence \citep[$\gamma=0.3$,][]{Burgers1948}, respectively \citep{Schober2012}. For a velocity field that is converged with resolution\footnote{\citet{Kortgen2017} and \citet{Jin2017} find this convergence to occur in the ISM for $\Dres \lesssim 0.1$~pc.}, upper and lower limits for $\gamma$ can be obtained by assuming ideal Kolmogorov turbulence ($\text{Re} \propto \Dres^{-4/3}$, \citealt{Kritsuk2011}) or a linear scaling of the hydrodynamical viscosity with resolution ($\text{Re} \propto \Dres^{-1/3}$, \citealt{Rieder2017a,Vazza2018}), yielding $\Gamma_\varepsilon \propto \Dres^{-2/3}$ and $\Gamma_\varepsilon \propto \Dres^{-1/3}$, respectively. 

As the fiducial scenario, we illustrate by a dashed red line in Fig.~\ref{fig:GrowthComparison} the scaling $\Gamma_\varepsilon \propto \text{Re}^{1/2} \propto \Dres^{-1/2}$ (i.e. Kolmogorov turbulence + viscosity $\nu \propto \Dres$; see also \citealt{Beresnyak2019}), allowing it to become negative due to the aforementioned dissipation observed during the feedback phase. We note that the \qEul~runs follow well this scaling, indicating that these simulations capture well turbulent amplification. This is particularly important during the accretion phase, as turbulent amplification is a well-known mechanism to generate saturated magnetic fields starting from weak primordial magnetic fields on timescales that could be as short as several hundreds of Myr \citep{Schlickeiser2012}. \citet{Balsara2004b} explore magnetic energy growth in 200~pc turbulent boxes using a $2^\text{nd}$ order Godunov scheme such as the one used here. Extrapolating our growth rate during the accretion phase to the resolutions in their turbulent boxes (i.e. $0.7 - 1.5$~pc), the quasi-Lagrangian AMR runs fall short of their amplification ($\Gamma_\varepsilon\sim10\, \text{Gyr}^{-1}$ for 0.7~pc) whereas the \qEul~runs extrapolation is comparable, on the order of $\Gamma_\varepsilon \sim 60\, \text{Gyr}^{-1}$ if we assumed a $0.7\,\pc$ resolution. \citet{Bendre2015} also explore magnetic amplification in the ISM using the {\sc nirvana} code to run non-ideal MHD simulations with a similar resolution to the one employed here ($\Dres \sim 8$~pc). They find comparable growth rates of $\Gamma_\varepsilon \sim 3 - 8\, \text{Gyr}^{-1}$, although we note that they explore amplification towards saturation whereas we remain in the kinematic regime. The quasi-Lagrangian refinement runs appear to follow a different proportionality with $\text{Re}$ when $\Dres < 80$~pc, with a scaling resembling the $\Gamma_\varepsilon \propto \text{Re}^{0.3}$ obtained by \citet{Federrath2011}. We include dashed lines with this alternative proportionality for the \qLag~(green dashed) and \qLweak~(blue dashed) simulations.

When comparing our \qEul~with the \qLag~and \qLweak~runs, the scaling observed with refinement markedly favours the new refinement method both in terms of agreement with the expected growth rate scaling \citep{Beresnyak2019}, as well as in terms of reaching higher amplification rates. As in previous work \citep{Martin-Alvarez2018}, amplification is faster at very high redshifts, when accretion is predominant, potentially driving turbulence directly \citep{Klessen2010}, by promoting SN feedback \citep{Hopkins2013}, or through the promotion of gravitational instabilities \citep{Elmegreen2010,Krumholz2016}. Importantly, this 'accretion phase' is the most crucial when bridging the gap between weak primordial seeds and $\muG$ galaxy magnetisations, especially due to the detection of magnetic fields of such strengths at high redshift \citep{Bernet2008}. Assuming realistic ISM viscosities and diffusivities, our results, in agreement with previous work \citep[e.g.][]{Pakmor2014,Rieder2016,Federrath2016}, suggest that turbulent dynamo amplification should saturate at $z \gg 6$, and potentially a few 100~Myr after galaxy formation. This allows in turn for other processes to maintain and reorganise magnetic fields during the remainder of the evolution of galaxies \citep{Chamandy2013,Moss2013}.

\subsection{Turbulent dynamo amplification}

\subsubsection{Comparison with adiabatic magnetic compression}

\begin{figure*}
    \centering
    \includegraphics[width=2\columnwidth]{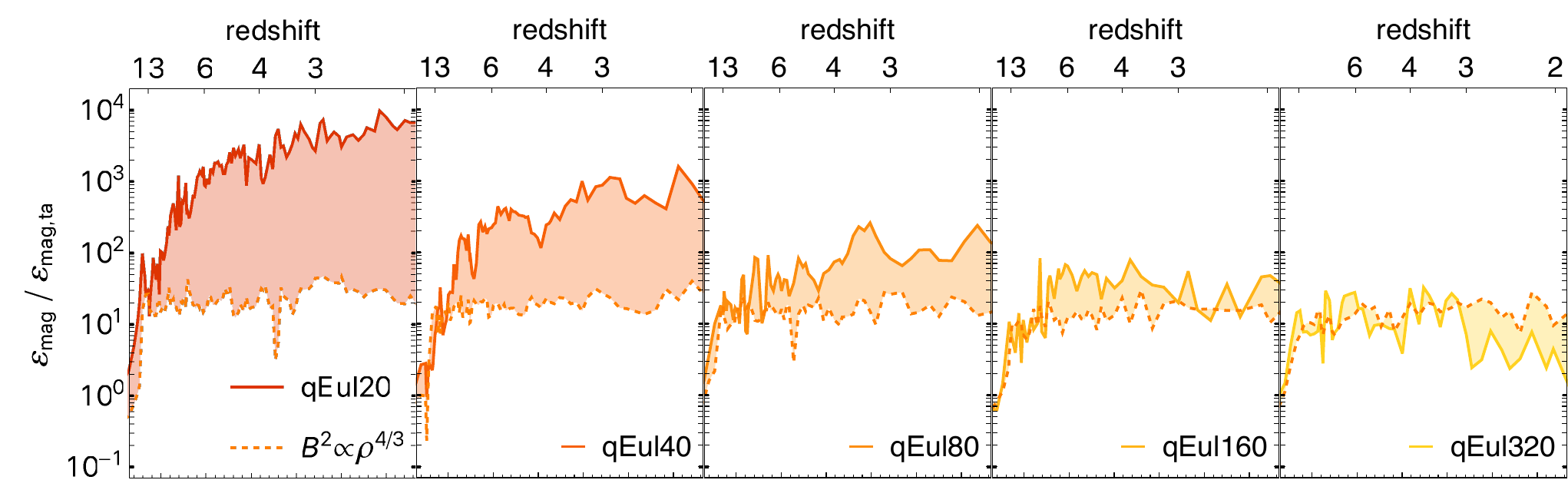}\\
    \includegraphics[width=2\columnwidth]{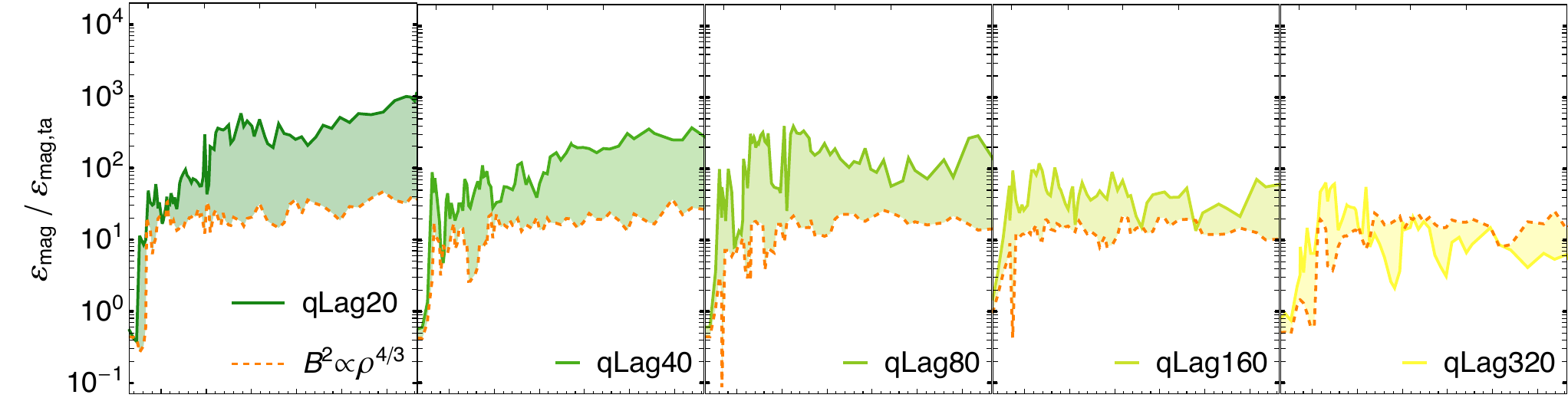}\\
    \includegraphics[width=2\columnwidth]{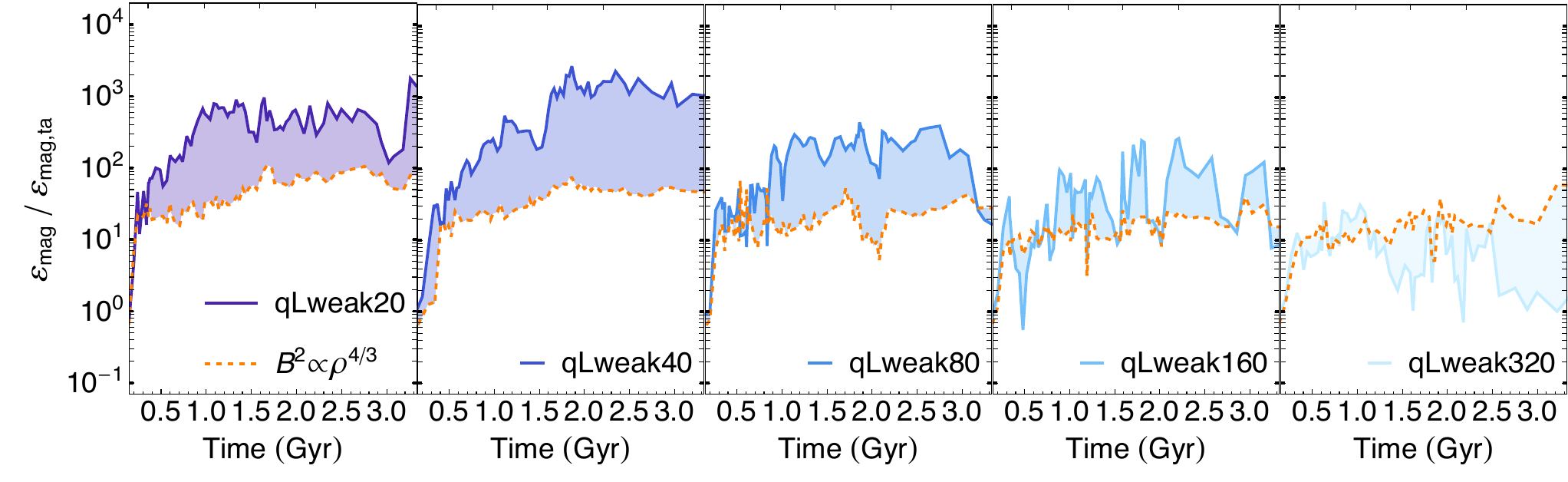}\\
    \caption{Specific magnetic energy measurement (solid lines) and isotropic magnetic field adiabatic compression estimate (i.e. $B^2\propto \rho_\text{gas}^{4/3}$, see text description for $\Cmag$; dashed lines) in the galactic region (\galregion). In accordance with turbulent amplification, runs with better resolution display larger growths and separation above their compressional estimates \citep[e.g. Fig.~9 by][]{Federrath2011}. This increase is significantly enhanced for the \qEul~runs (top row) when compared with their \qLag~(middle row) and \qLweak~analogues (bottom row).}
    \label{fig:Compression}
\end{figure*}

\begin{figure*}
    \centering
    \includegraphics[width=2.1\columnwidth]{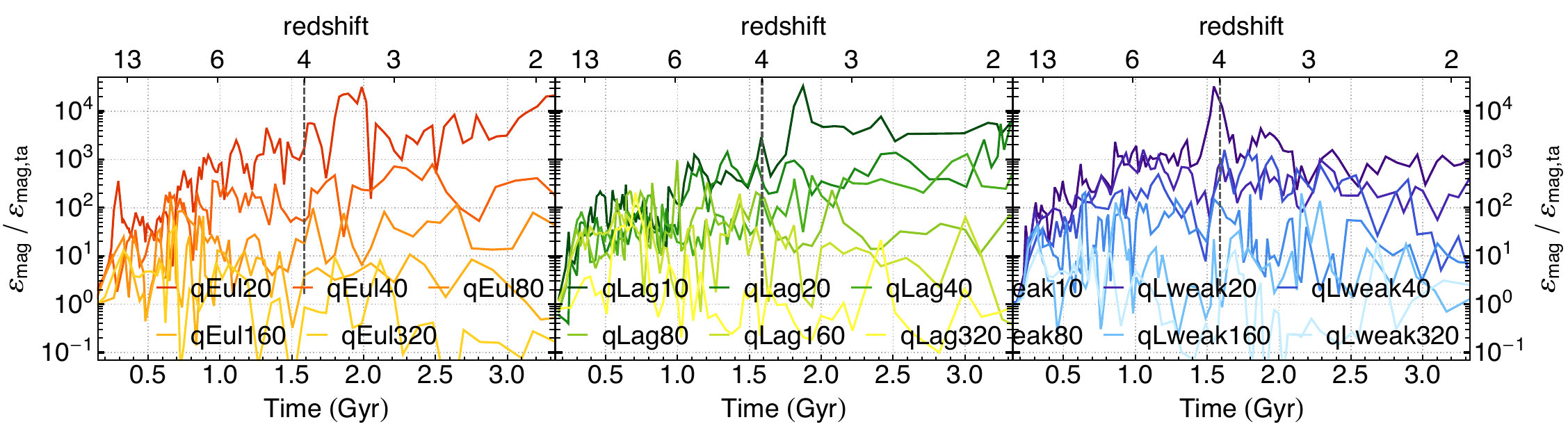}\\
    \caption{Time evolution of the specific magnetic energy within the central $r < 1\,\kpc$ of the galactic region. Turbulence in the central region sustains magnetic energy amplification for longer times. This is particularly notable for the \qEulVeinte~simulation, with amplification persisting throughout most of the feedback phase.}
    \label{fig:Emag1kpc}
\end{figure*}

We now focus on understanding how our refinement scheme affects other turbulent amplification signatures. One important contributor to energy growth during the early stages of the formation of a galaxy is the compression of magnetic field lines. As described, adiabatic-compression amplification predicts a scaling of the magnetic field strength with gas density $B \propto \rho_\text{gas}^{2/3}$. We use this proportionality to estimate the expected magnetic energy growth if it was to evolve affected exclusively by this compressional process. For an individual cell $i$ the corresponding magnetic energy would be only a function of its current density
\begin{equation}
	\epsilon_{\Cmag\text{,i}} (t) = \langle \epsilon_\text{mag,gal} (t_0)  \rangle
	\text{dx}_\text{i}^3
    \left( \frac{\rho_\text{gas,i}(t)}{ \langle \rho_\text{gas,gal}(t_0) \rangle } \right)^{4/3} \,.
	\label{eq:CompEstimate}
\end{equation}	
Here $\epsilon_\text{mag}$ is the magnetic energy density, and $\langle X_\text{gal} (t_0) \rangle$ indicates an average of quantity $X$ over the whole galactic region at $t_0$. We obtain the final estimate for adiabatic isotropic evolution of the specific magnetic energy by summing the contribution from each cell ($V_\text{cell}\epsilon_{\Cmag\text{,i}}$) to the total magnetic energy $M_\text{gas} \Cmag$, and divide it by the total gas mass in the region to obtain $\Cmag$. We select $t_0$ as the time of minimum specific magnetic energy. This corresponds to the turnaround of density perturbations, prior to their collapse. We compare this estimate (dashed lines) with the evolution of the specific magnetic energy (solid lines) in each run in Fig.~\ref{fig:Compression}. The adiabatic-compression estimates are similar across all simulations. However, they appear to undergo some mild time evolution for the quasi-Lagrangian AMR runs, slowly increasing with time and growing in the interval $z \in [13, 2]$ by a factor of $\sim2$ (\qLag) or $\sim3$ (\qLweak), depending on the SN feedback. Conversely, the estimates are remarkably constant for the \qEul~runs. When comparing $\Cmag$ with the specific magnetic energy in each run, higher resolutions lead to a larger deviation from the compression estimate, as expected for dynamo amplification \citep{Federrath2011,Sur2012,Pakmor2017}. At the lowest resolutions probed, the specific magnetic energy is instead reduced progressively and decreases below the compressional estimate. This latter effect follows from two considerations: numerical resistivity dissipating magnetic energy and the accretion of gas with lower specific magnetic energy, which underwent an additional magnetisation decrease prior to accretion due to cosmological expansion diluting magnetic energy. As noted by comparing the panels in Fig.~\ref{fig:sEmComparison}, at a fixed $\Dres$, $\emag$ is always higher in the \qEul~simulations. Combined with lower and more stable values for the compressional estimate, this leads to a larger separation between the values of $\emag$ and $\Cmag$, signaling notably more significant non-compressional amplification in these simulations. This is also in agreement with our discussion of the gas density and magnetic field PDFs in Section~\ref{ss:QErefinementISM}. At $z = 2$, specifically for $\Dres = 20$~pc, \qLweakVeinte~has an $\emag$ a factor of $\sim20$ above $\Cmag$, whereas this goes up almost to $10^3$ in \qEulVeinte.

\subsubsection{Amplification in the central region of the galaxy}

Due to the prevalence of SN events, a deeper local gravitational potential and a higher dispersion support of the stellar component, central regions of galaxies show a higher degree of turbulence. We review the  specific magnetic energy growth within the central $r < 1\,\kpc$ of the galactic region, where turbulence will sustain magnetic amplification for longer. Fig.~\ref{fig:Emag1kpc} shows the specific magnetic energy in a fixed physical size sphere. Panels show during the feedback phase ($z \lesssim 4$) the magnetic energy per unit mass to present some mild growth in the \qLagVeinte~and \qLagDiez~simulations, and continued amplification in the more turbulent \qEulVeinte~simulation.

\begin{figure}
    \centering
    \includegraphics[width=\columnwidth]{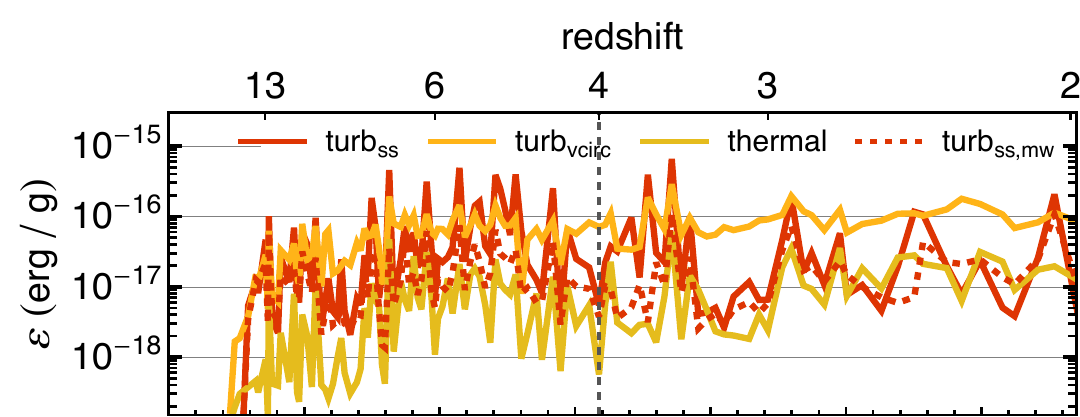}\\
    \includegraphics[width=\columnwidth]{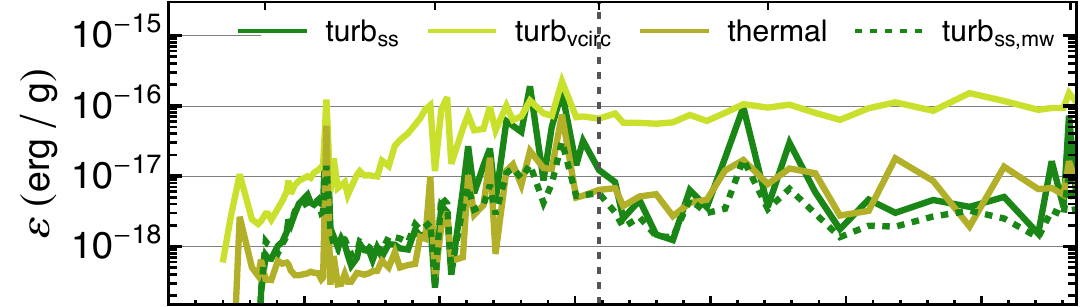}\\
    \includegraphics[width=\columnwidth]{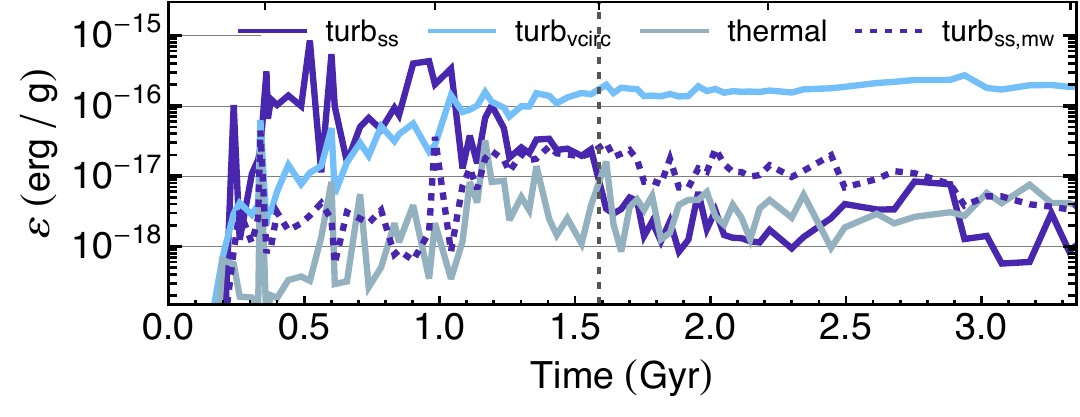}\\
    \caption{Turbulent and thermal energy comparison for the galactic region (\galregion) in each of our $\Dres = 20\,\pc$ simulations. From top to bottom, we show \qEulVeinte, \qLagVeinte, and \qLweakVeinte~, respectively. Plot displays our small-scale turbulence estimate volume ($\text{turb}_\text{ss}$, solid dark lines) and mass ($\text{turb}_\text{ss,mw}$, dashed dark lines) weighted, a simple estimate for the global turbulence ($\text{turb}_\text{circ}$, solid clear lines) and the thermal energy (solid pale lines). Small-scale turbulence dominates in \qEulVeinte~during the accretion phase, reinforcing the claim that additional amplification during this period is driven by a higher degree of small-scale turbulence.}
    \label{fig:EnergiesComparison}
\end{figure}

\subsubsection{A turbulence-dominated ISM}
\label{sss:TurbISM}

\label{sss:WarmphaseTurb}
\begin{figure}
    \centering
    \includegraphics[width=\columnwidth]{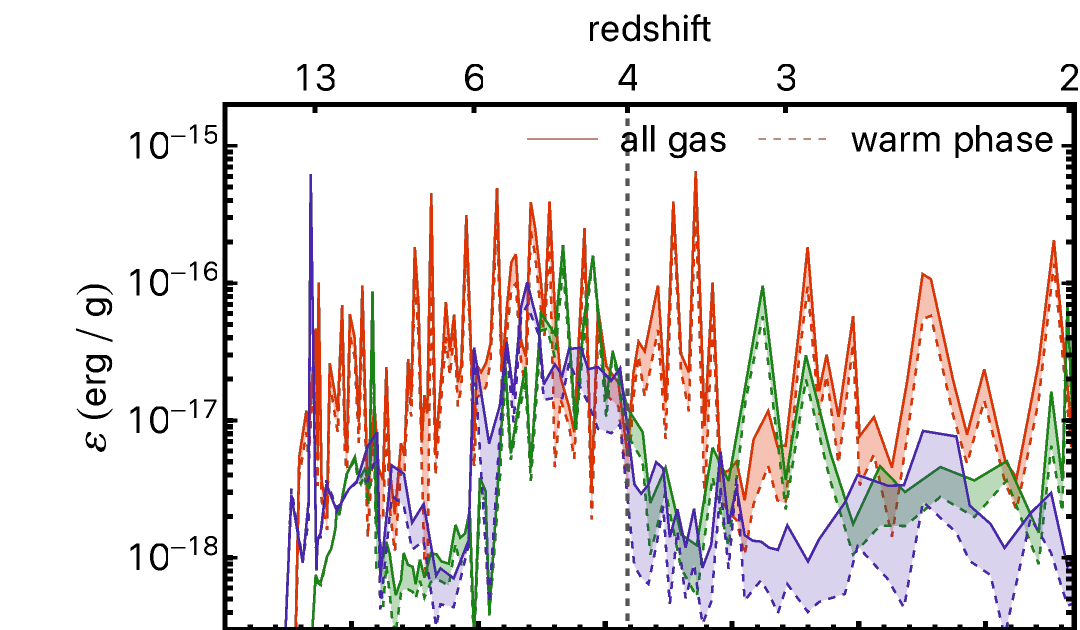}\\
    \includegraphics[width=\columnwidth]{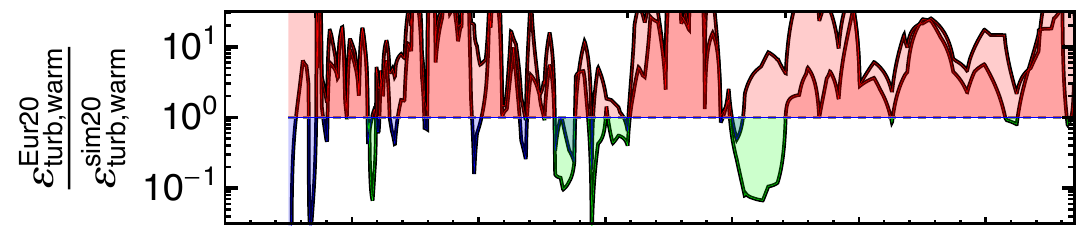}\\
    \includegraphics[width=\columnwidth]{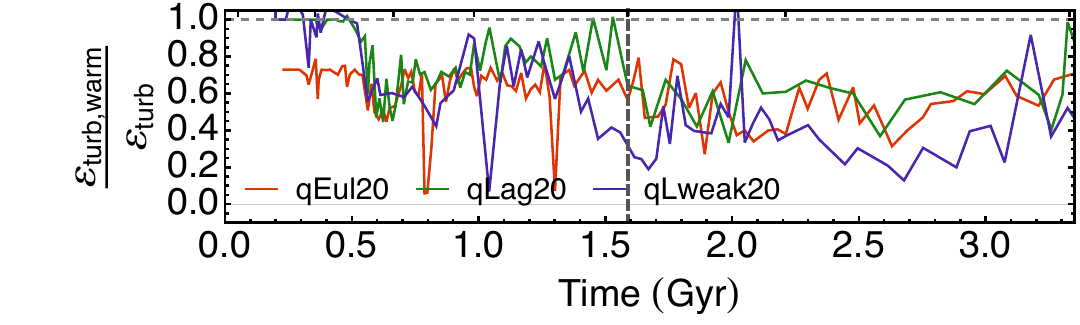}\\
    \caption{{\bf (Top panel)} All-phases specific turbulent energy (solid lines) and warm phase-only specific turbulent energy (dashed lines) evolution in the galactic region (\galregion) for the \qEulVeinte~(red), \qLagVeinte~(green) and \qLweakVeinte~(blue) runs. During the accretion phase, \qEulVeinte~has higher specific turbulence, particularly in the warm phase of the galaxy. During the feedback phase, all simulations portray similar amounts of total turbulent energy, but the runs with stronger SN feedback have a higher specific turbulent energy in the warm phase, as shown in the middle panel. {\bf (Middle panel)} \qEulVeinte~to \qLagVeinte~and \qEulVeinte~to \qLweakVeinte~warm phase specific turbulent energy ratio, with {\it sim20} replaced by \qLagVeinte~(green line) and \qLweakVeinte~(blue line). Turbulence in the warm phase of \qEulVeinte~is higher than in the other two runs. {\bf (Bottom panel)} Ratio of warm phase to total specific turbulent energy for each of the three simulations. \qEulVeinte~and \qLagVeinte~contain a stable fraction of warm to total specific turbulent energy whereas this ratio decreases for \qLagVeinte~throughout the feedback phase.}
    \label{fig:WarmTurbulence}
\end{figure}
Fig.~\ref{fig:EnergiesComparison} compares the turbulent and thermal energies in the ISM of the simulated galaxies (\qEulVeinte, top; \qLagVeinte, middle; \qLweakVeinte, bottom). We include an additional set of dashed lines ($\text{turb}_\text{vcirc}$) that approximate the global turbulent energy by assuming that the turbulent velocity for the gas in a cell $i$ is $v_\text{turb,i} = \left|\vec{v}_i\right| - v_\text{circ}$. This latter measurement provides a simple order-of-magnitude estimate of the turbulence combining all galactic scales, which we use to compute the specific energy $\varepsilon_\text{turb,vcirc}$. Comparing the dotted lines across all three simulations reveals no important differences of the large-scale turbulence. However, some changes appear when taking into consideration our measurements of small-scale turbulence. While all simulations have comparable small-scale turbulent energy density $\varepsilon_\text{turb,ss}$ during the feedback phase, \qEulVeinte~displays a higher turbulent energy during the accretion phase, correlating well with the time period when this simulation presents a faster magnetic amplification rate. Interestingly, this simulation displays a more fluctuating time profile for $\varepsilon_\text{turb,ss}$, suggesting that the \qEul~scheme better captures turbulence time-variability whereas the quasi-Lagrangian approach may be losing significant turbulent substructure. When comparing these turbulent energies with the specific thermal energy during the accretion phase, we find \qEulVeinte~to have a turbulence-dominated ISM whereas for the other runs, the thermal and turbulent energies are comparable. During the feedback phase, all runs have approximate equipartition between their small-scale turbulence and thermal energies, suggesting that SN feedback is driving turbulence in this later period of galaxy evolution. The mass-weighted and volume-weighted small-scale turbulent energy measures are comparable in all runs. The mass-weighted estimate remains slightly below the volume-weighted one, except for the \qLweak~runs. The $\rho_\text{gas}$ and $v_\text{turb}$ distributions shown in Fig.~\ref{fig:PDFs} reveal that this higher energy is derived from the higher densities in the \qLweakVeinte~simulation, which also displays a lower turbulent velocity distribution for $v_\text{turb,mw}$ than the other runs.

\subsubsection{Magnetic amplification in the warm phase of the ISM}
Here we study the turbulence in our $\Dres = 20\,\pc$ simulations, and we show that the additional amplification observed in \qEulVeinte~correlates with a higher degree of turbulence in its warm phase. The top panel of Fig.~\ref{fig:WarmTurbulence} shows the specific small-scale (i.e. computed as described in Section~\ref{ss:SSturbulence}) turbulent energy in the galactic region. Solid lines correspond to the specific energy across all phases, whereas dashed lines correspond to its value considering solely gas in the warm gas phase. While $\varepsilon_\text{turb}$ as measured in all-phases appears higher in \qEulVeinte, the most striking difference between the two refinement schemes is the oscillatory nature of the quasi-Eulerian prescription, particularly during the accretion phase. \qEulVeinte~varies from values comparable to \qLagVeinte~and \qLweakVeinte~to sharp increases of about 1 dex. While these peaks are also present to some extent in the other runs, they are less prominent and less frequent. Turbulence in the warm phase is higher in \qEulVeinte~than its quasi-Lagrangian counterparts at virtually all times. This is more explicitly shown in the middle panel of Fig.~\ref{fig:WarmTurbulence}, which depicts the ratio of warm phase turbulence between \qEulVeinte~and each of the \qLagVeinte~and \qLweakVeinte~runs. Finally, we observe a growing divergence between specific turbulence in the warm phase and all-phases during the feedback phase ($t \gtrsim 1.3$~Gyr), particularly clear for \qLweakVeinte. We show this in more detail in the bottom panel of Fig.~\ref{fig:WarmTurbulence}, where the ratio between the two is shown. This separation occurs to a lesser extent in \qEulVeinte~and \qLagVeinte, and correlates well with the lower amplification rates described in Fig.~\ref{fig:GrowthComparison}. 

\begin{figure*}
    \centering
    \includegraphics[width=\columnwidth]{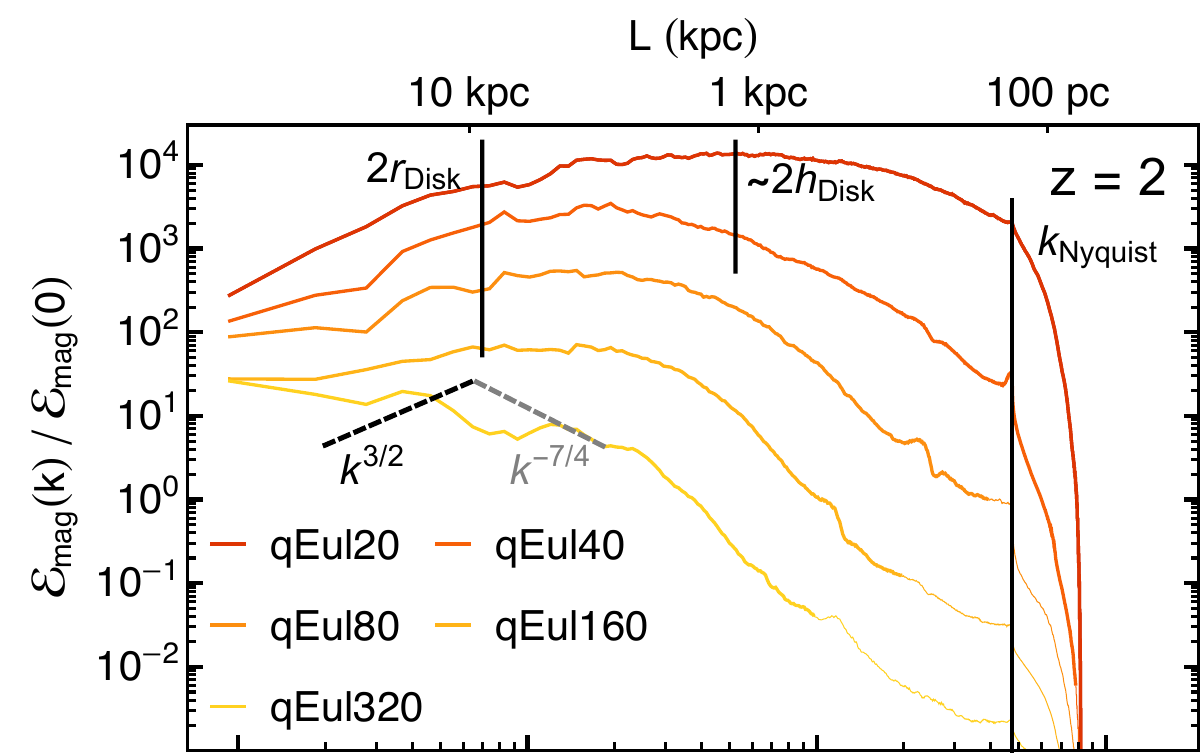}%
    \includegraphics[width=\columnwidth]{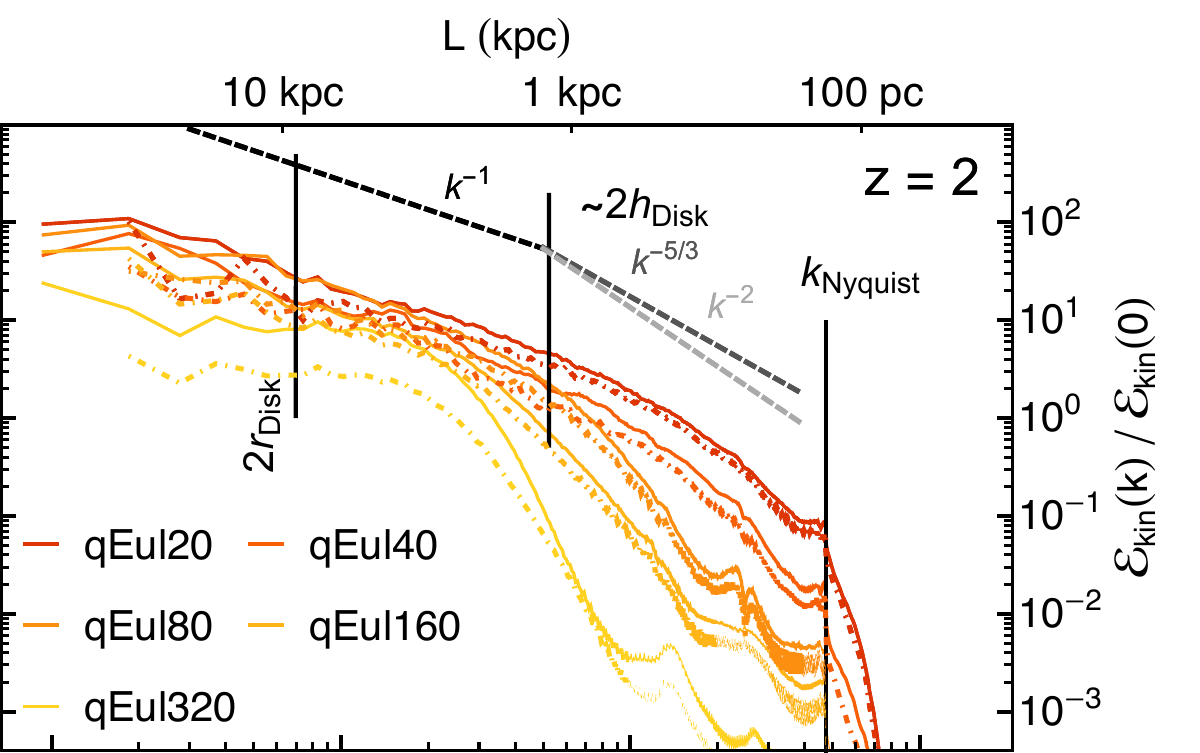}\\
    \includegraphics[width=\columnwidth]{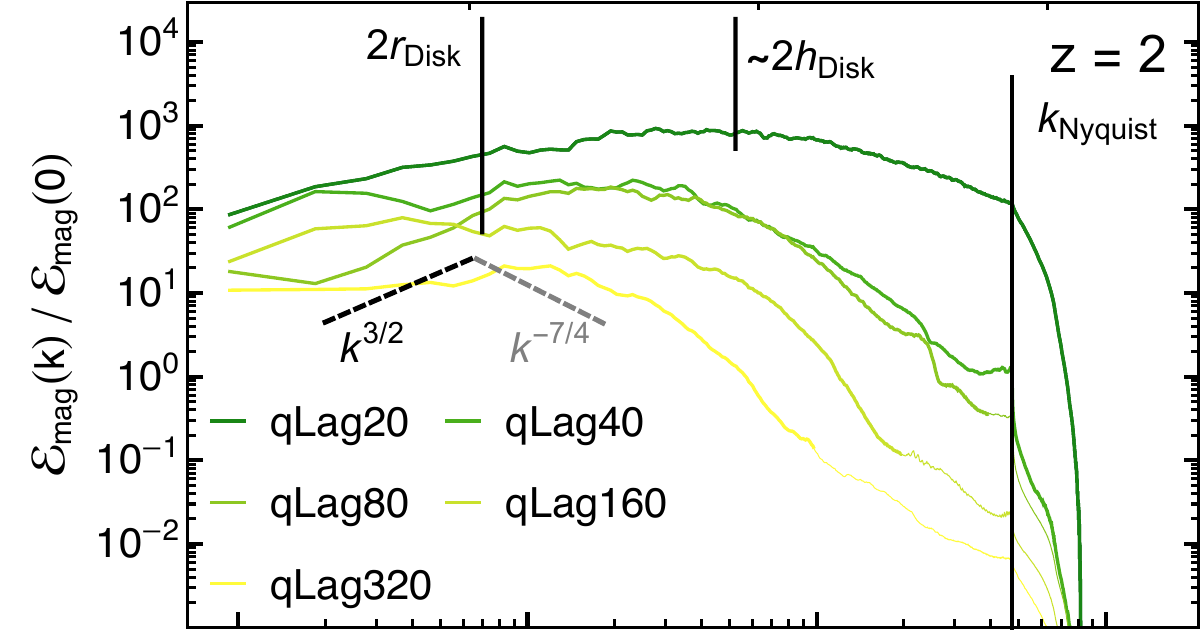}%
    \includegraphics[width=\columnwidth]{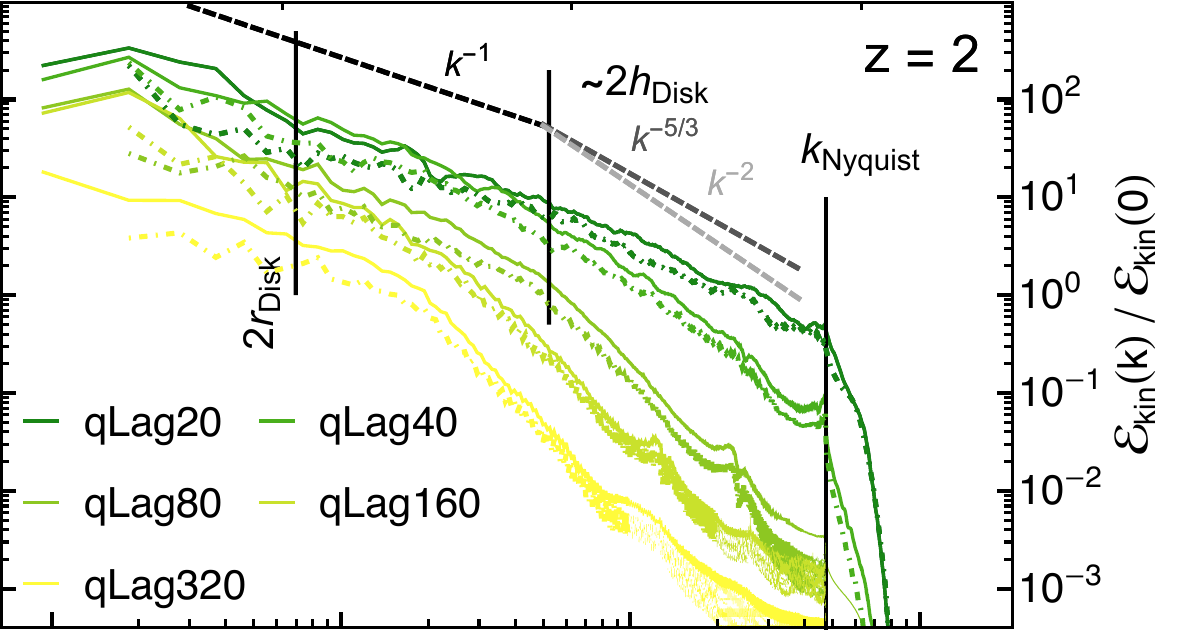}\\
    \includegraphics[width=\columnwidth]{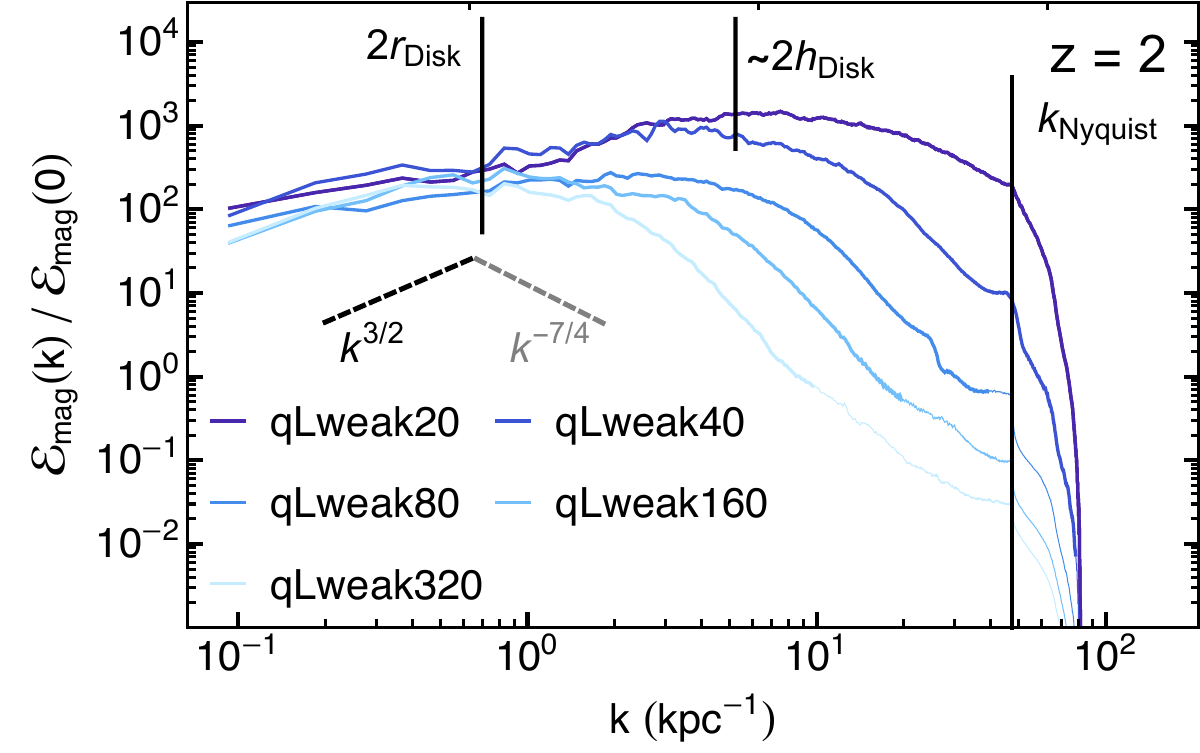}%
    \includegraphics[width=\columnwidth]{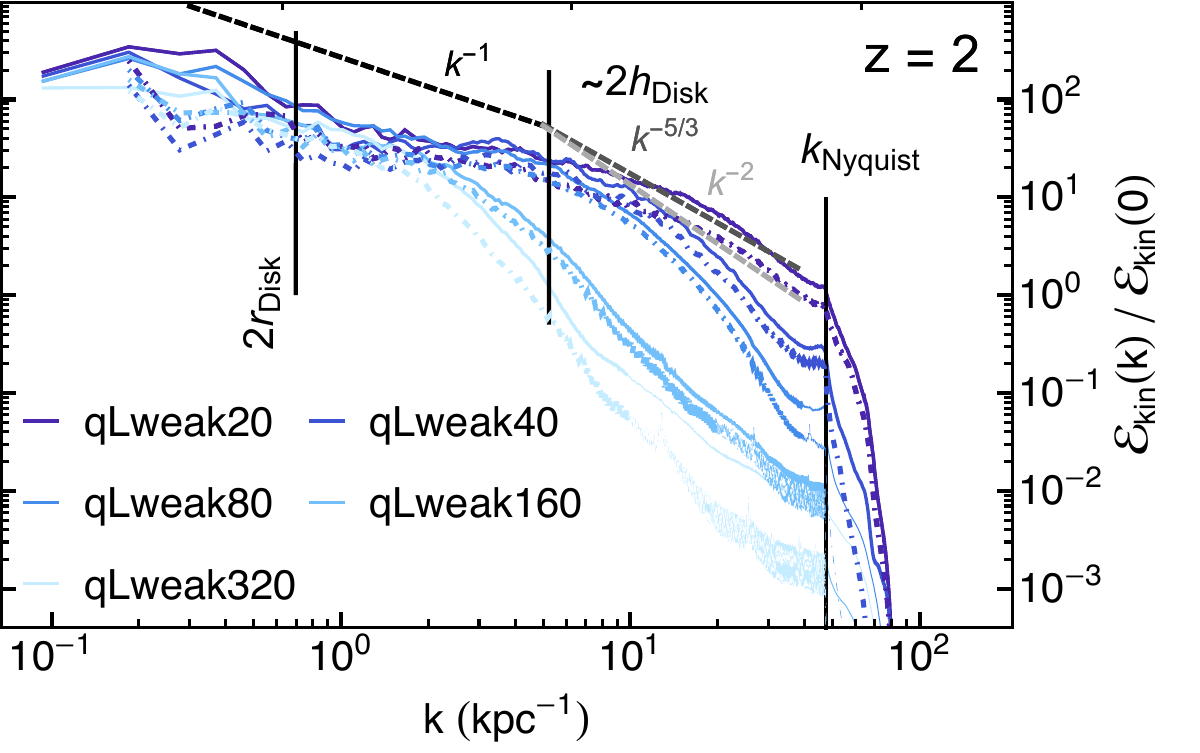}\\  
    \caption{Galaxy energy spectra for cubic boxes of 34~kpc per side (${\sim}0.4\, r_\text{DM}$) at $z = 2$, zero-padded. Left and right columns correspond to the magnetic and kinetic energy spectra, respectively. Each row shows the spectra of the \qEul~(top), \qLag~(middle) and \qLweak~(bottom) runs. Line shade goes from darker towards lighter as the resolution is decreased. For simulations with resolutions coarser than that of the FFT (i.e. 66~pc), line thickness is thinned beyond $k_\text{res} = \pi / \Dres$. We include physical length $L$ along the x-axis as top ticks. Vertical solid lines denote important scales for the galaxy. From lower to higher $k$ these are the galaxy gas disk size ($r_\text{Disk}$), gas disk thickness ($h_\text{Disk}$), and the FFT Nyquist frequency ($k_\text{Nyquist}$). Finally, we include as dot-dashed lines kinetic energy spectra from which we have approximately removed disk rotation (see text). Magnetic energy spectra with \qEul~refinement display more pronounced inverse-cascades, evidencing turbulent dynamo activity in our simulations.}
    \label{fig:AllSpectra}
\end{figure*}

Interestingly, the relative turbulence of the warm phase progressively decays in the \qLweak~simulations, whereas it only falls from $\sim 70$\% to $\sim 50$\% in the \qEul~and \qLag~runs. This suggests that stellar feedback is crucial in driving turbulence during this later stage of galaxy evolution, once most of the stellar mass and a rotationally supported disk emerges \citep{Martin-Alvarez2018}. The predominance of turbulence, particularly in the warm phase, supports the argument that, at the very least in our numerical simulations, the majority of the turbulent amplification in galaxy simulations takes place in the warm phase of the ISM.

\begin{figure}
    \centering
    \includegraphics[width=\columnwidth]{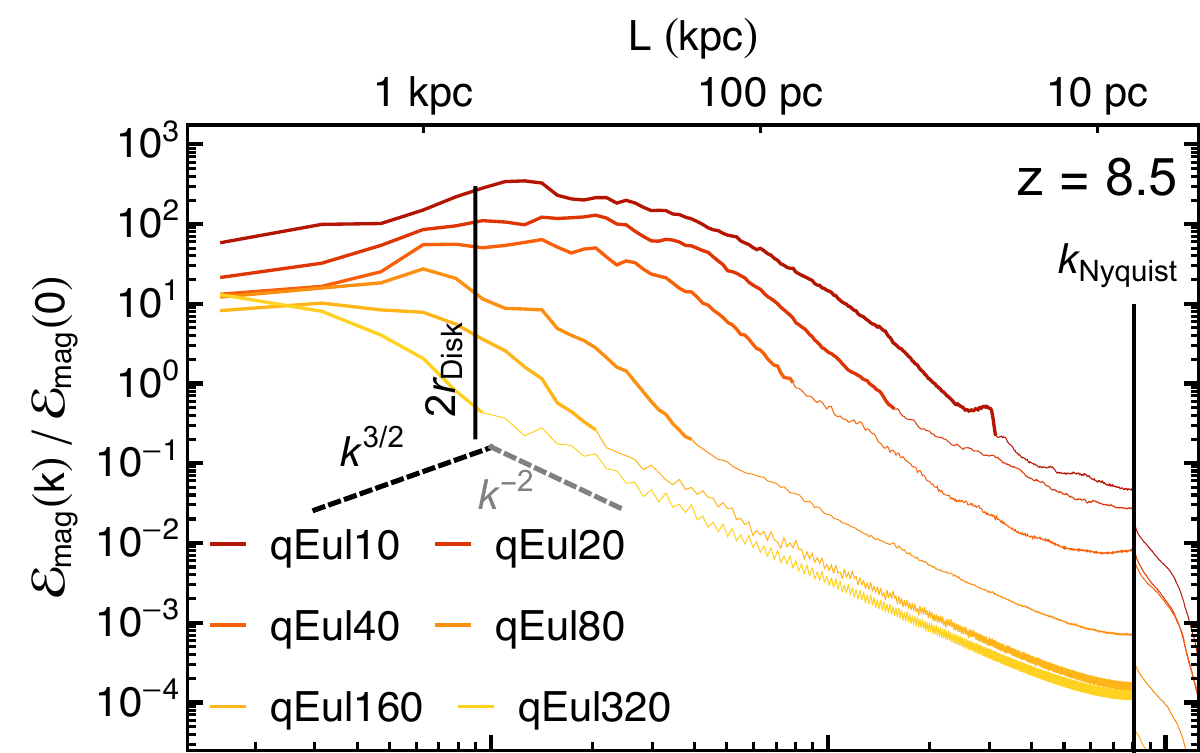}\\
    \includegraphics[width=\columnwidth]{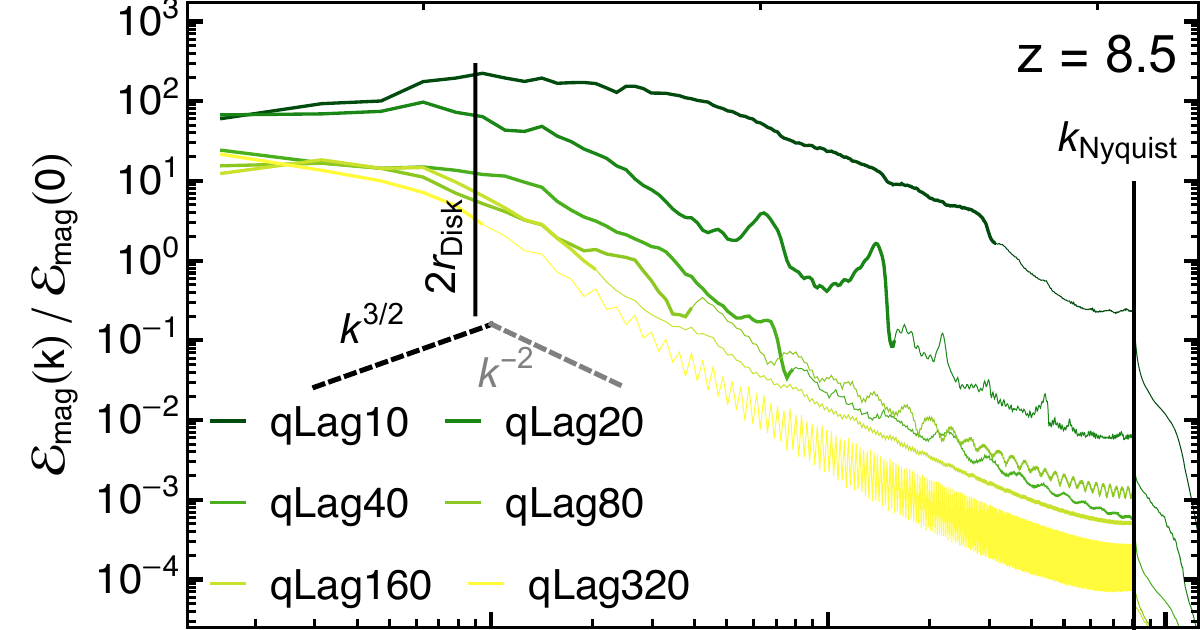}\\
    \includegraphics[width=\columnwidth]{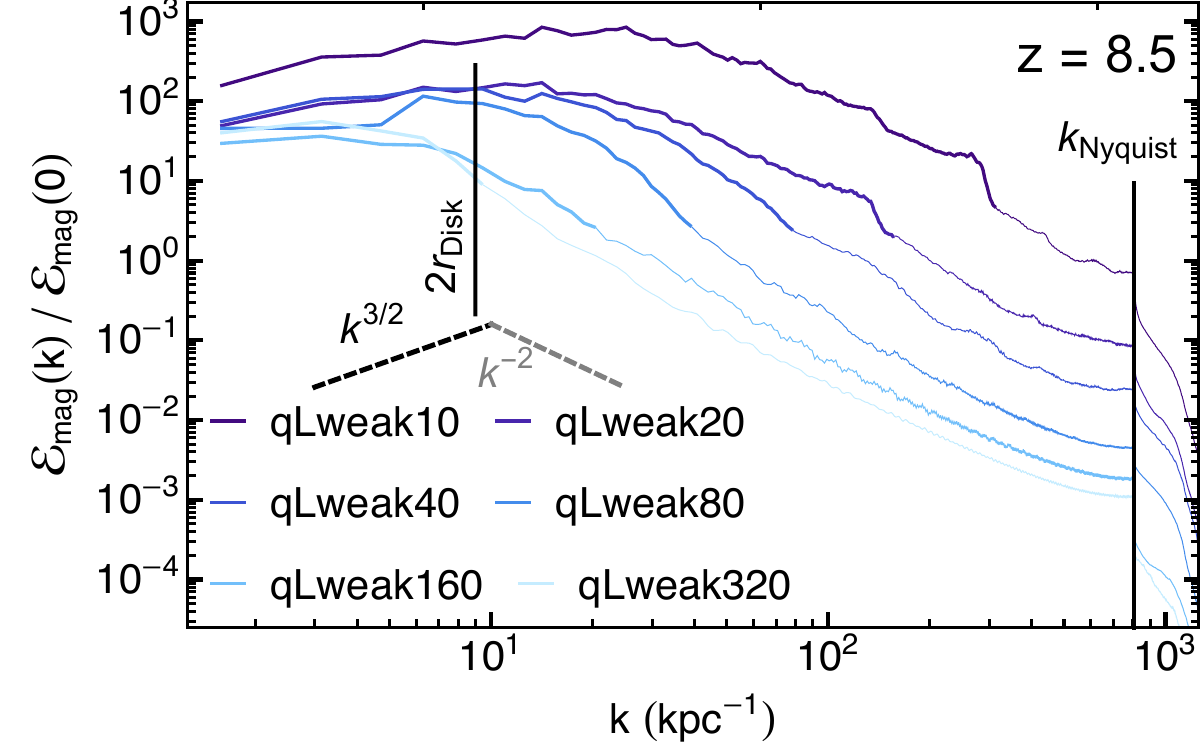}\\
    \caption{Magnetic energy spectra at $z = 8.5$ analogous to Fig.~\ref{fig:AllSpectra}, but now for regions 2~kpc on a side (${\sim}0.4\,r_\text{DM}$) with FFT boxes centred on the galaxy. The figure also includes spectra for our $\Dres = 10$~pc runs, which all have reached $z = 8.5$. As for the feedback phase, our runs with \qEul~refinement show a smoother and clearer inverse-cascade.}
    \label{fig:EmagSpectra}
\end{figure}

\begin{figure}
    \centering
    \includegraphics[width=\columnwidth]{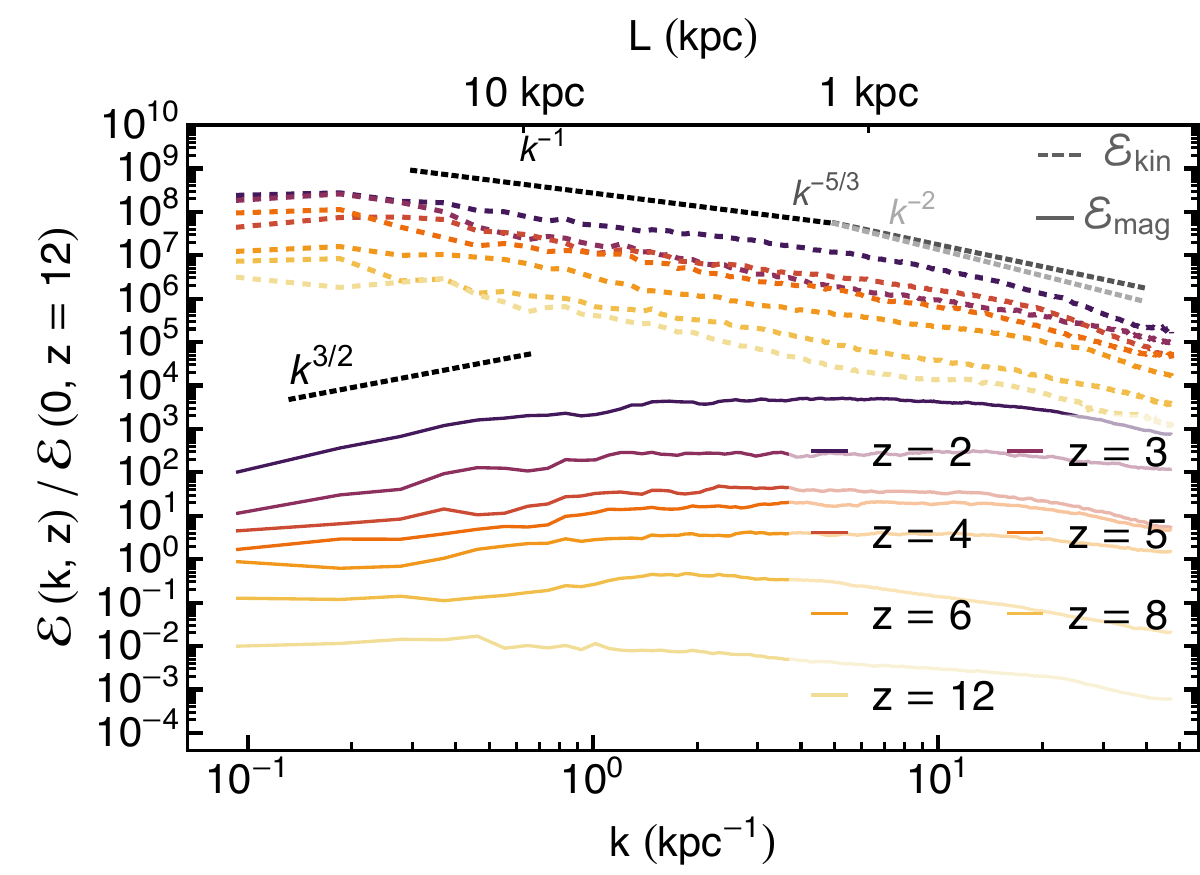}\\
    \caption{Evolution of the kinetic (dashed lines) and magnetic energy spectra (solid lines) for the galaxy in the \qEulVeinte~simulation. Magnetic energy power at scales of $\sim 1\,\kpc$ grows approximately 6 dex during the studied interval.}
    \label{fig:SpectraEvo}
\end{figure}

\begin{figure}
    \centering
    \includegraphics[width=\columnwidth]{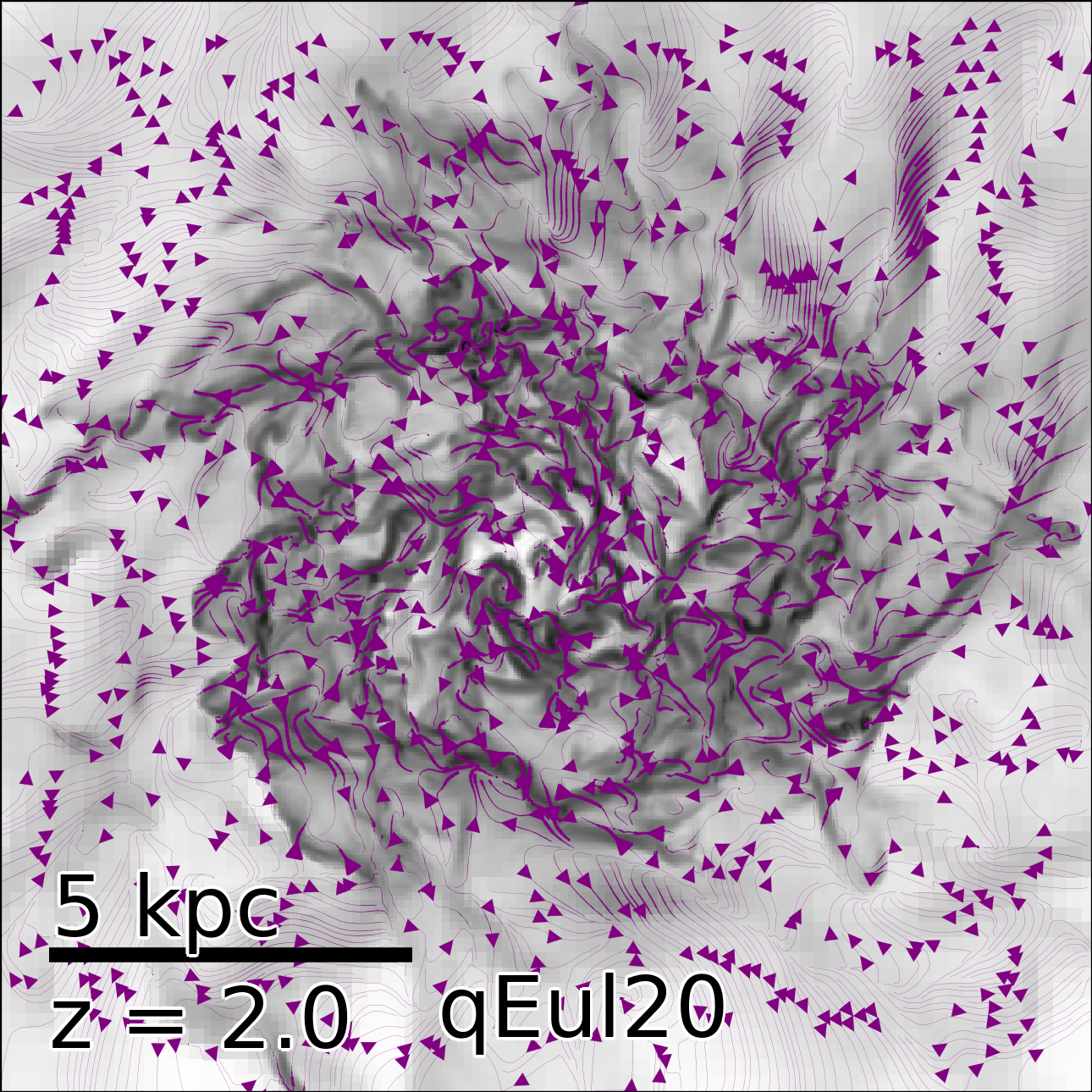}\\
    \includegraphics[width=\columnwidth]{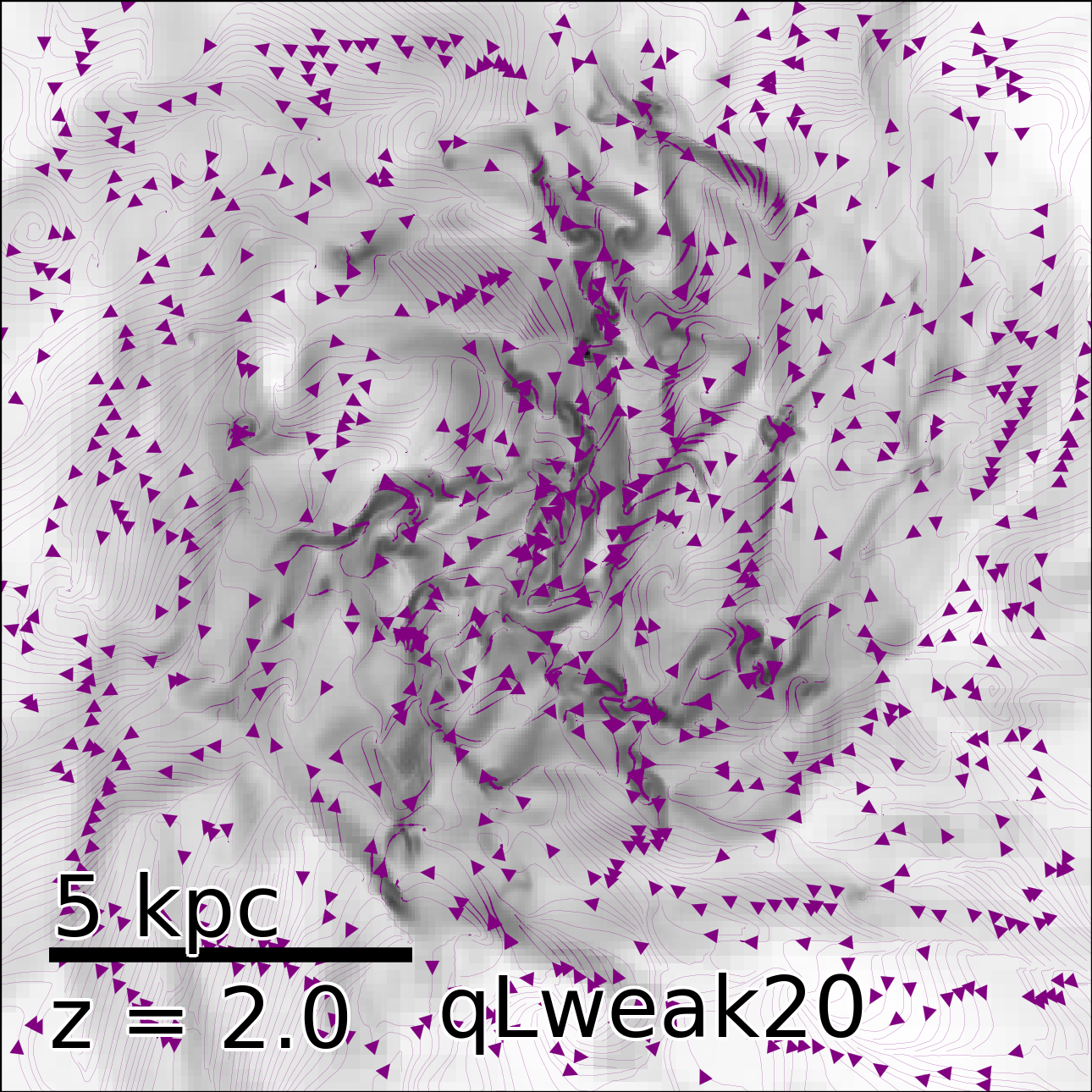}\\
    \caption{Magnetic field lines plotted over the magnetic field strength for the \qEulVeinte~(top) and \qLweakVeinte~(bottom) simulations. The simulation with our quasi-Eulerian refinement strategy displays a more turbulent structure of its magnetic field lines whereas the \qLweakVeinte~simulation illustrates a more organised structure with magnetic fields coherent on galactic scales.}
    \label{fig:MagLines}
\end{figure}

\subsubsection{Magnetic and turbulent energies spectral study}
\label{ss:SpectralStudy}

A well-known signature of small-scale turbulent dynamo amplification is the presence of an inverse cascade in the magnetic energy spectrum $\mathcal{E}_\text{mag}$. Among scalings of $\mathcal{E}_\text{mag} \propto k^\alpha$ with $\alpha \geq 0$, the case most characteristic of turbulent amplification is that in which Kolmogorov turbulence (with turbulent energy spectrum $\mathcal{E}_\text{kin} \propto k^{-5/3}$) yields $\alpha = 3/2$ \citep{Kazantsev1968}. This inverse-cascade emerges from magnetic amplification below the viscous scale $k_\nu$ (where $\nu$ is the numerical viscosity) but above the magnetic dissipation scale $k_\eta$ \citep{Schekochihin2002}. Numerical simulations such as the ones studied here cannot typically capture most of these sub-viscous scales due to their computationally limited spatial resolution and low Prandtl number (typically $Pm = \nu / \eta \sim 2$). Thus, we expect amplification to take place at the smallest scales resolved in our simulations. We assess $\mathcal{E}_\text{mag}$ and $\mathcal{E}_\text{kin}$ in all our runs.

We compute our energy spectra using a Fast Fourier transform (FFT\footnote{We employ the FFTW library (\href{http://www.fftw.org/}{http://www.fftw.org/}).}) of uniformly discretised cubic boxes centred on the galaxy onto which the AMR grid data is interpolated. Each of these cubes has a physical size of $L_\text{FFT} \sim 2\, \left(0.2\,r_\text{DM}\right)$ resolved with 512 cells per side. This cube is then zero-padded up to 1024 cells per side. Appendix \ref{ap:0pad} shows our magnetic energy spectra without zero-padding, briefly discussing the impact of such padding. Our FFT assumes periodic boundary conditions of this box. Appendix B in \citet{Martin-Alvarez2018} discussed these boundary conditions and the influence of galaxy morphology on the resulting spectra. 

To facilitate comparison amongst our different runs, all spectra are normalised to their $k = 0$ frequency. We show in Fig.~\ref{fig:AllSpectra} energy spectra at $z = 2$. These have FFT resolutions of $\Delta x_\text{FFT} \sim $~66 pc ($L_\text{FFT} (z = 2) =$ 34~kpc). Rows show runs for increasing $\Dres$ for \qEul~(top), \qLag~(middle) and \qLweak~(bottom) simulations. The left and right columns show magnetic and kinetic energy spectra, respectively, with line colouring shifting from higher to lower resolution as the colour shade becomes lighter. In addition to the kinetic energy spectra, we include as dot-dashed lines 'turbulent' energy spectra, for which we modify the toroidal component $v_\phi$ of the velocity in each AMR grid cell to substract disk rotation $v_\phi = v_{\phi, 0} - v_\text{circ} (r)$. These coordinates are defined aligning $v_z$ with the angular momentum of the galaxy. Finally, at scales below twice the resolution of a simulation ($L_\text{res} = 2 \Dres = 2 \pi / k_\text{res}$), line thickness is thinned to indicate unresolved scales.

By $z = 2$, the galaxy has formed a rotationally supported disk. We show with vertical lines in each plot the approximate size of the gas disk ($2 r_\text{Disk}$) and its thickness ($2 h_\text{Disk}$) in the $\Dres = 20\,\pc$ runs. While $r_\text{Disk}$ is relatively unchanged with resolution, we find $h_\text{Disk}$ increases as the resolution is degraded. Finally, we include an additional vertical black line at the Nyquist frequency ($k_\text{Nyquist}$). In the left column, runs with $\Dres \leq 80\,\pc$ have magnetic energy spectra that showcase inverse-cascades similar to those found in turbulence simulations \citep[e.g.][]{Federrath2014}, rising up to their maximum at scales slightly below the thickness of the disk. These are particularly prominent for the \qEul~runs. At $k > k_\text{peak}$, the spectra smoothly decay down to $k_\text{Nyquist}$.

The right column of Fig.~\ref{fig:AllSpectra} shows the kinetic energy spectra. We focus on the solid lines first. While all sets of simulations present roughly three ranges, these appear better reproduced by the \qEul~runs. We find an energy containing range extending to approximately across the length of the disk. At scales $k > \pi / h_\text{Disk}$, a Kolmogorov-like turbulent cascade appears with scaling $\alpha = -5/3$ for the \qEul~and \qLag~runs, but intermediate between Kolmogorov and Burgers ($\alpha = -2$) for \qLweak. Amongst the \qLweak~runs, \qLweakVeinte~has the most Kolmogorov-like spectrum. At scales of a few 100~pc ($k \sim 60\, \kpc^{-1}$), most energy spectra decay as $\alpha \sim k^{-11/4}$. The kinetic energy spectra from which we have approximately removed disk rotation (dot-dashed lines) portray a very similar picture to the standard kinetic energy spectra (solid lines) at scales smaller than disk thickness ($k > \pi / h_\text{Disk}$). However, toward lower $k$, dashed lines diverge from their solid counterparts, indicating that the reduction of kinetic energy corresponds to the rotation of the disk predominantly at these galactic scales. Interestingly, the \qLweak~runs have the lowest ratio of turbulent to kinetic spectra at large k. This points towards a lower degree of turbulence in these runs and more organised motions. The separation between both types of lines is reduced once again towards the very lowest $k$ values ($k < 0.3\, \kpc^{-1}$), as these are probing scales beyond the size of the galaxy.

We re-explore the magnetic energy spectra at redshift $z = 8.5$ in Fig.~\ref{fig:EmagSpectra}, for which \qEulDiez~is also available. These FFT boxes feature $\Delta x_\text{FFT} (z = 8.5) \sim $ 4~pc (and size $L_\text{FFT} (z = 8.5) =$ 2~kpc), zero-padded to 1024 cells per side. At this stage, the $2 r_\text{Disk}$ vertical black line is indicative of the approximate size of the entire system, which has not yet developed a rotationally supported disk. For runs with $\Dres \leq 80\,\pc$, magnetic energy spectra for \qEul,\qLag~and \qLweak~runs display a weak inverse cascade to scales of about $k \sim 20\,\kpc^{-1}$. Nonetheless, there is a more pronounced accumulation of energy in the \qEul~runs, with a steeper decay towards larger $k$ values. At $k > k_\text{res}$, all spectra decay as $\mathcal{E}_\text{mag} \propto k^{-2}$. The highest quasi-Lagrangian AMR refinement runs present bumps at scales comparable to the AMR level transitions, with a particularly prominent accumulation of energy at scales $k \lesssim k_\text{res}$. On the other hand, this only mildly occurs for \qEulDiez~in the \qEul~runs. Magnetic energy spectra appear to be in agreement with the behaviour observed for growth factors in Fig.~\ref{fig:GrowthComparison}, both during the accretion (Fig.~\ref{fig:EmagSpectra}) and feedback phases (Fig.~\ref{fig:AllSpectra}). 

We finally study the evolution of the kinetic and magnetic energy spectra of our \qEulVeinte~simulation in Fig.~\ref{fig:SpectraEvo}, to show that magnetic energy spectra develop an inverse-cascade and gain more power with time. We generate these using cubes with a fixed physical size of $L_\text{FFT} \sim 2\, \left(0.2\,r_\text{DM} (z = 2)\right)$ resolved with 512 cells per side, zero-padded to 1024 cells per side. We normalise kinetic and magnetic energy spectra to their $k = 0$ value at $z = 12$, to illustrate their growth. Power in kinetic energy spectra increases $\sim 2$ dex from $z = 12$ to $z = 2$. As the turbulence remains approximately constant (Fig.~\ref{fig:EnergiesComparison}), we expect this to be dominated by the galaxy growing in size and filling up the FFT box. Kinetic energy spectra maintain approximately constant shapes through their evolution, with a higher gain of power at small scales. Magnetic energy spectra gain a larger amount of power through this time period, with their large scale power increasing approximately $\sim 4$ dex. Our galaxy develops an inverse-cascade shortly after $z \sim 12$, with its onset displacing from $k \sim 1\,\kpc^{-1}$ towards scales $k < 0.1\,\kpc^{-1}$. Magnetic power at scales of $\sim 1\,\kpc$ ($k \sim 10\,\kpc^{-1}$) grows more than 6 dex, suggesting turbulent amplification.

\subsubsection{A glance at the structure of the magnetic field}
Finally, we perform a visual inspection of the magnetic field structure in our \qEulVeinte~and \qLweakVeinte~galaxies at $z = 2$. We show magnetic field streamlines overplotted over the magnetic field strength in Fig.~\ref{fig:MagLines}. Thicker lines depict locally stronger magnetic fields. \qEulVeinte~displays organised magnetic fields around the galaxy that become increasingly turbulent at distances shorter than $\sim 5\,\kpc$. Such a turbulent structure is in agreement with turbulent amplification. For comparison, we also show the \qLweakVeinte~galaxy. This simulation has a much lower growth factor (Fig.~\ref{fig:GrowthComparison}) and negligible amplification at this stage of the feedback phase. Magnetic field lines for this galaxy display an organised structure at galactic scales, and only some turbulent lines within the central region of the projection.

\section{Conclusions} 
\label{s:Conclusions}
In this manuscript, we have studied the evolution of the turbulent and magnetic components of a Milky Way-like galaxy comparing a conventional quasi-Lagrangian AMR refinement strategy with a new density-threshold, quasi-Eulerian refinement method that resolves the entire galaxy with an approximately uniform grid. Our simulations were generated using our own modified version of the {\sc ramses-mhd} code \citep{Teyssier2002,Fromang2006}, which computes the evolution of the magnetic field using a CT method, divergence-less down to numerical precision by construction. We explore a set of simulations with this new refinement strategy (\qEul) and two sets with the fiducial quasi-Lagrangian AMR. The first of the two employs an equal SN strength as the \qEul~set (\qLag), while the second one probes factor of 0.5 weaker SN feedback (\qLweak). 

Departing from an extremely weak magnetic field at the beginning of the simulations, we follow the evolution of the magnetic and (small-scale) turbulent energies in the galaxy. Our main findings are that:
\begin{enumerate}
    \item our quasi-Eulerian refinement strategy significantly increases the specific small-scale turbulent energy in the galaxy, especially within the warm phase.
    \item switching from a regular quasi-Lagrangian AMR strategy to our new quasi-Eulerian refinement at a fixed maximal spatial resolution leads to an increase of $\sim$1 dex in the specific magnetic energy in the galaxy (Fig.~\ref{fig:sEmComparison}).
    \item at a given spatial resolution, the growth of the magnetic energy is always found to be higher for \qEul~runs than for their corresponding \qLag~and \qLweak~simulations (Fig.~\ref{fig:GrowthComparison}). Encouragingly, growth rates also show better agreement with turbulent box simulations in terms of their scaling with spatial resolution, especially during the accretion phase.
    \item the additional magnetic amplification in \qEul~runs takes place even though the ISM is less dense and clumpy (Fig.~\ref{fig:PDFs}). This combined with the comparison of magnetic energy with the expected adiabatic compression estimate (Fig.~\ref{fig:Compression}) indicates that the extra amplification observed is not compressional in nature, but produced through stretching of magnetic field lines.
    \item the \qEulVeinte~run attains a large increase of warm-phase turbulence when compared with \qLagVeinte~and \qLweakVeinte~(Fig.~\ref{fig:WarmTurbulence}). The additional amplification is most likely the product of turbulent dynamo action occurring in the warm gas phase, due to its significant volume-filling fraction.
    \item the magnetic energy spectra are consistent with the turbulent dynamo amplification process regardless of the refinement method adopted. However, at $z = 2$, the \qEul~runs have a more clear concentration of magnetic energy at scales of $\sim 1$~kpc with an inverse cascade towards larger scales (Fig.~\ref{fig:AllSpectra}). 
\end{enumerate}

Due to their limited resolution, cosmological MHD simulations of turbulent dynamo amplification in galaxies employing CT schemes are far from reaching the magnetic field growth rates expected in nature. Nonetheless, refinement schemes such as the one used by our \qEul~simulations are a promising approach for future numerical experiments, opening a pathway towards the expected amplification with increasing resolution. As galaxy formation simulations mature into the new era where more realistic galaxies are generated, sophisticated models are being developed to accurately capture important processes such as star formation or stellar feedback at sub-galactic scales. In a similar manner, unresolved turbulent magnetic field amplification can be modelled with sub-grid methods such as that implemented for {\sc ramses} by \citet{Liu2021}. Nonetheless, with simulations commencing to better resolve the multi-phase ISM and capturing the physics of turbulence and magnetic fields, it is equally important to adopt more sophisticated refinement strategies. These are bound to make a significant difference in the modelling of turbulent and magnetic properties of galaxies, as well as to provide us with a better understanding of the kinematically complex structure within the volume-dominant warm and hot phases of the ISM.

\section*{Acknowledgements}
We kindly thank the referee for insightful comments and suggestions that contributed to improve the quality of this manuscript. This work was supported by the ERC Starting Grant 638707 'Black holes and their host galaxies: co-evolution across cosmic time'. This work is part of the Horizon-UK project, which used the DiRAC Complexity system, operated by the University of Leicester IT Services, which forms part of the STFC DiRAC HPC Facility (\href{www.dirac.ac.uk}{www.dirac.ac.uk}). This equipment is funded by BIS National E-Infrastructure capital grant ST/K000373/1 and STFC DiRAC Operations grant ST/K0003259/1. DiRAC is part of the National E-Infrastructure. The authors also acknowledge the usage of the FFTW library: \href{http://www.fftw.org/}{http://www.fftw.org/}.

\section*{Data availability}
The data employed in this manuscript is to be shared upon reasonable request contacting the corresponding author.


\bibliographystyle{mnras}
\bibliography{./references}

\begin{thebibliography}{}
\makeatletter
\relax
\def\mn@urlcharsother{\let\do\@makeother \do\$\do\&\do\#\do\^\do\_\do\%\do\~}
\def\mn@doi{\begingroup\mn@urlcharsother \@ifnextchar [ {\mn@doi@}
  {\mn@doi@[]}}
\def\mn@doi@[#1]#2{\def\@tempa{#1}\ifx\@tempa\@empty \href
  {http://dx.doi.org/#2} {doi:#2}\else \href {http://dx.doi.org/#2} {#1}\fi
  \endgroup}
\def\mn@eprint#1#2{\mn@eprint@#1:#2::\@nil}
\def\mn@eprint@arXiv#1{\href {http://arxiv.org/abs/#1} {{\tt arXiv:#1}}}
\def\mn@eprint@dblp#1{\href {http://dblp.uni-trier.de/rec/bibtex/#1.xml}
  {dblp:#1}}
\def\mn@eprint@#1:#2:#3:#4\@nil{\def\@tempa {#1}\def\@tempb {#2}\def\@tempc
  {#3}\ifx \@tempc \@empty \let \@tempc \@tempb \let \@tempb \@tempa \fi \ifx
  \@tempb \@empty \def\@tempb {arXiv}\fi \@ifundefined
  {mn@eprint@\@tempb}{\@tempb:\@tempc}{\expandafter \expandafter \csname
  mn@eprint@\@tempb\endcsname \expandafter{\@tempc}}}

\bibitem[\protect\citeauthoryear{Alves~Batista \& Saveliev}{Alves~Batista \&
  Saveliev}{2021}]{Alves-Batista2021}
Alves~Batista R.,  Saveliev A.,  2021, \mn@doi [Universe]
  {10.3390/universe7070223}, 7, 223

\bibitem[\protect\citeauthoryear{Attia, Teyssier, Katz, Kimm, Martin-Alvarez,
  Ocvirk  \& Rosdahl}{Attia et~al.}{2021}]{Attia2021}
Attia O.,  Teyssier R.,  Katz H.,  Kimm T.,  Martin-Alvarez S.,  Ocvirk P.,
  Rosdahl J.,  2021, MNRAS, 000, 1

\bibitem[\protect\citeauthoryear{Balsara \& Kim}{Balsara \&
  Kim}{2004}]{Balsara2004}
Balsara D.~S.,  Kim J.,  2004, \mn@doi [The Astrophysical Journal]
  {10.1086/381051}, 602, 1079

\bibitem[\protect\citeauthoryear{Balsara, Kim, Mac~Low  \& Mathews}{Balsara
  et~al.}{2004}]{Balsara2004b}
Balsara D.~S.,  Kim J.,  Mac~Low M.,   Mathews G.~J.,  2004, \mn@doi [The
  Astrophysical Journal] {10.1086/425297}, 617, 339

\bibitem[\protect\citeauthoryear{Beck}{Beck}{2007}]{Beck2007}
Beck R.,  2007, \mn@doi [Astronomy and Astrophysics]
  {10.1051/0004-6361:20066988}, 470, 539

\bibitem[\protect\citeauthoryear{Beck}{Beck}{2015}]{Beck2015}
Beck R.,  2015, \mn@doi [Astronomy {\&} Astrophysics]
  {10.1051/0004-6361/201425572}, 578, A93

\bibitem[\protect\citeauthoryear{Bendre, Gressel  \& Elstner}{Bendre
  et~al.}{2015}]{Bendre2015}
Bendre A.,  Gressel O.,   Elstner D.,  2015, \mn@doi [Astronomische
  Nachrichten] {10.1002/asna.201512211}, 336, 991

\bibitem[\protect\citeauthoryear{Bennett \& Sijacki}{Bennett \&
  Sijacki}{2020}]{Bennett2020}
Bennett J.~S.,  Sijacki D.,  2020, \mn@doi [Monthly Notices of the Royal
  Astronomical Society] {10.1093/mnras/staa2835}, 499, 597

\bibitem[\protect\citeauthoryear{Beresnyak}{Beresnyak}{2019}]{Beresnyak2019}
Beresnyak A.,  2019, ] {10.1007/s41115-019-0005-8}

\bibitem[\protect\citeauthoryear{Bernet, Miniati, Lilly, Kronberg  \&
  Dessauges-Zavadsky}{Bernet et~al.}{2008}]{Bernet2008}
Bernet M.~L.,  Miniati F.,  Lilly S.~J.,  Kronberg P.~P.,   Dessauges-Zavadsky
  M.,  2008, \mn@doi [Nature] {10.1038/nature07105}, 454, 302

\bibitem[\protect\citeauthoryear{Bhat \& Subramanian}{Bhat \&
  Subramanian}{2013}]{Bhat2013}
Bhat P.,  Subramanian K.,  2013, in Proceedings of the International
  Astronomical Union. Narnia, p.~400, \mn@doi{10.1017/S1743921314011697}, \url
  {http://academic.oup.com/mnras/article/429/3/2469/1008313/Fluctuation-dynamos-and-their-Faraday-rotation}

\bibitem[\protect\citeauthoryear{Bray \& Scaife}{Bray \&
  Scaife}{2018}]{Bray2018}
Bray J.~D.,  Scaife A. M.~M.,  2018, \mn@doi [The Astrophysical Journal]
  {10.3847/1538-4357/aac777}, 861, 3

\bibitem[\protect\citeauthoryear{Broderick, Chang  \& Pfrommer}{Broderick
  et~al.}{2012}]{Broderick2012}
Broderick A.~E.,  Chang P.,   Pfrommer C.,  2012, \mn@doi [The Astrophysical
  Journal] {10.1088/0004-637X/752/1/22}, 752, 22

\bibitem[\protect\citeauthoryear{Broderick, Tiede, Chang, Lamberts, Pfrommer,
  Puchwein, Shalaby  \& Werhahn}{Broderick et~al.}{2018}]{Broderick2018}
Broderick A.~E.,  Tiede P.,  Chang P.,  Lamberts A.,  Pfrommer C.,  Puchwein
  E.,  Shalaby M.,   Werhahn M.,  2018, \mn@doi [The Astrophysical Journal]
  {10.3847/1538-4357/aae5f2}, 868, 87

\bibitem[\protect\citeauthoryear{Burgers}{Burgers}{1948}]{Burgers1948}
Burgers J.~M.,  1948, \mn@doi [Advances in Applied Mechanics]
  {10.1016/S0065-2156(08)70100-5}, 1, 171

\bibitem[\protect\citeauthoryear{Butsky, Zrake, Kim, Yang  \& Abel}{Butsky
  et~al.}{2017}]{Butsky2017}
Butsky I.,  Zrake J.,  Kim J.-h.,  Yang H.-I.,   Abel T.,  2017, \mn@doi [The
  Astrophysical Journal] {10.3847/1538-4357/aa799f}, 843, 113

\bibitem[\protect\citeauthoryear{Chamandy, Subramanian  \& Shukurov}{Chamandy
  et~al.}{2013}]{Chamandy2013}
Chamandy L.,  Subramanian K.,   Shukurov A.,  2013, \mn@doi [Monthly Notices of
  the Royal Astronomical Society] {10.1093/mnras/sts297}, 428, 3569

\bibitem[\protect\citeauthoryear{Dedner, Kemm, Kr{\"{o}}ner, Munz, Schnitzer
  \& Wesenberg}{Dedner et~al.}{2002}]{Dedner2002}
Dedner A.,  Kemm F.,  Kr{\"{o}}ner D.,  Munz C.-D.,  Schnitzer T.,   Wesenberg
  M.,  2002, \mn@doi [Journal of Computational Physics]
  {10.1006/JCPH.2001.6961}, 175, 645

\bibitem[\protect\citeauthoryear{Dubois \& Teyssier}{Dubois \&
  Teyssier}{2010}]{Dubois2010}
Dubois Y.,  Teyssier R.,  2010, \mn@doi [Astronomy {\&} Astrophysics]
  {10.1051/0004-6361/200913014}, 523, A72

\bibitem[\protect\citeauthoryear{Dunkley et~al.,}{Dunkley
  et~al.}{2009}]{Dunkley2009}
Dunkley J.,  et~al., 2009, \mn@doi [Astrophysical Journal, Supplement Series]
  {10.1088/0067-0049/180/2/306}, 180, 306

\bibitem[\protect\citeauthoryear{Elmegreen \& Burkert}{Elmegreen \&
  Burkert}{2010}]{Elmegreen2010}
Elmegreen B.~G.,  Burkert A.,  2010, \mn@doi [Astrophysical Journal]
  {10.1088/0004-637X/712/1/294}, 712, 294

\bibitem[\protect\citeauthoryear{Evirgen, Gent, Shukurov, Fletcher  \&
  Bushby}{Evirgen et~al.}{2017}]{Evirgen2017}
Evirgen C.~C.,  Gent F.~A.,  Shukurov A.,  Fletcher A.,   Bushby P.,  2017,
  \mn@doi [Monthly Notices of the Royal Astronomical Society: Letters]
  {10.1093/mnrasl/slw196}, 464, L105

\bibitem[\protect\citeauthoryear{Evirgen, Gent, Shukurov, Fletcher  \&
  Bushby}{Evirgen et~al.}{2019}]{Evirgen2019}
Evirgen C.~C.,  Gent F.~A.,  Shukurov A.,  Fletcher A.,   Bushby P.~J.,  2019,
  \mn@doi [Monthly Notices of the Royal Astronomical Society]
  {10.1093/mnras/stz2084}, 488, 5065

\bibitem[\protect\citeauthoryear{Federrath}{Federrath}{2016}]{Federrath2016}
Federrath C.,  2016, \mn@doi [Journal of Plasma Physics]
  {10.1017/s0022377816001069}, 82, 535820601

\bibitem[\protect\citeauthoryear{Federrath \& Klessen}{Federrath \&
  Klessen}{2012}]{Federrath2012}
Federrath C.,  Klessen R.~S.,  2012, \mn@doi [Astrophysical Journal]
  {10.1088/0004-637X/761/2/156}, 761, 156

\bibitem[\protect\citeauthoryear{Federrath, Sur, Schleicher, Banerjee  \&
  Klessen}{Federrath et~al.}{2011}]{Federrath2011}
Federrath C.,  Sur S.,  Schleicher D.~R.,  Banerjee R.,   Klessen R.~S.,  2011,
  \mn@doi [Astrophysical Journal] {10.1088/0004-637X/731/1/62}, 731, 62

\bibitem[\protect\citeauthoryear{Federrath, Schober, Bovino  \&
  Schleicher}{Federrath et~al.}{2014}]{Federrath2014}
Federrath C.,  Schober J.,  Bovino S.,   Schleicher D.~R.,  2014, \mn@doi
  [Astrophysical Journal Letters] {10.1088/2041-8205/797/2/L19}, 797, L19

\bibitem[\protect\citeauthoryear{Ferland, Korista, Verner, Ferguson, Kingdon
  \& Verner}{Ferland et~al.}{1998}]{Ferland1998}
Ferland G.~J.,  Korista K.~T.,  Verner D.~A.,  Ferguson J.~W.,  Kingdon J.~B.,
   Verner E.~M.,  1998, \mn@doi [Publications of the Astronomical Society of
  the Pacific] {10.1086/316190}, 110, 761

\bibitem[\protect\citeauthoryear{F{\"{o}}rster~Schreiber
  et~al.,}{F{\"{o}}rster~Schreiber et~al.}{2009}]{ForsterSchreiber2009}
F{\"{o}}rster~Schreiber N.~M.,  et~al., 2009, \mn@doi [Astrophysical Journal]
  {10.1088/0004-637X/706/2/1364}, 706, 1364

\bibitem[\protect\citeauthoryear{Fromang, Hennebelle  \& Teyssier}{Fromang
  et~al.}{2006}]{Fromang2006}
Fromang S.,  Hennebelle P.,   Teyssier R.,  2006, \mn@doi [Astronomy {\&}
  Astrophysics] {10.1051/0004-6361:20065371}, 457, 371

\bibitem[\protect\citeauthoryear{Garaldi, Pakmor  \& Springel}{Garaldi
  et~al.}{2021}]{Garaldi2021}
Garaldi E.,  Pakmor R.,   Springel V.,  2021, \mn@doi [Monthly Notices of the
  Royal Astronomical Society] {10.1093/mnras/stab086}, 502, 5726

\bibitem[\protect\citeauthoryear{Gent}{Gent}{2012}]{Gent2012}
Gent F.,  2012, Ph.D. Thesis, p.~153

\bibitem[\protect\citeauthoryear{Gressel, Elstner  \& Ziegler}{Gressel
  et~al.}{2013}]{Gressel2013}
Gressel O.,  Elstner D.,   Ziegler U.,  2013, \mn@doi [Astronomy {\&}
  Astrophysics] {10.1051/0004-6361/201322349}, 560, A93

\bibitem[\protect\citeauthoryear{Gr{\o}nnow, Tepper-Garc{\'{i}}a  \&
  Bland-Hawthorn}{Gr{\o}nnow et~al.}{2018}]{Gronnow2018}
Gr{\o}nnow A.,  Tepper-Garc{\'{i}}a T.,   Bland-Hawthorn J.,  2018, \mn@doi
  [The Astrophysical Journal] {10.3847/1538-4357/aada0e}, 865, 64

\bibitem[\protect\citeauthoryear{Haardt \& Madau}{Haardt \&
  Madau}{1996}]{Haardt1996}
Haardt F.,  Madau P.,  1996, \mn@doi [The Astrophysical Journal]
  {10.1086/177035}, 461, 20

\bibitem[\protect\citeauthoryear{Hopkins}{Hopkins}{2016}]{Hopkins2016}
Hopkins P.~F.,  2016, \mn@doi [Monthly Notices of the Royal Astronomical
  Society] {10.1093/mnras/stw1578}, 462, 576

\bibitem[\protect\citeauthoryear{Hopkins, Kere{\v{s}}  \& Murray}{Hopkins
  et~al.}{2013}]{Hopkins2013}
Hopkins P.~F.,  Kere{\v{s}} D.,   Murray N.,  2013, \mn@doi [Monthly Notices of
  the Royal Astronomical Society] {10.1093/mnras/stt472}, 432, 2639

\bibitem[\protect\citeauthoryear{Iffrig \& Hennebelle}{Iffrig \&
  Hennebelle}{2017}]{Iffrig2017}
Iffrig O.,  Hennebelle P.,  2017, \mn@doi [Astronomy {\&} Astrophysics]
  {10.1051/0004-6361/201630290}, 604, A70

\bibitem[\protect\citeauthoryear{Inoue \& Yoshida}{Inoue \&
  Yoshida}{2019}]{Inoue2019}
Inoue S.,  Yoshida N.,  2019, \mn@doi [Monthly Notices of the Royal
  Astronomical Society] {10.1093/mnras/stz584}, 485, 3024

\bibitem[\protect\citeauthoryear{Ji, Oh  \& McCourt}{Ji et~al.}{2018}]{Ji2018}
Ji S.,  Oh O.~P.,   McCourt M.,  2018, \mn@doi [Monthly Notices of the Royal
  Astronomical Society] {10.1093/mnras/sty293}, 476, 852

\bibitem[\protect\citeauthoryear{Jin, Salim, Federrath, Tasker, Habe  \&
  Kainulainen}{Jin et~al.}{2017}]{Jin2017}
Jin K.,  Salim D.~M.,  Federrath C.,  Tasker E.~J.,  Habe A.,   Kainulainen
  J.~T.,  2017, \mn@doi [MNRAS] {10.1093/mnras/stx737}, 469, 383

\bibitem[\protect\citeauthoryear{Kandus, Kunze  \& Tsagas}{Kandus
  et~al.}{2011}]{Kandus2011}
Kandus A.,  Kunze K.~E.,   Tsagas C.~G.,  2011, \mn@doi [Physics Reports]
  {10.1016/j.physrep.2011.03.001}, 505, 1

\bibitem[\protect\citeauthoryear{Katz, Martin-Alvarez, Devriendt, Slyz  \&
  Kimm}{Katz et~al.}{2019}]{KMA2019}
Katz H.,  Martin-Alvarez S.,  Devriendt J.,  Slyz A.,   Kimm T.,  2019, \mn@doi
  [Monthly Notices of the Royal Astronomical Society] {10.1093/mnras/stz055},
  484, 2620

\bibitem[\protect\citeauthoryear{Katz et~al.,}{Katz et~al.}{2021}]{KMA2021}
Katz H.,  et~al., 2021, \mn@doi [Monthly Notices of the Royal Astronomical
  Society] {10.1093/mnras/stab2148}, 507, 1254

\bibitem[\protect\citeauthoryear{Kazantsev}{Kazantsev}{1968}]{Kazantsev1968}
Kazantsev A.~P.,  1968, Soviet Physics JETP, 26, 1031

\bibitem[\protect\citeauthoryear{Kimm \& Cen}{Kimm \& Cen}{2014}]{Kimm2014}
Kimm T.,  Cen R.,  2014, \mn@doi [Astrophysical Journal]
  {10.1088/0004-637X/788/2/121}, 788, 121

\bibitem[\protect\citeauthoryear{Kimm, Cen, Devriendt, Dubois  \& Slyz}{Kimm
  et~al.}{2015}]{Kimm2015}
Kimm T.,  Cen R.,  Devriendt J.,  Dubois Y.,   Slyz A.,  2015, \mn@doi [Monthly
  Notices of the Royal Astronomical Society] {10.1093/mnras/stv1211}, 451, 2900

\bibitem[\protect\citeauthoryear{Kimm, Katz, Haehnelt, Rosdahl, Devriendt  \&
  Slyz}{Kimm et~al.}{2017}]{Kimm2017}
Kimm T.,  Katz H.,  Haehnelt M.,  Rosdahl J.,  Devriendt J.,   Slyz A.,  2017,
  \mn@doi [Monthly Notices of the Royal Astronomical Society]
  {10.1093/mnras/stx052}, 466, stx052

\bibitem[\protect\citeauthoryear{Klessen \& Hennebelle}{Klessen \&
  Hennebelle}{2010}]{Klessen2010}
Klessen R.~S.,  Hennebelle P.,  2010, \mn@doi [Astronomy and Astrophysics]
  {10.1051/0004-6361/200913780}, 520, A17

\bibitem[\protect\citeauthoryear{Kolmogorov}{Kolmogorov}{1941}]{Kolmogorov1941}
Kolmogorov A.,  1941, Akademiia Nauk SSSR Doklady, 30, 301

\bibitem[\protect\citeauthoryear{K{\"{o}}rtgen, Federrath  \&
  Banerjee}{K{\"{o}}rtgen et~al.}{2017}]{Kortgen2017}
K{\"{o}}rtgen B.,  Federrath C.,   Banerjee R.,  2017, \mn@doi [Monthly Notices
  of the Royal Astronomical Society] {10.1093/mnras/stx2208}, 472, 2496

\bibitem[\protect\citeauthoryear{K{\"{o}}rtgen, Banerjee, Pudritz  \&
  Schmidt}{K{\"{o}}rtgen et~al.}{2019}]{Kortgen2019}
K{\"{o}}rtgen B.,  Banerjee R.,  Pudritz R.~E.,   Schmidt W.,  2019, \mn@doi
  [Monthly Notices of the Royal Astronomical Society] {10.1093/mnras/stz2491},
  489, 5004

\bibitem[\protect\citeauthoryear{Kritsuk et~al.,}{Kritsuk
  et~al.}{2011}]{Kritsuk2011}
Kritsuk A.~G.,  et~al., 2011, \mn@doi [Astrophysical Journal]
  {10.1088/0004-637X/737/1/13}, 737, 13

\bibitem[\protect\citeauthoryear{Kroupa}{Kroupa}{2001}]{Kroupa2001}
Kroupa P.,  2001, \mn@doi [Monthly Notices of the Royal Astronomical Society]
  {10.1046/j.1365-8711.2001.04022.x}, 322, 231

\bibitem[\protect\citeauthoryear{Krumholz \& Burkhart}{Krumholz \&
  Burkhart}{2016}]{Krumholz2016}
Krumholz M.~R.,  Burkhart B.,  2016, \mn@doi [Monthly Notices of the Royal
  Astronomical Society] {10.1093/mnras/stw434}, 458, 1671

\bibitem[\protect\citeauthoryear{Lesch \& Chiba}{Lesch \&
  Chiba}{1995}]{Lesch1995}
Lesch H.,  Chiba M.,  1995, Astronomy and Astrophysics, 297, 305

\bibitem[\protect\citeauthoryear{Liu, Kretschmer  \& Teyssier}{Liu
  et~al.}{2021}]{Liu2021}
Liu Y.,  Kretschmer M.,   Teyssier R.,  2021, MNRAS, 000, 1

\bibitem[\protect\citeauthoryear{Marinacci \& Vogelsberger}{Marinacci \&
  Vogelsberger}{2016}]{Marinacci2016}
Marinacci F.,  Vogelsberger M.,  2016, \mn@doi [Monthly Notices of the Royal
  Astronomical Society: Letters] {10.1093/mnrasl/slv176}, 456, L69

\bibitem[\protect\citeauthoryear{Martin-Alvarez, Devriendt, Slyz  \&
  Teyssier}{Martin-Alvarez et~al.}{2018}]{Martin-Alvarez2018}
Martin-Alvarez S.,  Devriendt J.,  Slyz A.,   Teyssier R.,  2018, \mn@doi
  [Monthly Notices of the Royal Astronomical Society] {10.1093/mnras/sty1623},
  479, 3343

\bibitem[\protect\citeauthoryear{Martin-Alvarez, Slyz, Devriendt  \&
  G{\'{o}}mez-Guijarro}{Martin-Alvarez et~al.}{2020}]{Martin-Alvarez2020}
Martin-Alvarez S.,  Slyz A.,  Devriendt J.,   G{\'{o}}mez-Guijarro C.,  2020,
  \mn@doi [Monthly Notices of the Royal Astronomical Society]
  {10.1093/mnras/staa1438}, 495, 4475

\bibitem[\protect\citeauthoryear{Martin-Alvarez, Katz, Sijacki, Devriendt  \&
  Slyz}{Martin-Alvarez et~al.}{2021}]{Martin-Alvarez2021}
Martin-Alvarez S.,  Katz H.,  Sijacki D.,  Devriendt J.,   Slyz A.,  2021,
  \mn@doi [MNRAS] {10.1093/mnras/stab968}, 504, 2517

\bibitem[\protect\citeauthoryear{Mocz, Pakmor, Springel, Vogelsberger,
  Marinacci  \& Hernquist}{Mocz et~al.}{2016}]{Mocz2016}
Mocz P.,  Pakmor R.,  Springel V.,  Vogelsberger M.,  Marinacci F.,   Hernquist
  L.,  2016, \mn@doi [Monthly Notices of the Royal Astronomical Society]
  {10.1093/mnras/stw2004}, 463, 477

\bibitem[\protect\citeauthoryear{Moss, Beck, Sokoloff, Stepanov, Krause  \&
  Arshakian}{Moss et~al.}{2013}]{Moss2013}
Moss D.,  Beck R.,  Sokoloff D.,  Stepanov R.,  Krause M.,   Arshakian T.~G.,
  2013, \mn@doi [Astronomy and Astrophysics] {10.1051/0004-6361/201321296},
  556, A147

\bibitem[\protect\citeauthoryear{Neronov \& Vovk}{Neronov \&
  Vovk}{2010}]{Neronov2010}
Neronov A.,  Vovk L.,  2010, \mn@doi [Science] {10.1126/science.1184192}, 328,
  73

\bibitem[\protect\citeauthoryear{Padoan \& Nordlund}{Padoan \&
  Nordlund}{2011}]{Padoan2011}
Padoan P.,  Nordlund A.,  2011, \mn@doi [Astrophysical Journal]
  {10.1088/0004-637X/730/1/40}, 730, 40

\bibitem[\protect\citeauthoryear{Pakmor \& Springel}{Pakmor \&
  Springel}{2013}]{Pakmor2013}
Pakmor R.,  Springel V.,  2013, \mn@doi [Monthly Notices of the Royal
  Astronomical Society] {10.1093/mnras/stt428}, 432, 176

\bibitem[\protect\citeauthoryear{Pakmor, Marinacci  \& Springel}{Pakmor
  et~al.}{2014}]{Pakmor2014}
Pakmor R.,  Marinacci F.,   Springel V.,  2014, \mn@doi [Astrophysical Journal
  Letters] {10.1088/2041-8205/783/1/L20}, 783, L20

\bibitem[\protect\citeauthoryear{Pakmor et~al.,}{Pakmor
  et~al.}{2017}]{Pakmor2017}
Pakmor R.,  et~al., 2017, \mn@doi [Monthly Notices of the Royal Astronomical
  Society] {10.1093/mnras/stx1074}, 469, 3185

\bibitem[\protect\citeauthoryear{Pillepich et~al.,}{Pillepich
  et~al.}{2018}]{Pillepich2018a}
Pillepich A.,  et~al., 2018, \mn@doi [Monthly Notices of the Royal Astronomical
  Society] {10.1093/mnras/stx2656}, 473, 4077

\bibitem[\protect\citeauthoryear{{Planck Collaboration}}{{Planck
  Collaboration}}{2015}]{PlanckCollaboration2015}
{Planck Collaboration} 2015, \mn@doi [Astronomy {\&} Astrophysics]
  {10.1051/0004-6361/201525821}, 594, A19

\bibitem[\protect\citeauthoryear{Powell, Roe, Linde, Gombosi  \&
  De~Zeeuw}{Powell et~al.}{1999}]{Powell1999}
Powell K.~G.,  Roe P.~L.,  Linde T.~J.,  Gombosi T.~I.,   De~Zeeuw D.~L.,
  1999, \mn@doi [Journal of Computational Physics] {10.1006/JCPH.1999.6299},
  154, 284

\bibitem[\protect\citeauthoryear{Powell, Slyz  \& Devriendt}{Powell
  et~al.}{2011}]{Powell2011}
Powell L.~C.,  Slyz A.,   Devriendt J.,  2011, \mn@doi [Monthly Notices of the
  Royal Astronomical Society] {10.1111/j.1365-2966.2011.18668.x}, 414, 3671

\bibitem[\protect\citeauthoryear{Power, Navarro, Jenkins, Frenk, White,
  Springel, Stadel  \& Quinn}{Power et~al.}{2003}]{Power2003}
Power C.,  Navarro J.~F.,  Jenkins A.,  Frenk C.~S.,  White S. D.~M.,  Springel
  V.,  Stadel J.,   Quinn T.,  2003, \mn@doi [Monthly Notices of the Royal
  Astronomical Society] {10.1046/j.1365-8711.2003.05925.x}, 338, 14

\bibitem[\protect\citeauthoryear{Pudritz \& Silk}{Pudritz \&
  Silk}{2002}]{Pudritz1989}
Pudritz R.~E.,  Silk J.,  2002, \mn@doi [The Astrophysical Journal]
  {10.1086/167625}, 342, 650

\bibitem[\protect\citeauthoryear{Rasera \& Teyssier}{Rasera \&
  Teyssier}{2006}]{Rasera2006}
Rasera Y.,  Teyssier R.,  2006, \mn@doi [Astronomy {\&} Astrophysics]
  {10.1051/0004-6361:20053116}, 445, 1

\bibitem[\protect\citeauthoryear{Rieder \& Teyssier}{Rieder \&
  Teyssier}{2016}]{Rieder2016}
Rieder M.,  Teyssier R.,  2016, \mn@doi [Monthly Notices of the Royal
  Astronomical Society] {10.1093/mnras/stv2985}, 457, 1722

\bibitem[\protect\citeauthoryear{Rieder \& Teyssier}{Rieder \&
  Teyssier}{2017a}]{Rieder2017a}
Rieder M.,  Teyssier R.,  2017a, \mn@doi [Monthly Notices of the Royal
  Astronomical Society] {10.1093/mnras/stx1670}, 471, 2674

\bibitem[\protect\citeauthoryear{Rieder \& Teyssier}{Rieder \&
  Teyssier}{2017b}]{Rieder2017b}
Rieder M.,  Teyssier R.,  2017b, \mn@doi [Monthly Notices of the Royal
  Astronomical Society] {10.1093/MNRAS/STX2276}, 472, 4368

\bibitem[\protect\citeauthoryear{Rosen \& Bregman}{Rosen \&
  Bregman}{1995}]{Rosen1995}
Rosen A.,  Bregman J.~N.,  1995, \mn@doi [The Astrophysical Journal]
  {10.1086/175303}, 440, 634

\bibitem[\protect\citeauthoryear{Safarzadeh \& Loeb}{Safarzadeh \&
  Loeb}{2019}]{Safarzadeh2019}
Safarzadeh M.,  Loeb A.,  2019

\bibitem[\protect\citeauthoryear{Sanati, Revaz, Schober, Kunze  \&
  Jablonka}{Sanati et~al.}{2020}]{Sanati2020}
Sanati M.,  Revaz Y.,  Schober J.,  Kunze K.~E.,   Jablonka P.,  2020, \mn@doi
  [Astronomy and Astrophysics] {10.1051/0004-6361/202038382}, 643, 54

\bibitem[\protect\citeauthoryear{Schekochihin, Cowley, Hammett, Maron  \&
  McWilliams}{Schekochihin et~al.}{2002}]{Schekochihin2002}
Schekochihin A.~A.,  Cowley S.~C.,  Hammett G.~W.,  Maron J.~L.,   McWilliams
  J.~C.,  2002, \mn@doi [New Journal of Physics] {10.1088/1367-2630/4/1/384},
  4, 84

\bibitem[\protect\citeauthoryear{Schlickeiser}{Schlickeiser}{2012}]{Schlickeiser2012}
Schlickeiser R.,  2012, \mn@doi [Physical Review Letters]
  {10.1103/PhysRevLett.109.261101}, 109, 261101

\bibitem[\protect\citeauthoryear{Schmidt}{Schmidt}{1959}]{Schmidt1959}
Schmidt M.,  1959, \mn@doi [The Astrophysical Journal] {10.1086/146614}, 129,
  243

\bibitem[\protect\citeauthoryear{Schober, Schleicher, Bovino  \&
  Klessen}{Schober et~al.}{2012}]{Schober2012}
Schober J.,  Schleicher D.,  Bovino S.,   Klessen R.~S.,  2012, \mn@doi
  [Physical Review E - Statistical, Nonlinear, and Soft Matter Physics]
  {10.1103/PhysRevE.86.066412}, 86, 066412

\bibitem[\protect\citeauthoryear{Schober, Rogachevskii, Brandenburg, Boyarsky,
  Fr{\"{o}}hlich, Ruchayskiy  \& Kleeorin}{Schober et~al.}{2018}]{Schober2018}
Schober J.,  Rogachevskii I.,  Brandenburg A.,  Boyarsky A.,  Fr{\"{o}}hlich
  J.,  Ruchayskiy O.,   Kleeorin N.,  2018, \mn@doi [The Astrophysical Journal]
  {10.3847/1538-4357/aaba75}, 858, 124

\bibitem[\protect\citeauthoryear{Stasyszyn, Dolag  \& Beck}{Stasyszyn
  et~al.}{2013}]{Stasyszyn2013}
Stasyszyn F.~A.,  Dolag K.,   Beck A.~M.,  2013, \mn@doi [Monthly Notices of
  the Royal Astronomical Society] {10.1093/mnras/sts018}, 428, 13

\bibitem[\protect\citeauthoryear{Steinwandel, Dolag, Lesch, Moster, Burkert  \&
  Prieto}{Steinwandel et~al.}{2019}]{Steinwandel2019}
Steinwandel U.~P.,  Dolag K.,  Lesch H.,  Moster B.~P.,  Burkert A.,   Prieto
  A.,  2019

\bibitem[\protect\citeauthoryear{Su, Hopkins, Hayward, Faucher-Gigu{\`{e}}re,
  Kere{\v{s}}, Ma  \& Robles}{Su et~al.}{2017}]{Su2017}
Su K.~Y.,  Hopkins P.~F.,  Hayward C.~C.,  Faucher-Gigu{\`{e}}re C.~A.,
  Kere{\v{s}} D.,  Ma X.,   Robles V.~H.,  2017, \mn@doi [Monthly Notices of
  the Royal Astronomical Society] {10.1093/MNRAS/STX1463}, 471, 144

\bibitem[\protect\citeauthoryear{Subramanian}{Subramanian}{2016}]{Subramanian2016}
Subramanian K.,  2016, \mn@doi [Reports on Progress in Physics]
  {10.1088/0034-4885/79/7/076901}, 79, 076901

\bibitem[\protect\citeauthoryear{Sur, Federrath, Schleicher, Banerjee  \&
  Klessen}{Sur et~al.}{2012}]{Sur2012}
Sur S.,  Federrath C.,  Schleicher D. R.~G.,  Banerjee R.,   Klessen R.~S.,
  2012, \mn@doi [Monthly Notices of the Royal Astronomical Society]
  {10.1111/j.1365-2966.2012.21100.x}, 423, 3148

\bibitem[\protect\citeauthoryear{Teyssier}{Teyssier}{2002}]{Teyssier2002}
Teyssier R.,  2002, \mn@doi [Astronomy {\&} Astrophysics]
  {10.1051/0004-6361:20011817}, 385, 337

\bibitem[\protect\citeauthoryear{Teyssier, Fromang  \& Dormy}{Teyssier
  et~al.}{2006}]{Teyssier2006}
Teyssier R.,  Fromang S.,   Dormy E.,  2006, \mn@doi [Journal of Computational
  Physics] {10.1016/j.jcp.2006.01.042}, 218, 44

\bibitem[\protect\citeauthoryear{T{\'{o}}th}{T{\'{o}}th}{2000}]{Toth2000}
T{\'{o}}th G.,  2000, \mn@doi [Journal of Computational Physics]
  {10.1006/jcph.2000.6519}, 161, 605

\bibitem[\protect\citeauthoryear{Trebitsch, Blaizot, Rosdahl, Devriendt  \&
  Slyz}{Trebitsch et~al.}{2017}]{Trebitsch2017}
Trebitsch M.,  Blaizot J.,  Rosdahl J.,  Devriendt J.,   Slyz A.,  2017,
  \mn@doi [Monthly Notices of the Royal Astronomical Society]
  {10.1093/mnras/stx1060}, 470, 224

\bibitem[\protect\citeauthoryear{Tweed, Devriendt, Blaizot, Colombi  \&
  Slyz}{Tweed et~al.}{2009}]{Tweed2009}
Tweed D.,  Devriendt J.,  Blaizot J.,  Colombi S.,   Slyz A.,  2009, \mn@doi
  [Astronomy {\&} Astrophysics] {10.1051/0004-6361/200911787}, 506, 647

\bibitem[\protect\citeauthoryear{Vazza, Br{\"{u}}ggen, Gheller  \& Wang}{Vazza
  et~al.}{2014}]{Vazza2014}
Vazza F.,  Br{\"{u}}ggen M.,  Gheller C.,   Wang P.,  2014, \mn@doi [Monthly
  Notices of the Royal Astronomical Society] {10.1093/mnras/stu1896}, 445, 3706

\bibitem[\protect\citeauthoryear{Vazza, Br{\"{u}}ggen, Gheller, Hackstein,
  Wittor  \& Hinz}{Vazza et~al.}{2017}]{Vazza2017}
Vazza F.,  Br{\"{u}}ggen M.,  Gheller C.,  Hackstein S.,  Wittor D.,   Hinz
  P.~M.,  2017, \mn@doi [Classical and Quantum Gravity]
  {10.1088/1361-6382/aa8e60}, 34, 234001

\bibitem[\protect\citeauthoryear{Vazza, Brunetti, Br{\"{u}}ggen  \&
  Bonafede}{Vazza et~al.}{2018}]{Vazza2018}
Vazza F.,  Brunetti G.,  Br{\"{u}}ggen M.,   Bonafede A.,  2018, \mn@doi
  [Monthly Notices of the Royal Astronomical Society] {10.1093/mnras/stx2830},
  474, 1672

\bibitem[\protect\citeauthoryear{Wang \& Abel}{Wang \& Abel}{2009}]{Wang2009}
Wang P.,  Abel T.,  2009, \mn@doi [The Astrophysical Journal]
  {10.1088/0004-637X/696/1/96}, 696, 96

\bibitem[\protect\citeauthoryear{Widrow}{Widrow}{2002}]{Widrow2002}
Widrow L.~M.,  2002, \mn@doi [Reviews of Modern Physics]
  {10.1103/RevModPhys.74.775}, 74, 775

\bibitem[\protect\citeauthoryear{Zamora-Avil{\'{e}}s, V{\'{a}}zquez-Semadeni,
  K{\"{o}}rtgen, Banerjee  \& Hartmann}{Zamora-Avil{\'{e}}s
  et~al.}{2018}]{Zamora-Aviles2018}
Zamora-Avil{\'{e}}s M.,  V{\'{a}}zquez-Semadeni E.,  K{\"{o}}rtgen B.,
  Banerjee R.,   Hartmann L.,  2018, \mn@doi [Monthly Notices of the Royal
  Astronomical Society] {10.1093/mnras/stx3080}, 474, 4824

\makeatother
\end{thebibliography}

\appendix
\section{Magnetic divergence constraint}\label{ap:Divergence}
\begin{figure}
    \centering
    \includegraphics[width=\columnwidth]{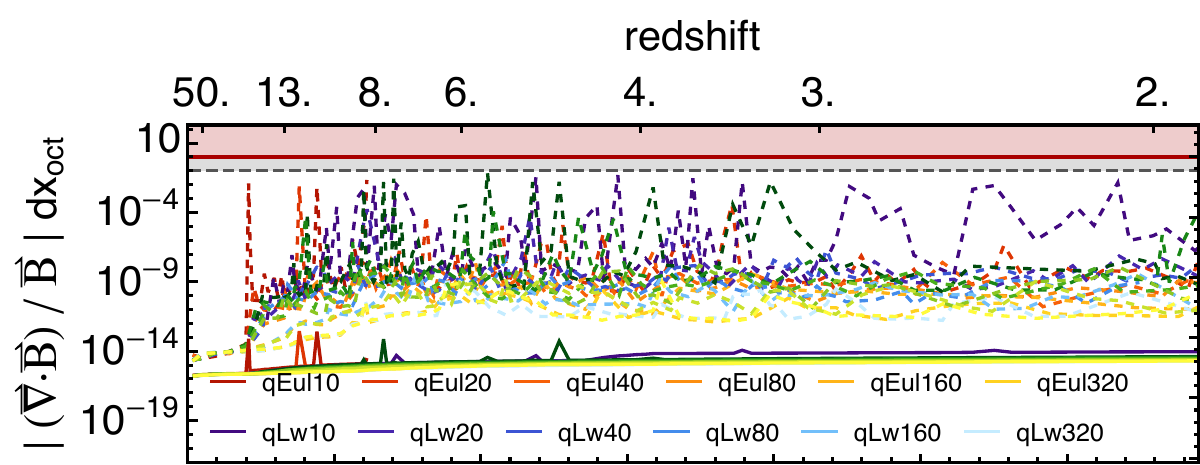}\\
    \includegraphics[width=\columnwidth]{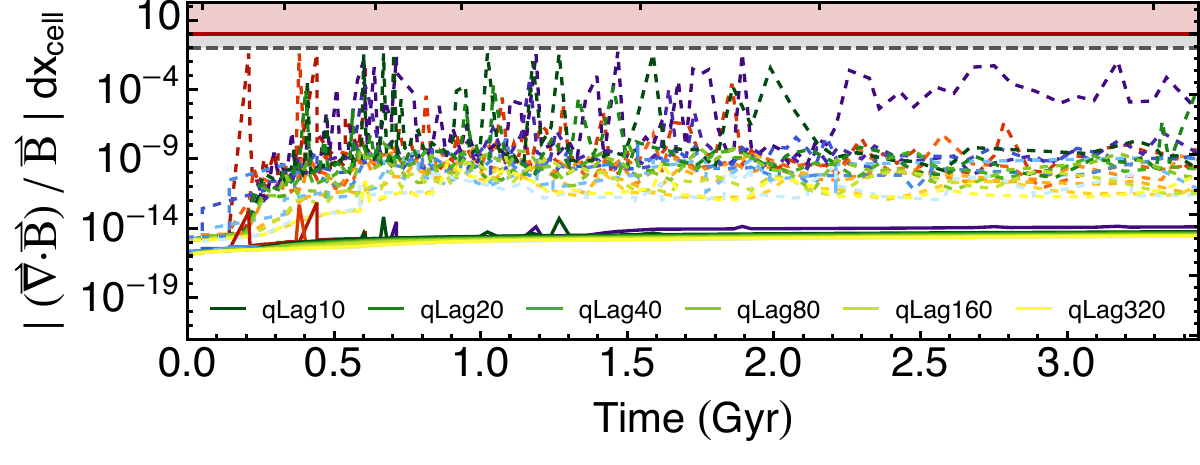}\\
    \caption{({\bf Top}) For all the octs (cubic groups of $2^3$~cells) in each of our simulations, solid (dashed) lines show the time evolution of the average (maximum) oct dimensionless ratio between the magnetic field divergence $|\vec{\nabla} \cdot \vec{B}|$ multiplied by resolution element size $\Delta x_\text{oct}$ and the local magnetic field $ |\vec{B}|$. For the computation of the octs divergence, we employ the cell-centred magnetic field of each comprising cell. ({\bf Bottom}) Same as for the top panel, but computing the relative divergence for all cells employing {\sc ramses} face-centred magnetic fields. The highest values for relative divergence are frequently found in cells with low $|\vec{B}|$. To avoid X- and O-points where $|\vec{B}| = 0$, we compute the total magnetic field for each cell in the bottom panel using an extended kernel of $1.5 \Delta x_\text{cell}$. For both panels, red solid and grey dashed lines delimit $|\vec{\nabla} \cdot \vec{B} / \vec{B}|\; \Delta x_\text{cell} =$ 1 and 0.1, respectively. Both face-centred and cell-centred magnetic fields maintain negligible ratios of magnetic divergence with respect to the local magnetic field. This is true for average and maximum values of this quantity in all cells and at all times.}
    \label{fig:Divergences}
\end{figure}

We briefly review the magnetic divergence of our simulations in Fig.~\ref{fig:Divergences}. Here we show the maximal (dashed) and average (solid) divergence multiplied by cell length ratio to magnetic field $|\vec{\nabla} \cdot \vec{B} / \vec{B}|\; \Delta x_\text{cell}$ for all cells in all simulations. The maximal relative divergence ratio in all the runs is virtually always below the percent level. We note that the divergence-less behaviour of RAMSES holds by construction for cell-centred magnetic fields. This is reflected in the top panel of Fig.~\ref{fig:Divergences}, where we show for each oct (i.e. group of $2^3$ cells) its divergence to field ratio using the cell-centred magnetic field components of each of its comprising 8 cells. The numerical magnetic divergence ratio value for all cells/octs with respect to their magnetic field are negligible at all times. This holds not only for the average value, but also for the maximum value found at all times amongst all the studied cells (e.g. $\sim10^{7}$~octs / $10^{8}$~cells for \qEulVeinte).

\section{The effect of zero-padding on cosmological magnetic energy spectra}\label{ap:0pad}

\begin{figure*}
    \centering
    \includegraphics[width=\columnwidth]{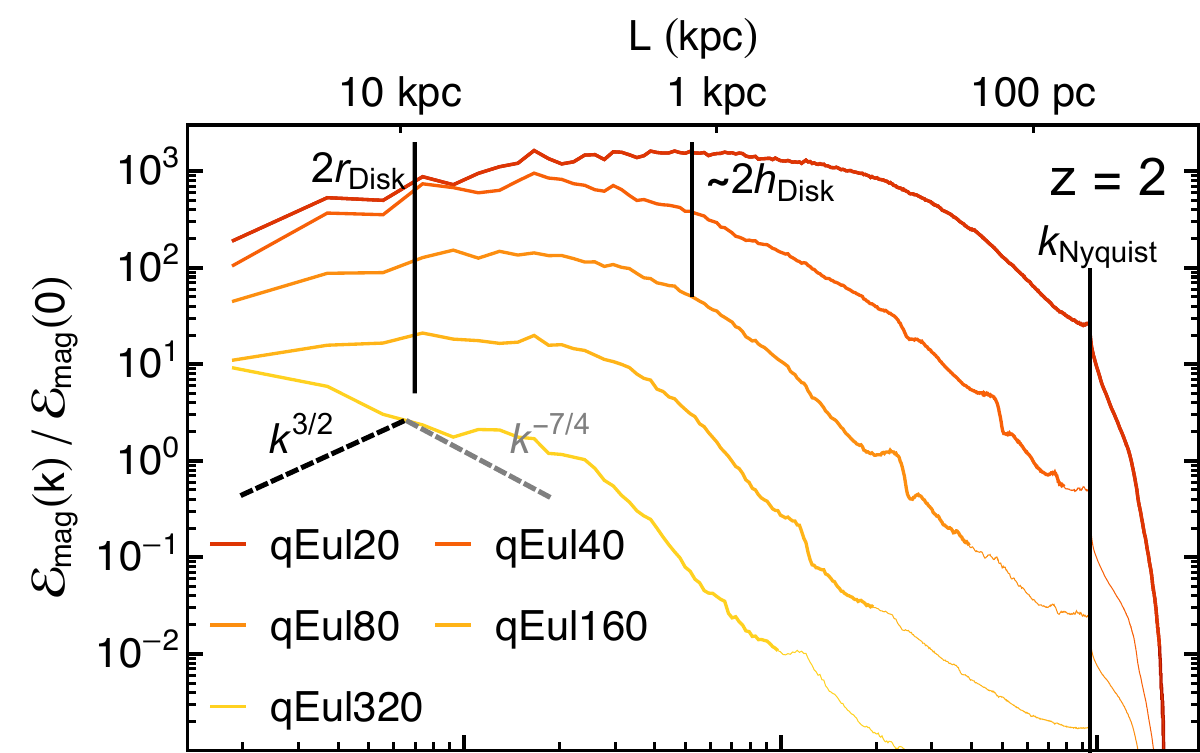}%
    \includegraphics[width=\columnwidth]{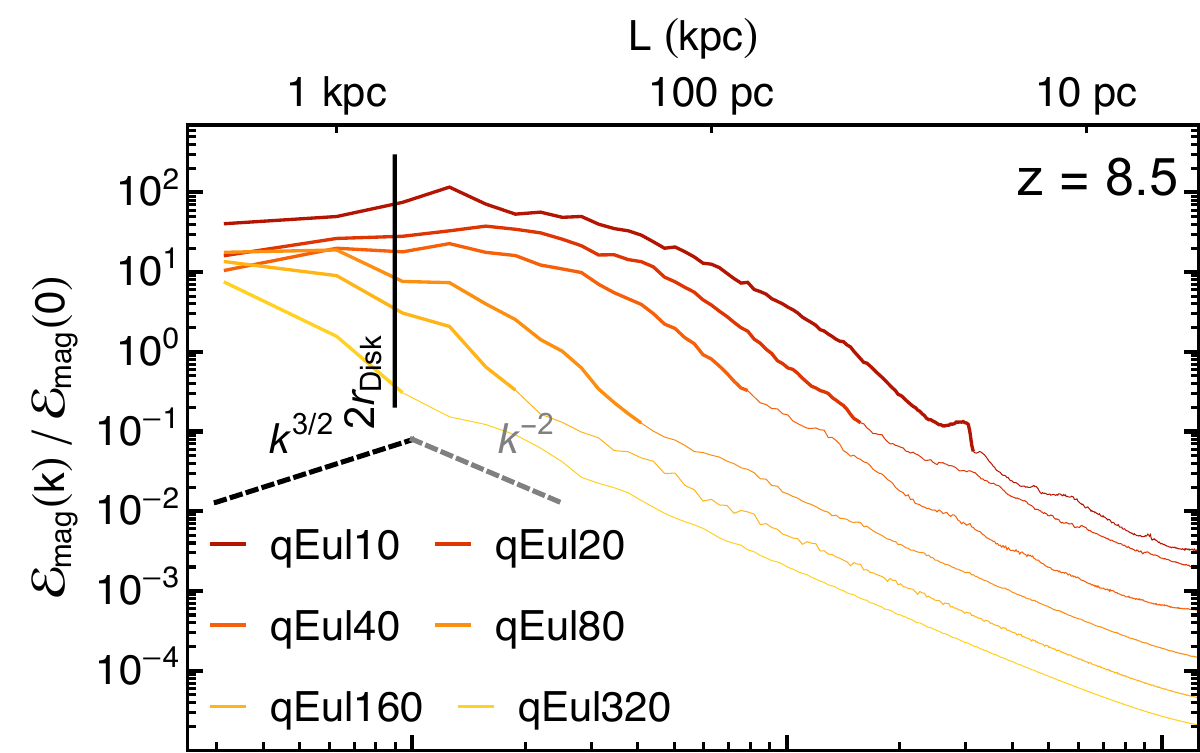}\\
    \includegraphics[width=\columnwidth]{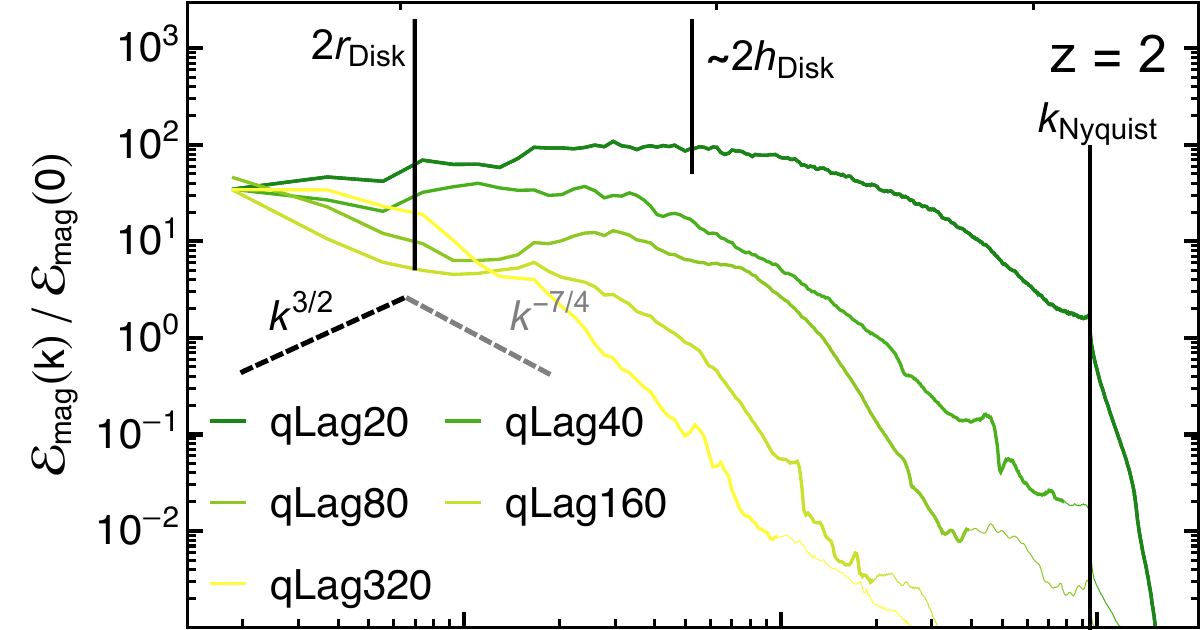}%
    \includegraphics[width=\columnwidth]{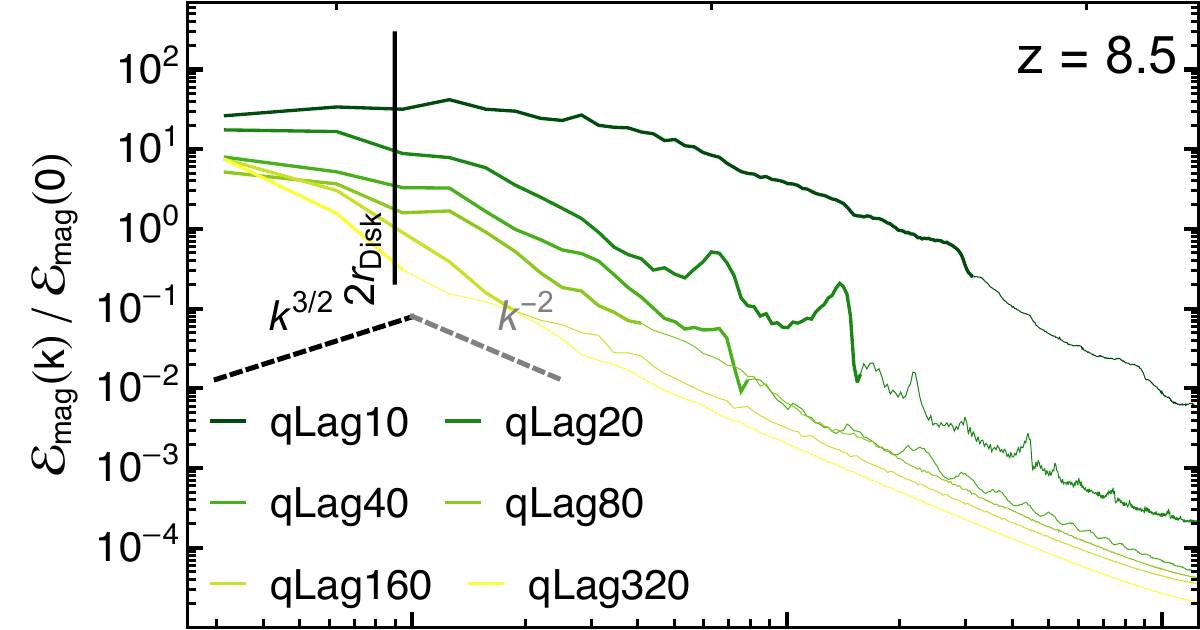}\\
    \includegraphics[width=\columnwidth]{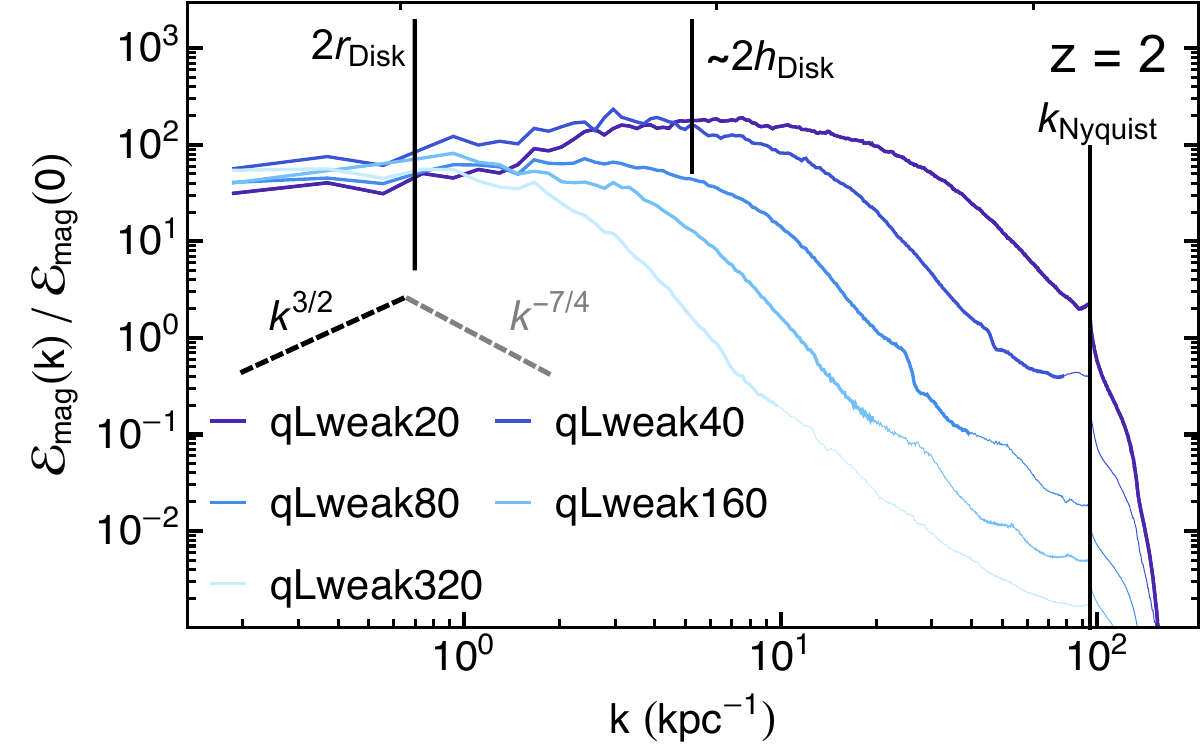}%
    \includegraphics[width=\columnwidth]{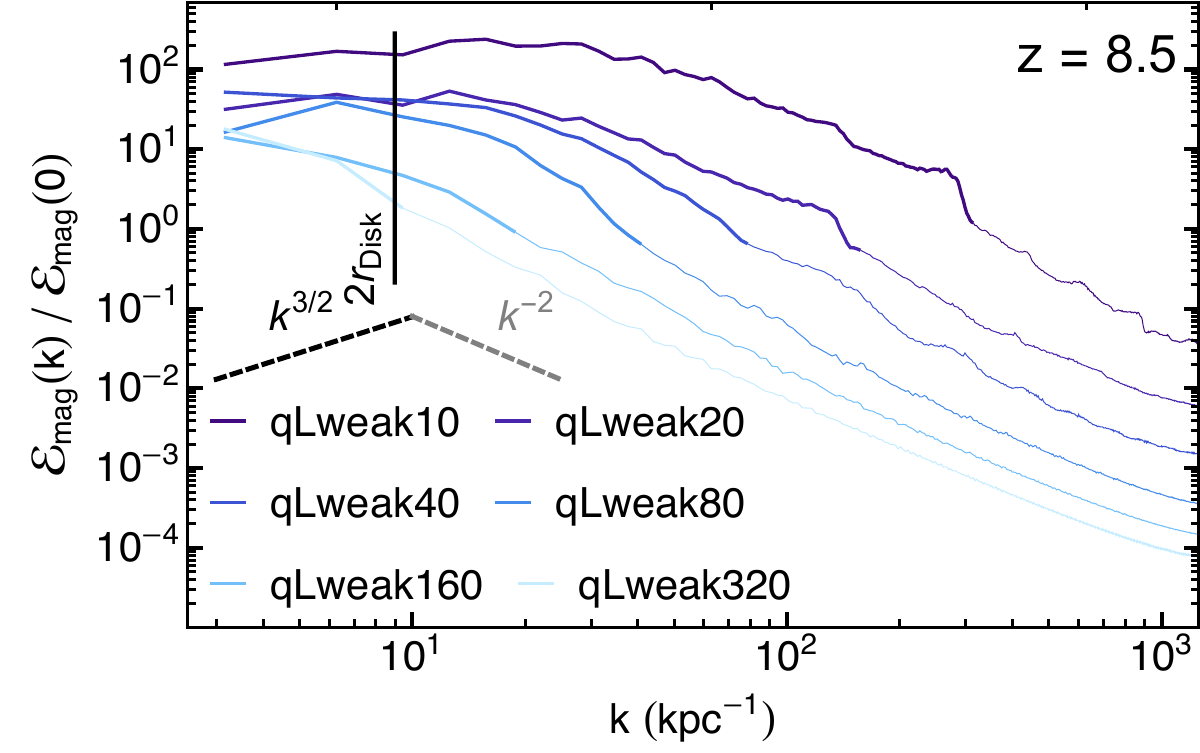}\\
    \caption{Galaxy magnetic energy spectra as in Fig.~\ref{fig:AllSpectra} (left column) and Fig.~\ref{fig:EmagSpectra}, but without employing 0-padding. The absence of 0-padding leads to flatter spectra towards larger scales, partially masking the inverse-cascade spectrum seen in Figs.~\ref{fig:AllSpectra} and~\ref{fig:EmagSpectra}.}
    \label{fig:0pad}
\end{figure*}
The inverse-cascade of power towards larger scales in magnetic energy spectra is a characteristic signature of turbulent dynamo amplification. An important consideration when exploring such spectra in cosmological simulations is whether to employ a zero-pad of the FFT domain \citep[as shown in the main text or by e.g.][]{Pakmor2017}. This padding will affect the shape of spectra at the largest scales of the transform. While the disk-like morphology of the galaxy provides some effective zero-padding, the periodicity assumed by the Fourier transform leads to a higher power at large scales that when the padding is included. Fig.~\ref{fig:0pad} shows our magnetic energy spectra at $z = 2$ (left column; Fig.~\ref{fig:AllSpectra}) and $z = 8.5$ (right column; Fig.~\ref{fig:EmagSpectra}) without the use of a zero-pad. The inverse-cascade appears more masked without a zero-pad. These spectra are obtained using the same physical regions as those presented in Section \ref{ss:SpectralStudy}, now using boxes with 1024 cells per side discretising the same region of interest (physical size $L_\text{FFT} \sim 2\, \left(0.2\,r_\text{DM}\right)$). Due to the impact observed due to zero-padding we recommend careful consideration when deciding whether a zero-pad is to be employed in the analysis of galaxies in cosmological simulations. 

\bsp	
\label{lastpage}
\end{document}